\documentclass[12pt]{article}
\usepackage{graphicx}
\usepackage{amsmath}
\allowdisplaybreaks
\usepackage{amssymb}
\usepackage{hyperref}

\usepackage{multirow}
\usepackage{cellspace}
\usepackage{caption}
\usepackage{mciteplus}
\graphicspath{{./}{./figures/}}

\newcommand{\MSbar}{\overline{\mbox{MS}}}

\newcommand{\ri}{\mbox{i}}
\newcommand{\HSESM}{HSESM}

\newcommand{\tb}{t_{\beta}}

\newcommand{\hl}{H_\mathrm{l}}
\newcommand{\hh}{H_\mathrm{h}}

\newcommand{\hlb}{H_\mathrm{l,B}}
\newcommand{\hhb}{H_\mathrm{h,B}}

\newcommand{\Mw}{M_W}
\newcommand{\Mz}{M_Z}
\newcommand{\mt}{M_t}
\newcommand{\Mhh}{M_{\hh}}
\newcommand{\Mhl}{M_{\hl}}

\newcommand{\ch}{c_{\tiny H}}
\newcommand{\chl}{c_{\tiny\hl}}
\newcommand{\chh}{c_{\tiny\hh}}

\begin{document}    

\begin{titlepage}
\noindent
\hfill December 22, 2022
\mbox{}

\vspace{0.5cm}
\begin{center}
\begin{Large}
\begin{center}
\begin{bf}
Electroweak corrections to {\boldmath{$g+g\to H_{l,h}$}} and 
{\boldmath{$H_{l,h} \to \gamma+\gamma$}} in the Higgs-singlet extension of 
the Standard model
\end{bf}
\end{center}
\end{Large}
  \vspace{0.8cm}

    \begin{large}
      Christian Sturm$\rm \, ^{a,\,}$%
      \footnote{\href{Christian.Sturm@physik.uni-wuerzburg.de}{Christian.Sturm@physik.uni-wuerzburg.de}}, 
      Benjamin Summ$\rm \, ^{a,\,}$%
      \footnote{\href{Benjamin.Summ@physik.uni-wuerzburg.de}{Benjamin.Summ@physik.uni-wuerzburg.de}} 
      and 
      Sandro Uccirati$\rm \, ^{b,\,}$%
      \footnote{\href{uccirati@to.infn.it}{uccirati@to.infn.it}}
  \end{large}
  \vskip .7cm
        {\small {\em 
           $\rm ^a$ 
            Universit{\"a}t W{\"u}rzburg,\\
            Institut f{\"u}r Theoretische Physik und Astrophysik,\\
            Lehrstuhl f{\"u}r Theoretische Physik II,\\
            Campus Hubland Nord,\\
            Emil-Hilb-Weg 22,\\
            D-97074 W{\"u}rzburg, \\
            Germany}}\\[0.2cm]
        {\small {\em 
            $\rm ^b$ 
            Universit{\`a} di Torino e INFN,\\
            10125 Torino,\\
            Italy}}\\
        \vspace{0.8cm}
\vspace*{4cm}
{\bf Abstract}
\end{center}
\begin{quotation}
\noindent
We calculate the next-to-leading order electroweak corrections for
Higgs-boson production in gluon fusion and the Higgs-boson decay into
two photons or gluons in the real Higgs-singlet extension of the
Standard model (HSESM). For the light Higgs-boson of the HSESM the
electroweak corrections for these processes are of the same order of
magnitude as in the Standard model. For the heavy Higgs-boson of the
HSESM the electroweak corrections can become large depending on the
considered scenario.
\end{quotation}
\end{titlepage}
%
%
\section{Introduction\label{sec:Introduction}}
The discovery of a Higgs boson at the Large Hadron
Collider~(LHC)~\cite{ATLAS:2012yve,*CMS:2012qbp} 
was a tremendous success and the beginning of detailed studies of its
properties. 
One important question is, whether the discovered Higgs boson is just
the Standard model~(SM) Higgs boson or whether it is part of a more
general Higgs sector. One of the simplest extensions of the SM Higgs
sector is the one where one adds an additional electroweak scalar singlet 
field~$S$ to the SM field
content~\cite{Silveira:1985rk,*McDonald:1993ex,Binoth:1996au,*Schabinger:2005ei,*Patt:2006fw,*Bowen:2007ia,*Barger:2007im}.

In general one can distinguish two types of Higgs-Singlet Extensions of
the SM~(\HSESM); the real singlet extension of the SM and the complex
singlet extension of the SM.
In the case of the real \HSESM\ one has one additional physical Higgs
boson~$H_h$ which we consider as heavy compared to the light observed Higgs
boson~$H_l$ with mass~$M_{H_l}\equiv M_h=125.25$~GeV~\cite{Zyla:2020zbs}. In the real
\HSESM\ one has only three new additional free parameters compared to the
SM, which can be expressed in terms of the new heavy Higgs-boson mass
$M_{H_h}$, the ratio of two vacuum expectation values, conventionally
denoted by $\tan\beta$, and a mixing angle~$\alpha$. The complex 
\HSESM\ can have in addition other two free parameters.
For certain configurations of these parameters the \HSESM\ can provide
candidates for dark matter.
In the following we will consider a real
scalar Higgs singlet $S$ which has a vacuum expectation value~(vev) as well as
a discrete $Z_2$ symmetry, $S\to-S$, so that terms which are odd in~$S$
do not appear in the potential. 
The 
\HSESM\ is being scrutinized by the ATLAS and CMS collaborations~\cite{
ATLAS:2015ciy,
*CMS:2015hra,
Aad:2019uzh} 
by deriving bounds and constraints on its new parameters.

Next-to-leading order~(NLO) electroweak~(EW) corrections to light and heavy
Higgs-boson production in Higgs strahlung and to Higgs-boson production
in vector-boson fusion as well as the NLO results on the
four-fermion~($f$) decays $H_{l,h}\to WW/ZZ\to 4f$ have been computed in
Refs.~\cite{Denner:2017vms,Altenkamp:2018bcs,Denner:2018opp}. 
Interference effects at the one-loop level for the $W^+W^-$ and $t\bar{t}$
decay modes with fully leptonic $WW$ decay have been studied in Ref.~%
\cite{Kauer:2019qei}. 
NLO electroweak corrections to the heavy-to-light Higgs-boson decay have
been determined in Ref.~\cite{Bojarski:2015kra}. 
The low energy behaviour of the \HSESM\ has been studied in
Ref.~\cite{Boggia:2016asg}.

Theoretical and experimental constrains and their impact
on the allowed parameter space as well as benchmark scenarios for
searches for an additional Higgs singlet have been studied in
Ref.~\cite{Robens:2021rkl,*Ilnicka:2018def,
  *Costa:2015llh,
  *Pruna:2013bma,Robens:2016xkb}.
Several benchmark scenarios have been summarized in the report of
the LHC Higgs Cross Section Working
Group~(HXSWG)~\cite{LHCHiggsCrossSectionWorkingGroup:2016ypw}.

Within this work we focus on the loop-induced Higgs-boson production and
decay processes. In the SM the complete electroweak corrections to
Higgs-boson production in gluon fusion and the Higgs-boson decay into
two photons are known since
long~\cite{Passarino:2007fp,Actis:2008ug,Actis:2008ts}. 
In the \HSESM\ theory predictions for cross sections of Higgs-boson
production via gluon fusion at higher order in perturbative
QCD can be obtained from the SM results, whereas higher order
electroweak corrections in this model are still unknown for this process.
In this paper we calculate the effect of the NLO electroweak
corrections on the production of a light and a heavy Higgs boson through
gluon~$(g)$ fusion, $g+g\to H_{l,h}$, in the real \HSESM.  Likewise
we calculate the NLO electroweak correction of the Higgs-boson decay
into two photons~($\gamma$) in the real \HSESM\ for the light and heavy
Higgs boson, $H_{l,h}\to\gamma+\gamma$, which are also still unknown. 
The few new parameters in the model under consideration will allow us then to provide even
scans over a wide range of the new input parameters, rather than
restricting ourselves to benchmark points only, which makes our results
more generally applicable, if further parameter regions will be
experimentally excluded. In addition we provide results for the
benchmark points collected in
Refs.~\cite{Robens:2016xkb,LHCHiggsCrossSectionWorkingGroup:2016ypw,Altenkamp:2018bcs},
which we will outline in more detail later in this work.

Electroweak corrections to loop-induced Higgs-boson production and decay 
processes can become large. This has been seen, for example,
in the calculation of the electroweak corrections to Higgs-boson
production through gluon fusion and the Higgs-boson decay into two
photons in a SM with a sequential fourth generation of heavy
fermions~\cite{Passarino:2011kv,Denner:2011vt}.  
Similarly the calculation of the two-loop, electroweak corrections to the
production of a light and a heavy neutral, scalar Higgs-boson through
the gluon fusion process in the Two-Higgs-Doublet
Model(2HDM)~\cite{Denner:2016etu,*Jenniches:2018xfn,Jenniches:2018zlb}
also showed that the corrections can be sizable.
The knowledge of the electroweak corrections in the \HSESM\ is thus
important. 
The outline of this paper is as follows. In Section~\ref{sec:model} we define
the \HSESM\ and its new parameters. 
Section~\ref{sec:calculation} contains the details of our calculations
and checks which we have performed.
In Section~\ref{sec:results} we present our results and discuss them.
Finally we close with our summary and conclusions in Section~\ref{sec:summary}.

\section{The Higgs-Singlet Extension of the Standard model\label{sec:model}}
The scalar potential of the Higgs-Singlet Extensions of the SM~(\HSESM),
which satisfies a $Z_2$ symmetry, $S\rightarrow -S$, is given by
\begin{align}
  \label{eq:thepotential}
  V_\text{\HSESM} = m_1^2 \Phi^\dagger \Phi
               + m_2^2 S^2+\frac{\lambda_1}{2}\left(\Phi^\dagger\Phi\right)^2
               +\frac{\lambda_2}{2}S^4
               +\lambda_3\Phi^\dagger \Phi S^2,
\end{align}
where we have adopted the conventions of Ref.~\cite{Denner:2017vms} with
the scalar doublet field~$\Phi$ and the scalar singlet field~$S$.
The parameters $m^2_1$, $m^2_2$ and $\lambda_i$ ($i=1,2,3$) of
the potential are all real.
For
\begin{equation}
  \label{eq:lambdaineq}
  \lambda_1>0,\quad \lambda_2>0\quad\text{and}\quad\lambda_{3}^2 < \lambda_1\*\lambda_2
\end{equation}
the potential has a global minimum with non-vanishing vacuum
expectation values~(vevs) of the scalar fields.
The last inequality follows from the requirement that the Hessian matrix
is positive definite at the extremum.
The Hessian matrix is
up to a global factor the mass (squared) matrix~$M^2_{ij}$ from
Eq.~(\ref{eq:massmatrix}) below. It is real symmetric and thus
orthogonal diagonalizable with real eigenvalues.\\
The Higgs doublet $\Phi$ and the Higgs singlet $S$ are parameterized as
\begin{equation}
  \label{eq:exparoundvev}
    \Phi = \begin{pmatrix}
    \phi^+ \\
    \frac{1}{\sqrt{2}}(v+\rho_1+\text{i}\eta)
    \end{pmatrix}, \qquad
    S=\frac{v_S+\rho_2}{\sqrt{2}},
\end{equation}
respectively, where $\eta$ and $\phi^{\pm}$  are the would-be
Goldstone-boson fields.  Here, $v$ and $v_S$ are vacuum expectation
values whose ratio is defined as $\tb\equiv\tan\beta = v_S/v$.
The limit $t_\beta\to0$ corresponds to the limit of a vanishing vev $v_S\to0$.
After spontaneous symmetry
breaking the real fields $\rho_1$ and $\rho_2$ mix to produce the mass
eigenstates $\hh$ and $\hl$ through a orthogonal matrix
\begin{align}
  \label{eq:mixinganglealpha}
    \begin{pmatrix}
    \rho_1 \\
    \rho_2
    \end{pmatrix} = \begin{pmatrix}
    \cos \alpha & -\sin \alpha \\
    \sin \alpha & \cos \alpha
    \end{pmatrix} 
    \begin{pmatrix}
    \hl \\ 
    \hh
    \end{pmatrix}.
\end{align}
This rotation to the mass eigenstates 
diagonalizes the mass (squared) matrix
\begin{equation}
  \label{eq:massmatrix}
    M^2_{ij} = \left.\frac{\partial^2 V_{\text{\HSESM}}}{\partial \rho_i \partial \rho_j}\right|_{\rho_{i,j}=0,\eta=0,\phi^\pm=0},\qquad (i,j=1,2),
\end{equation}
which has eigenvalues $\Mhl^2$ and $\Mhh^2$ that satisfy $\Mhl < \Mhh$. The
angle $\alpha$ is restricted to the interval $(-\pi/2,\pi/2]$. The real
  scalar fields satisfy the minimum conditions for the scalar potential
\begin{equation}
    \langle\rho_i\rangle=0,\quad(i=1,2).
\end{equation}
Using these conditions and the rotation into the basis of mass
eigenstates one can express the potential parameters through the
physical input parameters, i.e. the masses of the light and heavy Higgs
boson, $\Mhl$ and $\Mhh$ as well as the mixing angle~$\alpha$, the ratio
of the vevs $\tan\beta$ and the vev~$v$. The quartic couplings take the form
\begin{align}
    \label{eq:lambda1}
    \lambda_1 &= \frac{\Mhl^2}{v^2}\cos^2\alpha+\frac{\Mhh^2}{v^2}\sin^2\alpha,\\
    \label{eq:lambda2}
    \lambda_2 &= \frac{\Mhl^2}{v^2 \tan^2 \beta}\sin^2\alpha+\frac{\Mhh^2}{v^2 \tan^2 \beta}\cos^2\alpha,\\
    \label{eq:lambda3}
    \lambda_3 &= \frac{\Mhl^2-\Mhh^2}{2v^2\tan \beta}\sin (2\alpha).
\end{align}
Instead of the ratio of the vevs~$\tan\beta$ one can also use $\lambda_3$ as
an input parameter by expressing $\tan\beta$ in terms of
$\lambda_3$ with the help of Eq.~(\ref{eq:lambda3}).
As per usual the vev~$v$ is fixed
through its relation to the $W$-boson mass $\Mw$ and the weak isospin
gauge coupling $g$, which is the same in the \HSESM\ as in the SM
\begin{equation}
    \Mw = \frac{1}{2}gv.
\end{equation}
As can be seen from Eq.~\eqref{eq:lambda2} the coupling $\lambda_2$ is
quadratically enhanced(suppressed) for small(large) values of
$\tan \beta$. The coupling $\lambda_3$ is the only coupling of the Higgs
sector that depends on the sign of the mixing angle~$\alpha$.
If the mixing angle~$\alpha$ is negative(positive), the 
coupling~$\lambda_3$ of Eq.~(\ref{eq:lambda3}) is always positive(negative), since
$\Mhl<\Mhh$ and $\tan\beta>0$.
The inequalities in~(\ref{eq:lambdaineq}) are equivalent to requiring that the quadratic physical Higgs-boson 
masses are positive.
Inserting Eqs.~(\ref{eq:lambda1})-(\ref{eq:lambda2}) into the expressions for
$m_{1,2}$ derived from the extremal condition of the potential one finds 
\begin{eqnarray}
  \label{eq:m12}
  m_1^2&=&-{M^2_{\hl}\cos^2\alpha+M^2_{\hh}\sin^2\alpha\over2}
        +{M_{\hh}^2-M_{\hl}^2\over4}\sin(2\alpha)\tan\beta,\\
  \label{eq:m22}
  m_2^2&=&-{M^2_{\hh}\cos^2\alpha+M^2_{\hl}\sin^2\alpha\over2}+{M_{\hh}^2-M_{\hl}^2\over4}{\sin(2\alpha)\over\tan\beta}.
\end{eqnarray}
We have assumed to have two non-vanishing vevs, so that one can expect
that at least one of the two parameter $m_1^2$ and $m_2^2$ has to be
negative\footnote{
If we would assume that $m^2_1$ and $m^2_2$ are both positive and use 
$\lambda_3$ from Eq.~(\ref{eq:lambda3}), this would lead to 
\[
- 
\frac{(M_{\hh}^2-M_{\hl}^2)^{2}\sin^{2}(2\alpha)}
     {4v^{2}(M^2_{\hl}\cos^2\alpha+M^2_{\hh}\sin^2\alpha)}
<
\lambda_{3}
<
- \frac{M^2_{\hh}\cos^2\alpha+M^2_{\hl}\sin^2\alpha}{v^{2}}
\, .
\]
Multiplying this inequality by the positive factor 
$M_{\hl}^{2} \cos^{2}\alpha + M_{\hh}^{2} \sin^{2}\alpha$
one can see that the upper bound of $\lambda_3$ is smaller than the 
lower bound, leading to a contradiction.
},
however the other can also be positive.

Note that the doublet $\Phi$ couples to gauge bosons and fermions in
exactly the same way as in the SM and the only effect of the Higgs singlet is
that there is Higgs mixing. Therefore the tree-level couplings of $\hl$
and $\hh$ to gauge bosons and fermions are the same as those of the SM
Higgs boson, but scaled by the respective trigonometric function of the mixing
angle~$\alpha$. We also note that $\tb$ only appears in purely scalar
tree-level vertices, which are polynomials of degree at most two in
$1/\tb$. 
\section{Calculation\label{sec:calculation}}
\subsection{Generalities}
\begin{figure}[!ht]
  \begin{center}
    \begin{minipage}{7.8cm}
      \begin{center}
        \includegraphics[width=7.8cm]{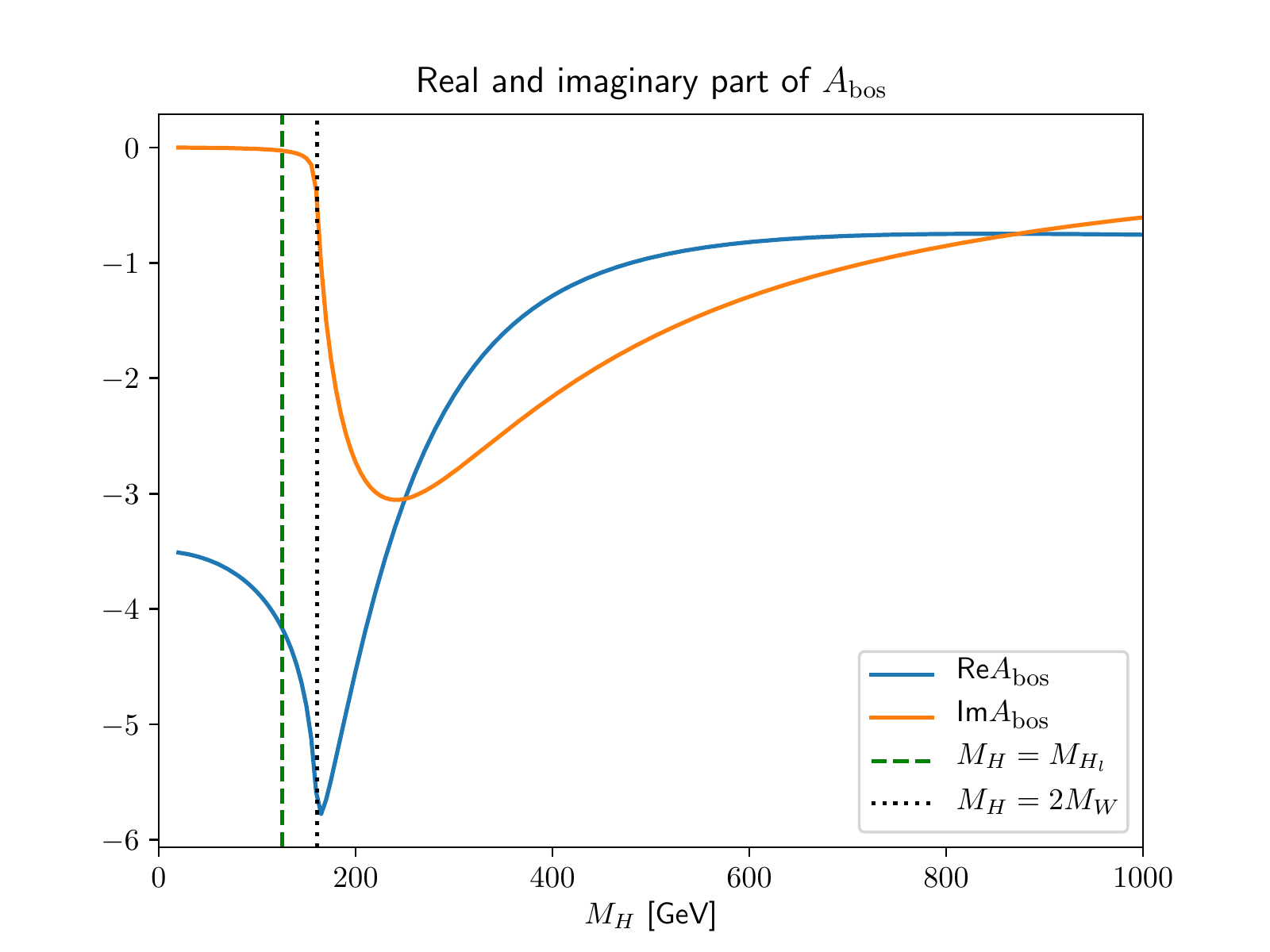}\\(a)
      \end{center}
    \end{minipage}
  \hspace*{0.1cm}
    \begin{minipage}{7.8cm}
      \begin{center}
        \includegraphics[width=7.8cm]{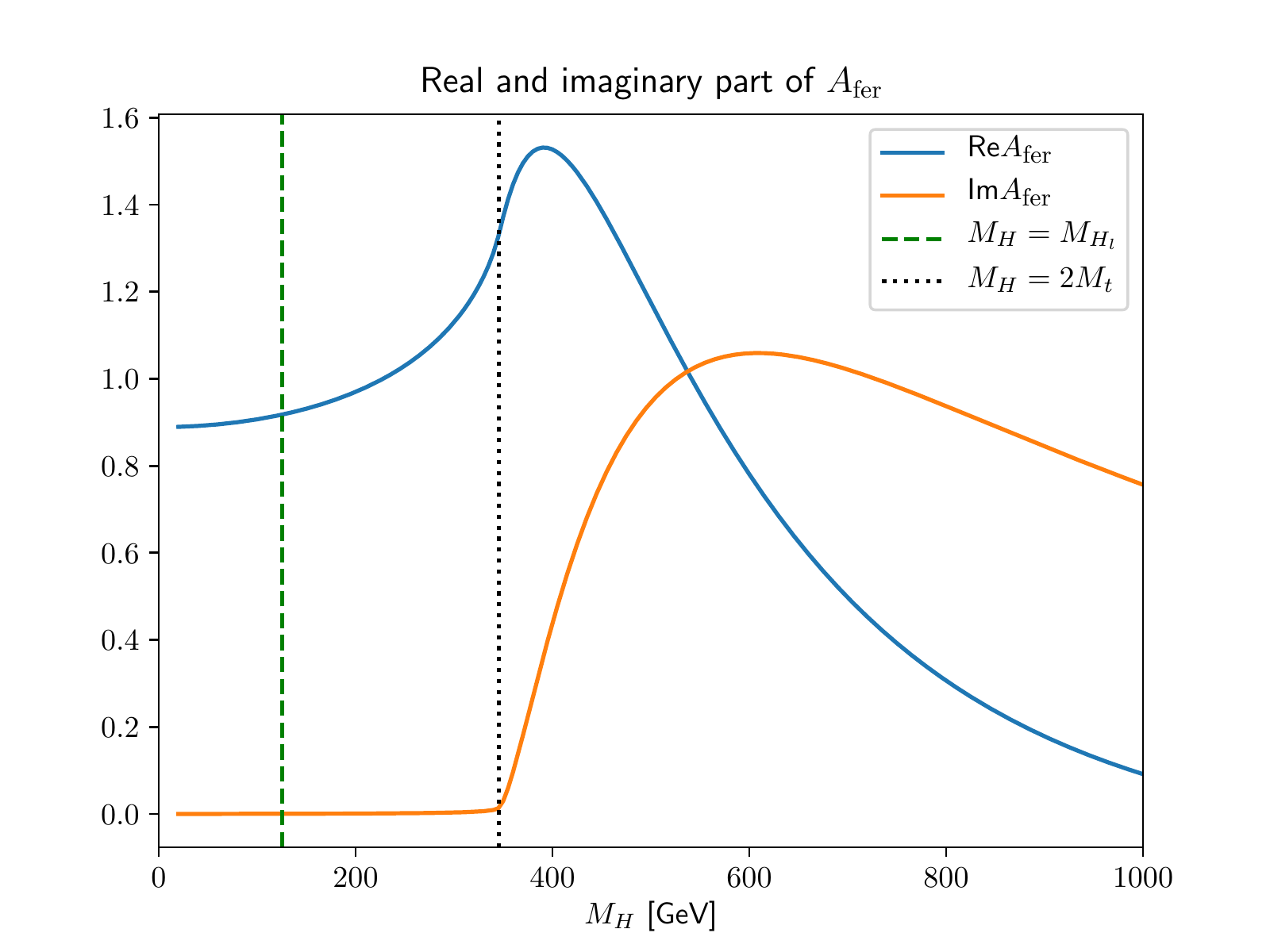}\\(b)
      \end{center}
    \end{minipage}
  \end{center}
  \caption{Real~(a) and imaginary~(b) part of the bosonic and fermionic
    contribution to $A^{\text{LO}}_{\gamma\gamma}$, respectively. The
    vertical dashed line denotes the location of the SM Higgs-boson
    $M_H=M_{H_l}=125.25$~GeV. 
      The vertical dotted lines indicate the location of the $WW$- and
      $t\bar{t}$-thresholds.  
    The dimensionless amplitudes are evaluated with complex
    masses as described in
    Section~\ref{sec:CalculationalDetails}.\label{fig:bosfermamps}}
\end{figure}

\begin{figure}[!ht]
  \begin{center}
    \begin{minipage}{7.8cm}
      \begin{center}
        \includegraphics[width=7.8cm]{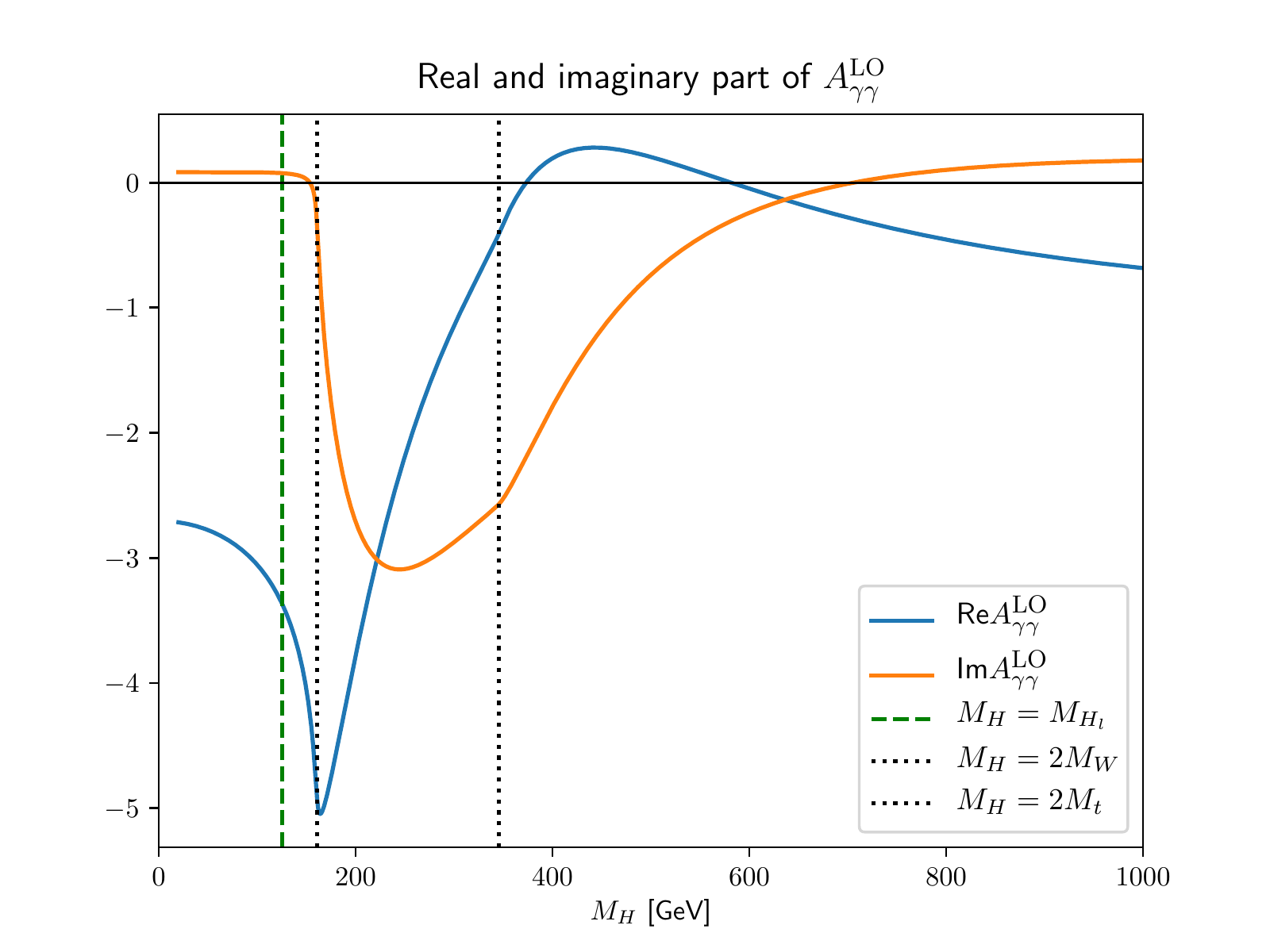}\\(a)
      \end{center}
    \end{minipage}
  \hspace*{0.1cm}
      \begin{minipage}{7.8cm}
      \begin{center}
        \includegraphics[width=7.8cm]{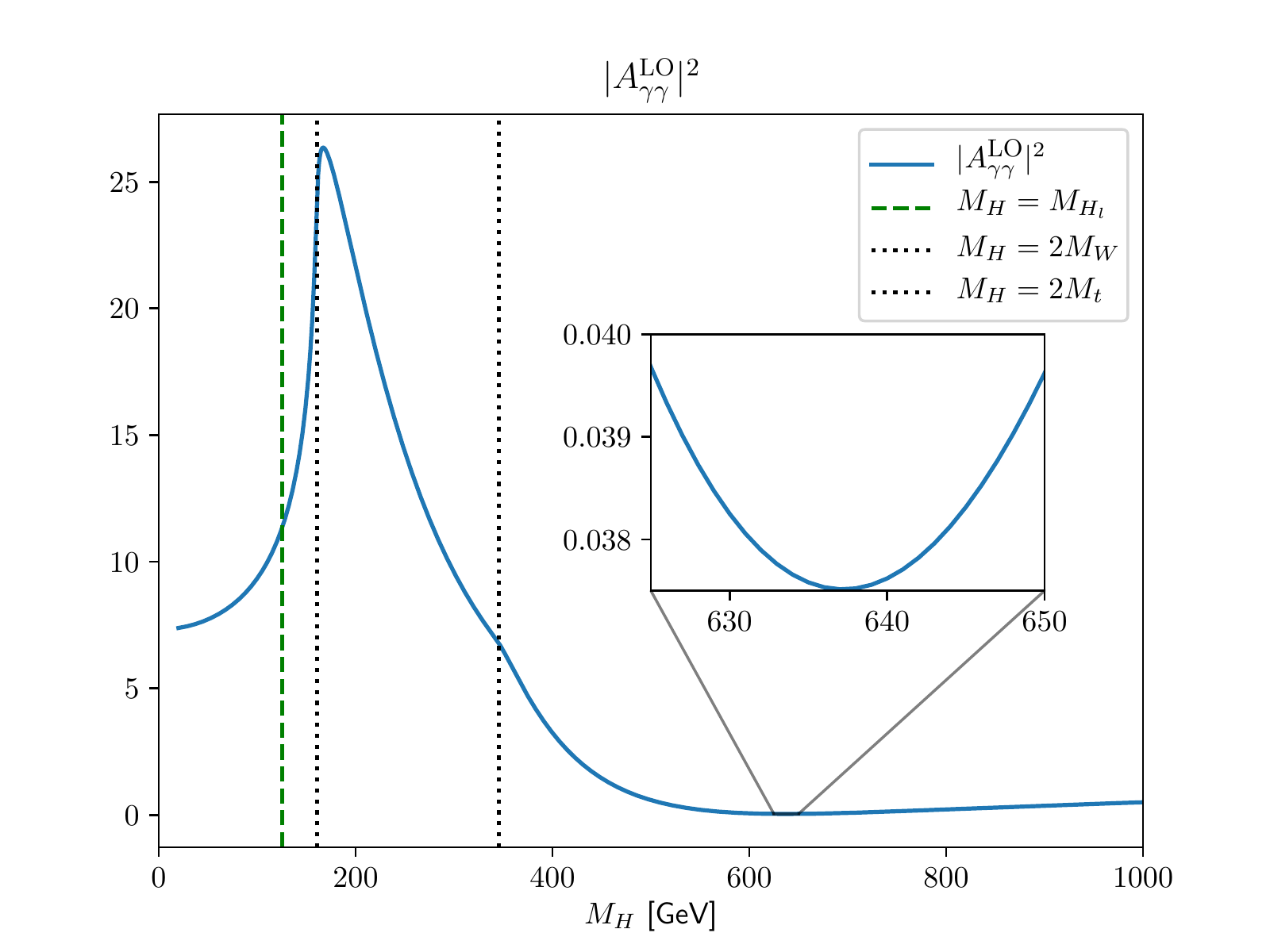}\\(b)
      \end{center}
    \end{minipage}
  \end{center}
    \caption{The real and imaginary part of the full dimensionless amplitude
      $A^{\text{LO}}_{\gamma\gamma}$ (left) and its squared modulus
      (right). The vertical dashed line denotes the location of the SM
      Higgs-boson $M_H=M_{H_l}=125.25$~GeV. 
      The vertical dotted lines indicate the location of the $WW$- and
      $t\bar{t}$-thresholds. 
      The dimensionless amplitudes are evaluated 
      with complex masses as described in
      Section~\ref{sec:CalculationalDetails}.\label{fig:ampsquared}} 
\end{figure}

\begin{figure}[!h]
  \begin{center}
    \begin{minipage}{7.8cm}
      \begin{center}
        \includegraphics[width=7.8cm]{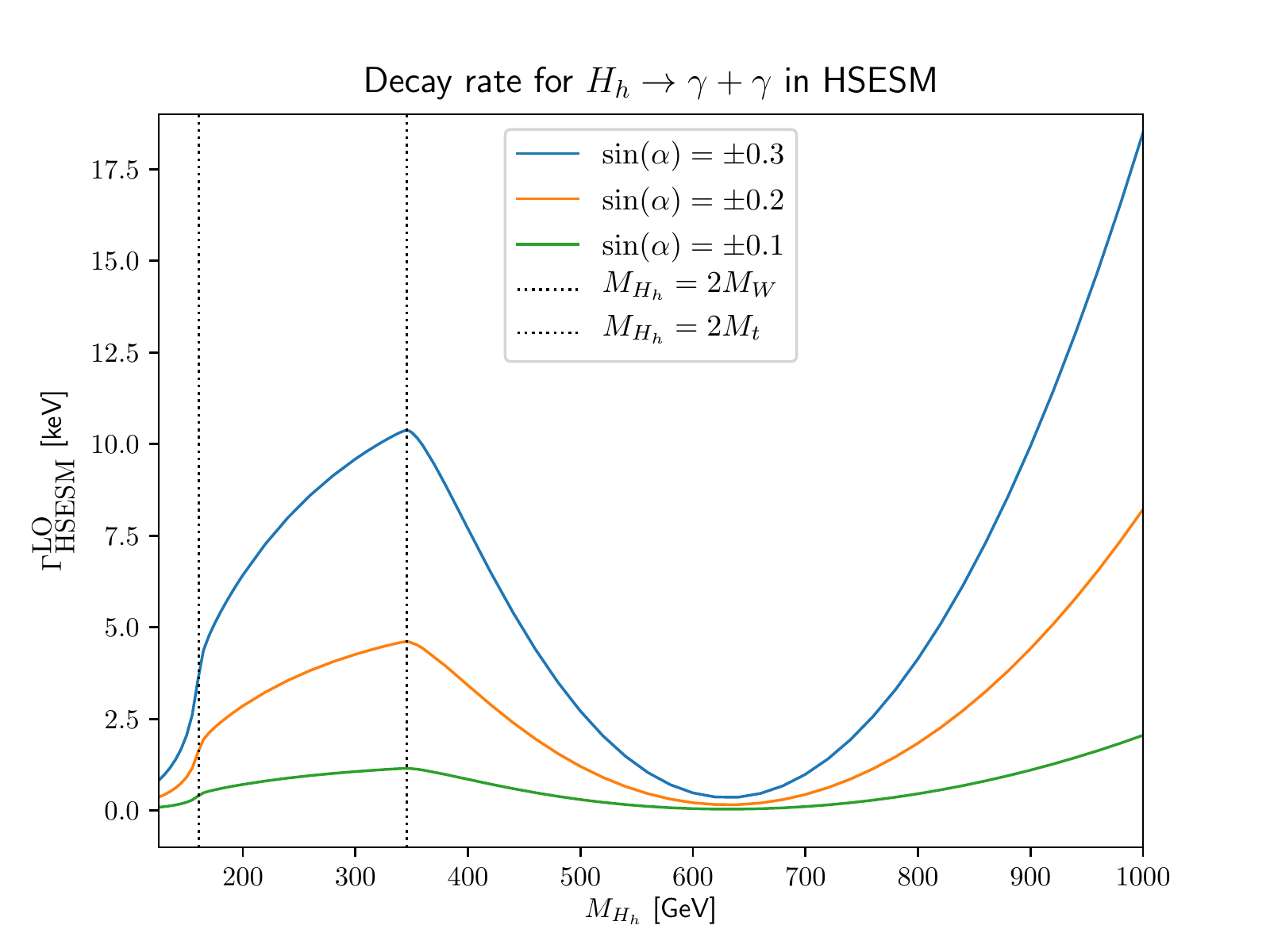}\\(a)
      \end{center}
    \end{minipage}
  \hspace*{0.1cm}
      \begin{minipage}{7.8cm}
      \begin{center}
        \includegraphics[width=7.8cm]{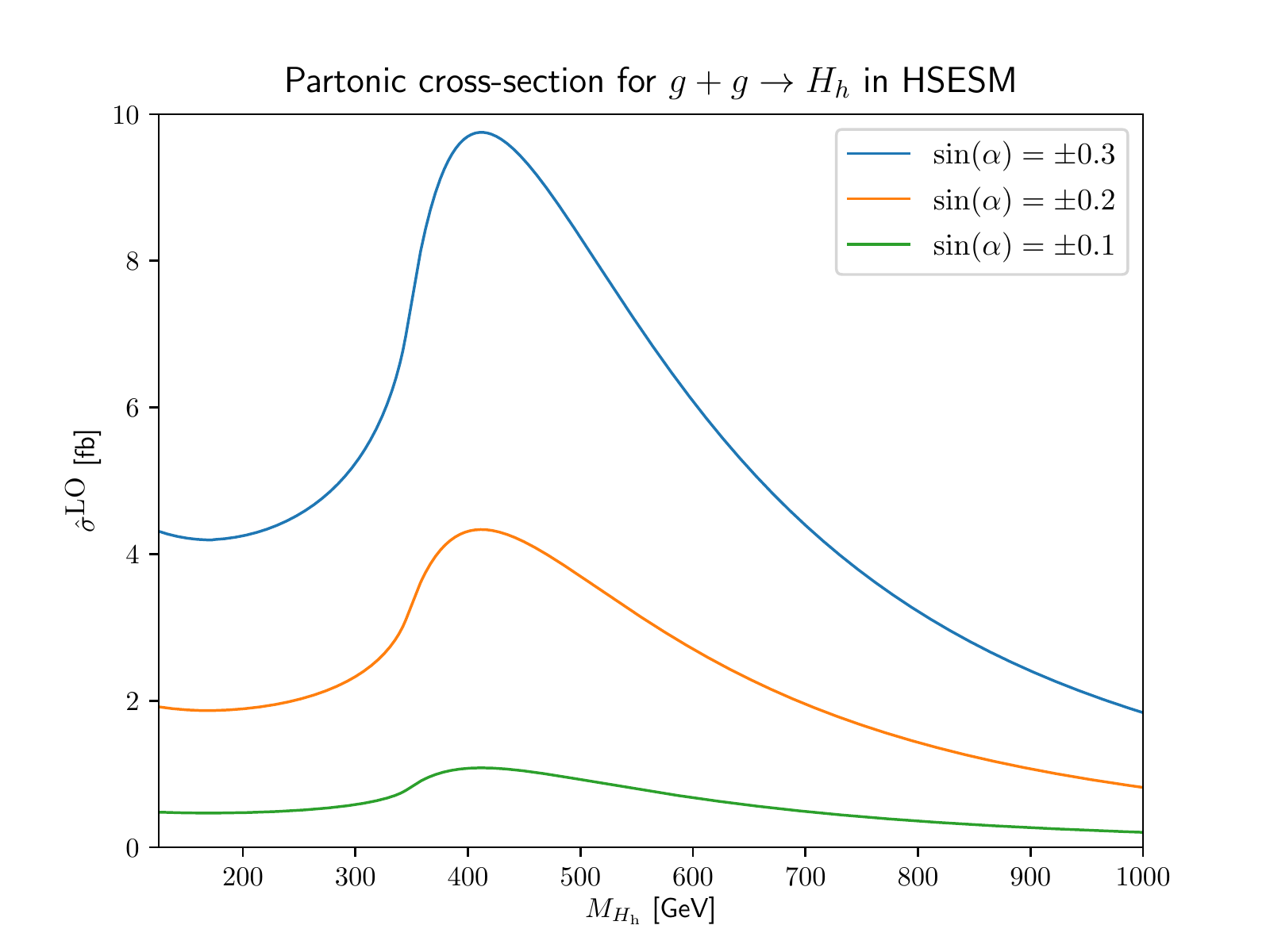}\\(b)
      \end{center}
    \end{minipage}
  \end{center}
    \caption{The LO partial decay width for $H_h \to \gamma+\gamma$ (a) and the
      LO partonic cross-section for $g+g \to H_h$
      (b).\label{fig:decaycross}}
\end{figure}

The leading-order~(LO) partial decay width~$\Gamma$ of a Higgs boson~$H$
of the \HSESM\ decaying into two photons~($\gamma$) is given by
\begin{equation}
  \Gamma^{\text{LO}}_{\text{\HSESM}}(H\to\gamma+\gamma)=
        {G_FM_H^3\alpha_{\text{em}}^2\over32\sqrt{2}\pi^3}\ch^2|A^{\text{LO}}_{\gamma\gamma}|^2,
\label{eq:decaywidth}
\end{equation}
where~$H$ stands here and in the following for either the light or the
heavy Higgs boson, $\hl$ or $\hh$; $G_F$ is the Fermi-coupling constant
and $\alpha_{\text{em}}$ the fine-structure constant. The coefficient $\ch$ is given by
\begin{equation}
  \label{eq:chl}
  \chl= \cos(\alpha),\qquad
  \chh= -\sin(\alpha),
\end{equation}
where the SM limit corresponds to $\sin\alpha\to0$, $\cos\alpha\to1$ and
is at LO independent of $\tan \beta$. Considering higher order
electroweak corrections the new Higgs sector decouples when sending in
addition $\tan \beta\to\infty$.
The dimensionless amplitude~$A^{\text{LO}}_{\gamma\gamma}$ can be
decomposed into a fermionic~$({\text{fer}})$ and bosonic~$(\text{bos})$
contribution $A^{\text{LO}}_{\gamma\gamma}=A_{\text{fer}}+A_{\text{bos}}$. The fermionic
contribution can again be subdivided into a part which arises from
leptons~$A_l$ and one which arises from quarks~$A_{q}$,
i.e. $A_{\text{fer}}=\sum_{l}Q_l^2A_l+N_c\sum_{q}Q_q^2A_{q}$, where
$Q_l$ and $Q_q$ are the electric charges of the fermions in units of the
elementary charge and $N_c$ is the number of colours. The sum runs over
all charged leptons~$l$ and quarks~$q$. We only consider the top quark
as massive and all other fermions as massless, so that only the term
containing $A_{\text{top}}$ will contribute to the dimensionless fermionic
amplitude~$A_{\text{fer}}$. Under these assumptions the
dimensionless fermionic and bosonic amplitudes read
\begin{eqnarray}
  \label{eq:Abos}
  A_{\text{bos}}(\tau_W)&=&
  -1-{3\over2\tau_W}\left[1+\left(2-{1\over\tau_W}\right)f(\tau_W)\right],\\
  A_{\text{top}}(\tau_t)&=&
  {1\over\tau_t}\left[1+\left(1-{1\over\tau_t}\right)f(\tau_t)\right],
  \label{eq:Atop}
\end{eqnarray}
with $\tau_p=M_H^2/(4M_p^2)$ ($p\in\{t,W\}$). The function~$f(\tau)$ is given by
\begin{equation}
\label{eq:ftau}
f(\tau_p)=
\begin{cases}
\arcsin^2\!\sqrt{\tau_p},& \mbox{if } \tau_p\leq 1,\\
-{1\over4}\left[\ln\left({1+\sqrt{1-1/\tau_p}\over1-\sqrt{1-1/\tau_p}}\right)-\ri\pi\right]^2,
& \mbox{if } \tau_p> 1.
\end{cases}
\end{equation}
The real and imaginary parts of the purely bosonic and fermionic dimensionless 
amplitudes of Eqs.~(\ref{eq:Abos}) and (\ref{eq:Atop}) have opposite signs in the whole
Higgs-boson mass range, see Fig.~\ref{fig:bosfermamps}.
The imaginary parts are zero below the $WW$ and $t\bar{t}$ threshold,
respectively, up to tiny contributions from the finite widths in the complex mass scheme.
\noindent
In particular for Higgs-boson masses above about $M_{H}\approx500$~GeV one can
observe strong cancellations between the dimensionless bosonic and fermionic
amplitude, 
so that the real and imaginary parts of the total dimensionless amplitude have modulus smaller than one, see
Fig.~\ref{fig:ampsquared}(a). 
Therefore the modulus squared of the total dimensionless amplitude is very small in this mass range, see Fig.~\ref{fig:ampsquared}(b).
Such cancellations, which lead to a small dimensionless LO amplitude, can cause the
relative NLO corrections to be huge, although the absolute size of the
NLO corrections is reasonable. We will observe exactly this situation in
Section~\ref{sec:gamgamHHSESM}.  A similar observation was made in
studying a sequential fourth generation of heavy
fermions~\cite{Denner:2011vt}.
In addition to these cancellations there is a minimum of the modulus squared of the dimensionless 
amplitude around $M_{H}\sim630-640$~GeV, see inset of
    Fig.~\ref{fig:ampsquared}(b), which is even more enhanced for the 
decay width of a heavy Higgs-boson, as can be seen in Fig.~\ref{fig:decaycross}(a).
The reason
for the enhancement is the overall factor of $M_H^3$ in
Eq.~(\ref{eq:decaywidth}) which 
causes the partial decay width to grow for large values of the heavy
Higgs-boson mass~$M_{\hh}$.
The leading-order cross section for the production of a Higgs boson~$H$
of the \HSESM\ through gluon fusion is given by
\begin{eqnarray}
\sigma_{\text{\HSESM}}^{\text{\tiny{LO}}}(g+g\to H)&=&
\frac{G_F M_H^2 \alpha_s^2}{128\sqrt{2}\pi} \, \ch^2 \,
|A^{\text{LO}}_{gg}|^2 \,\delta(s-M_H^2)\nonumber\\
&=&\hat{\sigma}^{\text{LO}}\,M_H^2\,\delta(s-M_H^2)\, ,
\label{eq:PartonicCrossSection}
\end{eqnarray}
with the strong coupling constant~$\alpha_s$ and the dimensionless amplitude
$A^{\text{LO}}_{gg}=A_{\text{top}}(\tau_t)$ from Eq.~(\ref{eq:Atop}).   
We consider again only the top-quark as a massive fermion and all
other quarks as massless.   Likewise the partial decay width for the
Higgs-boson decay into two gluons is given by 
\begin{equation}
  \label{eq:widthHgg}
  \Gamma_{\text{\HSESM}}^{\text{LO}}(H\to g+g)=
  \frac{G_F M_H^3 \alpha_s^2}{16\sqrt{2}\pi^3} \, \ch^2 \,
|A^{\text{LO}}_{gg}|^2 \,
.
\end{equation}
For the inclusion of the NLO electroweak corrections we write the dimensionless amplitudes as
\begin{equation}
A_z \; = \; A_z^{(1)} + g_W^{2} \* A_z^{(2)} + {\cal O}(g_W^{4}),
\qquad 
g_W^{2} \; = \; \frac{G_F M_W^2}{8\sqrt{2}\pi^2}
\, ,
\end{equation}
where $z$ denotes the process, either $z=\gamma\gamma$ or $z=gg$.  The
one-loop LO dimensionless amplitude is given by $A_z^{(1)}\equiv
A_z^{\text{LO}}$ and similarly $A_z^{(2)}$ is the two-loop dimensionless amplitude
containing the electroweak corrections. 
The NLO electroweak percentage corrections 
$\delta_{\text{EW}}$ are then given by
\begin{equation}
\label{eq:deltaEW}
\left|A_z\right|^2
\; = \;
\left|A_z^{(1)}\right|^2\left(1+\delta_{\text{EW}}\right)
\, ,
\qquad\mbox{with}\qquad
\delta_{\text{EW}}
\; = \;
\frac{2\,\text{Re}[g_W^{2} A_z^{(2)}\*A_z^{(1)\dagger}]}{|A_z^{(1)}|^2}
\, .
\end{equation}
Let us remark that in the determination of the electroweak percentage 
corrections of Eq.~(\ref{eq:deltaEW}) the overall factors of
Eq.~(\ref{eq:decaywidth}) and (\ref{eq:PartonicCrossSection}) cancel 
for both processes; in particular the overall Higgs-boson mass
dependence drops out. Hence~$\delta_{\text{EW}}$  as defined in
Eq.~(\ref{eq:deltaEW}) describes the electroweak percentage corrections
of the partial decay width and partonic cross section.
In Fig.~\ref{fig:decaycross} we show the LO partial decay
width of Eq.~(\ref{eq:decaywidth}) for the
process $H_h \to \gamma+\gamma$ as well as the partonic LO cross
section~$\hat{\sigma}^{\text{LO}}$ of Eq.~(\ref{eq:PartonicCrossSection}) for the 
process $g+g\to H_h$  as a function of the heavy Higgs-boson
mass for three different values of $\sin\alpha$ evaluated numerically
with the input parameters of Eqs.~(\ref{eq:inputparameters}).
For the process $g+g\to\hh$ we use a running strong coupling
constant~$\alpha_s$~\cite{Chetyrkin:2000yt}. In Fig.~\ref{fig:decaycross}(a) one can see a clear maximum at $\Mhh = 2\mt$ corresponding
to the $t\bar{t}$-threshold of the top-triangle and a kink at $\Mhh = 2M_W$, which corresponds to the $WW$-threshold present in
the bosonic contribution of the process~$\hh \to \gamma+\gamma$.

\subsection{Outline of the calculation\label{sec:CalculationalDetails}}
Feynrules~\cite{Alloul:2013bka,*Christensen:2008py} was employed to
generate all Feynman rules that are independent of gauge parameters, while 
the gauge-fixing vertices were implemented separately in the 
t'Hooft-Feynman gauge. 
All diagrams were generated using QGRAF~\cite{Nogueira:1991ex} and the resulting
expressions were manipulated using the in-house code QGS, which is based
on FORM~\cite{Vermaseren:2000nd,*Ruijl:2017dtg}. 
The program QGS is an extension of GraphShot(GS), already  used
for these same processes in the
SM~\cite{Passarino:2007fp,Actis:2008ug}.
QGS has been extended to work within
    the 2HDM~\cite{Jenniches:2018zlb} and now the \HSESM. 
The code, after performing the standard operations connected to the Dirac 
algebra with an anticommuting $\gamma_{5}$ (no anomalous diagrams are present 
for the processes under consideration), identifies the Lorentz structures of the 
amplitude by using projectors. This allows for a fast
computation of the amplitude by only considering those
Lorentz structures which enter in the squared amplitude of the physical
process at the end of the calculation. 
In a next 
step 
the obtained expressions are simplified by removing
reducible scalar products and using symmetries.  This allows to reduce
the amplitude to a combination of a small set of tensor integrals
contracted with external momenta, which can be considered as the master
integrals of the process under consideration.  The UV-divergent part of
these integrals is then extracted and canceled analytically against the
analogous part of the counterterms, which do not depend on the
specific renormalization scheme adopted. The inclusion of the finite part of the
counterterms, which we call finite renormalizations, will be discussed in
the next Section~\ref{sec:renormalization}. 
The obtained UV-finite amplitude still contains divergent integrals due to 
collinear singularities related to massless fermions. 
Since the electroweak corrections for both processes can not be divergent 
(there is no real emission which could compensate such singularities), we 
extract the divergent behaviour of each integral as logarithms in a small 
fictitious fermion mass and verify their cancellation 
analytically. 
The remaining loop integrals are then written in Feynman-parametric spaces in 
a form suited for numerical evaluation, following the techniques described 
in Refs.~\cite{Passarino:2002,*Ferroglia:2004,*Passarino:2006,Actis:2008ts}. 
The integrands of the obtained integrals are then collected in a Fortran 
library and the whole amplitude is evaluated numerically with the desired 
numerical accuracy using an in-house integrator based on the Korobov-Conroy 
number theoretic methods~\cite{korobov1957approximate,*korobov1963number,*conroy1967molecular}. 
Here we need to evaluate integrals up to dimension five numerically.

\subsection{Renormalization\label{sec:renormalization}}
For the renormalization procedure of the SM parameters of the two processes $g+g\to H_{l,h}$
and $H_{l,h}\to\gamma+\gamma$ we follow closely the SM case of Refs.~\cite{Passarino:2007fp,Actis:2008ts}.
For the renormalization of the $W$- and $Z$-boson masses as well as for
the top-quark mass we use the complex mass
scheme~\cite{Denner:1999gp,*Denner:2005fg,*Denner:2006ic}, where all
parameters of the theory which depend on $\Mw$, $\Mz$ and $\mt$
become complex.

Concentrating on the new parameters of the \HSESM\ discussed in 
Section~\ref{sec:model}, we see that there are the ratio of the vevs
$\tan\beta$ and the mixing angle~$\alpha$  which are not present in the SM. 
For a complete renormalization of the \HSESM\ both of these parameters have to be
renormalized. However, for the processes under consideration in this
work, it is sufficient to renormalize the mixing angle $\alpha$ since
$\tan\beta$ does not appear at LO.

A $\MSbar$ renormalization of these two parameters of the \HSESM\ requires a
proper treatment of the Higgs tadpoles in order to obtain
gauge-independent results for physical observables. The
Fleischer-Jegerlehner(FJ)-tadpole scheme was introduced for a consistent
treatment of the Higgs-tadpoles in the SM~\cite{Fleischer:1980ub}.  The
FJ tadpole scheme in the SM is equivalent to the $\beta_t$ scheme of
Ref.~\cite{Actis:2006ra}. The FJ tadpole scheme has been extended for a
general Higgs sector in Ref.~\cite{Denner:2016etu}, which we apply here
for the \HSESM. 
In Refs.~\cite{Dittmaier:2022maf,*Dittmaier:2022ivi} 
a new scheme for tadpole renormalization, dubbed gauge-invariant vacuum
expectation value scheme, was introduced being perturbatively stable. 
Its implementation is beyond the scope of the current work.

A $\MSbar$ renormalization of mixing angles can lead
to unnatural large corrections and can suffer from a large scale
dependence. Within this work we focus on the renormalization schemes for
the mixing angle~$\alpha$ in the \HSESM\ which were proposed in
Ref.~\cite{Denner:2018opp}. In particular we discuss three schemes
dubbed $ZZ$ scheme, $\overline{\Psi}\Psi$ scheme and OS scheme in the
following. 
First, the amplitudes for the decays of the
two Higgs-bosons of the \HSESM\ into two $Z$ bosons were used in
Ref.~\cite{Denner:2018opp} in order to
define a physical renormalization scheme. At LO they read
$\mathcal{M}^{(0)}_{H\to
  Z+Z}=-{c_H}M_We/(s_Wc_W^2)(\varepsilon^*_1\varepsilon^*_2)$, ($H=H_l$
or $H_h$) , where $e$ is
the elementary charge, $s_W$($c_W$) is the sine(cosine) of the weak
mixing angle, $\varepsilon_i$ ($i=1,2$) are the two polarization vectors of
the external $Z$ bosons and $c_H$ is defined in Eq.~(\ref{eq:chl}). After
all other parameters have been renormalized the renormalization
condition for the mixing angle~$\alpha$ requires that the ratio of
the two amplitudes 
\begin{equation}
   \label{eq:renormalizationconditionZZ}
   {\mathcal{M}_{H_h\to Z+Z}\over\mathcal{M}_{H_l\to Z+Z}}
   \stackrel{!}{=}
   -\frac{\sin(\alpha)}{\cos(\alpha)}
\end{equation}
is equal to its LO value~\cite{Denner:2018opp}, where we have adapted
the condition to the definition of the mixing angle used in our
work. Denoting bare quantities with a subscript $B$ the bare
mixing angle $\alpha_B$ is related to the renormalized one
through the counterterm $\delta \alpha$ as $\alpha_B = \alpha + \delta \alpha$.
Similarly, the bare and renormalized mass eigenstates of the heavy and
light Higgs boson fields of Eq.~(\ref{eq:mixinganglealpha}) are related
by the renormalization constants $\delta Z_{ij}$ ($i,j\in\{\hl,\hh\}$)
through 
\begin{equation}
  \label{eq:Higgsfieldrenorm}
  \left(
  \begin{matrix}
   \hhb \\ \hlb
  \end{matrix}
  \right)
  =
  \left(
  \begin{matrix}
    1+{1\over2}\delta Z_{\hh\hh}&{1\over2}\delta Z_{\hh\hl}\\
    {1\over2}\delta Z_{\hl\hh}&1+{1\over2}\delta Z_{\hl\hl}
  \end{matrix}
  \right)
  \left(
  \begin{matrix}
   \hh\\\hl
  \end{matrix}
  \right).
\end{equation}
Using the renormalization condition of Eq.~(\ref{eq:renormalizationconditionZZ}) the counterterm
of the mixing angle can be written as~\cite{Denner:2018opp}
\begin{equation}
  \label{eq:deltaalphaZZ}
  \delta \alpha = c_\alpha s_\alpha\left(\delta_{H_lZZ}-\delta_{H_hZZ}\right)
  +\frac{1}{2} c_\alpha s_\alpha \left(\delta Z_{\hl\hl}-\delta Z_{\hh\hh}\right)
  -\frac{1}{2}\left(\delta Z_{\hh\hl} s^2_\alpha-\delta Z_{\hl\hh} c^2_\alpha\right),
\end{equation}
where $c_\alpha = \cos(\alpha)$, $s_\alpha = \sin(\alpha)$ and $\delta_{H_iZZ}$
($i\in\{l,h\}$) are the unrenormalized relative one-loop corrections to the two
decays. In practical applications a choice of the polarization vectors has
to be made. If one considers the amplitudes of these two decays at LO
without the external polarization vectors $\varepsilon_i$, their Lorentz
structure consists only of the metric tensor. Considering higher order corrections
additional tensor structures arise, which depend on the external
momenta.
We have studied several physical renormalization schemes already in
Ref.~\cite{Jenniches:2018zlb} in the context of the 2HDM, where also the
Higgs boson decay into two $Z$ bosons was considered in a
renormalization condition. There we required as renormalization
condition that the coefficient of the metric tensor in the above tensor
decomposition of the amplitude is equal to its LO value.  Similarly as
in the 2HDM, we choose the polarization vectors of the external $Z$
bosons here in the \HSESM\ in such a way that all additional tensors
vanish and only the form factor of the metric tensor survives. We will
call this scheme $ZZ$ scheme in the following.

A second renormalization scheme was defined in
Ref.~\cite{Denner:2018opp} by adding a fermion singlet
field~$\Psi$ to the field content of the \HSESM\ and coupling it to
the singlet scalar through a Yukawa coupling which is sent to zero in order to recover the original theory.
As renormalization condition for the mixing angle~$\alpha$ one considers
here again the decays of the light and heavy Higgs bosons, this time
into a pair of the new singlet fields~$\Psi$, and again requires that the
ratio of the two amplitudes is equal to its LO value in the limit of
vanishing coupling:
\begin{equation}
  \label{eq:renormalizationconditionPsiPsi}
{\mathcal{M}_{H_h\to\Psi+\overline{\Psi}}\over\mathcal{M}_{H_l\to\Psi+\overline{\Psi}}}  
\stackrel{!}{=}{\over}
{\mathcal{M}^{(0)}_{H_h\to\Psi+\overline{\Psi}}\over\mathcal{M}^{(0)}_{H_l\to\Psi+\overline{\Psi}}}.  
\end{equation}
From this renormalization condition, the counterterm of the mixing angle
$\alpha$ can be calculated to be~\cite{Denner:2018opp}
\begin{equation}
  \label{eq:deltaalphaPsiPsi}
  \delta \alpha = -\frac{1}{2} c_\alpha s_\alpha \left(\delta Z_{\hl\hl}-\delta Z_{\hh\hh}\right)
  -\frac{1}{2}\left(\delta Z_{\hh\hl} c^2_\alpha-\delta Z_{\hl\hh} s^2_\alpha\right).
\end{equation}
This scheme will be denoted in the following as $\overline{\Psi}\Psi$
scheme. We note that, as opposed to the ZZ scheme, the counterterm of the
$\overline{\Psi}\Psi$ scheme does not depend on the unrenormalized one-loop corrections
of the defining decays. This is due to the fact that these vanish in the limit of vanishing
Yukawa coupling. 

The explicit counterterms for the mixing angle~$\alpha$ for the above
two schemes were already given in Ref.~\cite{Denner:2018opp}, which we adapted to
our conventions.
A drawback of the $ZZ$ and $\overline{\Psi}\Psi$ schemes is that they
introduce artificial thresholds for the processes $g+g\to H_l$ and
$H_l\to\gamma+\gamma$ when the heavy Higgs boson mass $M_{H_h}$ is equal to twice the
light Higgs boson mass $M_{H_l}$, as we will see explicitly in
Section~\ref{sec:results}. These thresholds originate from the 
wave function factor of the heavy Higgs boson, which enters the counterterm
of the mixing angle through the decay amplitude of the heavy Higgs boson. Interestingly,
the threshold singularities induced in this way, have opposite signs in the two schemes
as can be seen from Eqs.~(\ref{eq:deltaalphaZZ}) and (\ref{eq:deltaalphaPsiPsi}). This is
due to the fact that in the ZZ scheme the couplings are induced by $\rho_1$, which couples to
gauge bosons, whereas in the $\overline{\Psi}\Psi$ scheme the couplings are induced by $\rho_2$.

In the third and last scheme which we will consider in this paper the
counterterm~$\delta\alpha$ of the mixing angle~$\alpha$ is fixed through 
the non-diagonal field renormalization constants of the light and heavy
Higgs boson, avoiding the introduction of a threshold at $\Mhh = 2 \Mhl$ in
processes with a light external Higgs boson.
The counterterm~$\delta\alpha$ in this scheme is simply given by
\begin{equation}
  \label{eq:deltaalphaOS}
  \delta\alpha={\delta Z_{\hh\hl} -\delta Z_{\hl\hh}\over4}.
\end{equation}
This renormalization condition has been introduced for the 2HDM and the
\HSESM~\cite{Kanemura:2004mg,Denner:2018opp}. The non-diagonal field
renormalization constants can be expressed in terms of the non-diagonal,
on-shell Higgs-boson mixing self-energies in the background-field method.
We use here the conventional formalism which requires extra terms which
are given in appendix~B of Ref.~\cite{Denner:2018opp}. 
We will label this scheme in the following as on-shell(OS) scheme.

\subsection{Checks}
Several checks were performed to validate the set-up and the correctness
of the computation. The UV- and tadpole renormalization procedures were
validated by checking analytically that the amplitudes are UV-finite after 
renormalization. We implemented the FJ tadpole scheme for the \HSESM\
and verified that the physical counterterms are independent of gauge
parameters in a general $R_\xi$ gauge. The finite renormalization was
validated by checking that the dependence on the renormalization scale
$\mu$ cancels in the three renormalization schemes described earlier
in this section. The Feynman rules were checked by comparing the
amplitudes for the processes $H_l\to Z+Z$, $H_h\to Z+Z$ and 
$H_h\to H_l+H_l$ as well as for $g+g\to H_l$, $g+g\to H_h$, $H_l\to \gamma+\gamma$ and
$H_h\to\gamma+\gamma$ with Recola~\cite{Denner:2017vms,Denner:2017wsf}
at LO and at NLO for the non loop-induced processes. The processes $H_l\to Z+Z$ and $H_h\to Z+Z$
were compared in the $\MSbar$, OS and $\overline{\Psi}\Psi$ schemes, whereas the process $H_h\to H_l+H_l$
was checked in the $\MSbar$ scheme only, since it requires additional
renormalization of the parameter~$\tan\beta$ in the other schemes.
We checked that the amplitudes for the processes $H_l\to Z+Z$ and $H_h\to Z+Z$
fulfill indeed the renormalization condition of
Eq.~(\ref{eq:renormalizationconditionZZ}) at NLO in the $ZZ$ scheme. 
The appropriate IR behaviour of the amplitudes of the processes
$g+g\to\hl$, $\hl\to\gamma+\gamma$, $g+g\to\hh$
and $\hh\to\gamma+\gamma$ was guaranteed by checking that the
collinear singularities cancel as already described in Section~\ref{sec:CalculationalDetails}.
Furthermore, for all four processes it was checked that
the Ward identity is satisfied. 
For the processes $g+g\to\hl$ as well as $\hl\to\gamma+\gamma$
the corresponding amplitudes in the \HSESM\ were found to agree analytically with the
corresponding SM amplitudes in the SM limit of the \HSESM.
We checked our numerical integration by scaling all integrals through the 
introduction of a scaling parameter~$s$ 
for all four processes. The scaling parameter~$s$ cancels in the total
amplitude. The value of the part of the amplitude which contains
integrals of the same integration dimension, however, depends on the specific choice of $s$. These differences cancel in the sum
of all integration dimensions. We checked that the numerical value of the total amplitude is not affected by varying the value
of~$s$.

\section{Results and discussion\label{sec:results}}
In this section, we present the numerical results of phenomenologically
interesting scenarios for light and heavy Higgs-boson production in gluon 
fusion and for the Higgs-boson decays into two photons. 
First, we consider the benchmark points~(BPs) of Tab.~\ref{tab:BP}. 
The BPs are taken from the LHC Higgs cross section working
group report~\cite{LHCHiggsCrossSectionWorkingGroup:2016ypw} and 
were proposed originally in Ref.~\cite{Robens:2016xkb}.  We have adapted
the input parameters of the BPs in Tab.~\ref{tab:BP} to our conventions of
Section~\ref{sec:model}, i.e.\ in Ref.~\cite{Robens:2016xkb}
the mixing angle $\alpha$ has opposite sign and $\tan\beta$
is given by the reciprocal value compared to Tab.~\ref{tab:BP}. 
\begin{table}[!h]
  \centering
\begin{tabular}{|c||c|c|c|}
  \hline
   BP                   & $M_{\hh}$ [GeV]& $\sin(\alpha)$ & $1/\tan(\beta)$\\\hline\hline
  BHM200$\mp$\phantom{a}&  200          &$\mp$0.29&1.19 \\\hline
  BHM300$\mp$\phantom{a}&  300          &$\mp$0.31&0.79 \\\hline
  BHM400a$\mp$          &  400          &$\mp$0.26&0.58 \\\hline
  BHM400b\phantom{$\pm$}&  400          &$  +$0.26&0.59 \\\hline
  BHM500a$\mp$          &  500          &$\mp$0.24&0.46 \\\hline
  BHM500b\phantom{$\pm$}&  500          &$  +$0.24&0.47 \\\hline
  BHM600$\mp$\phantom{a}&  600          &$\mp$0.22&0.38 \\\hline
  BHM700a$\mp$          &  700          &$\mp$0.21&0.31 \\\hline
  BHM700b\phantom{$\pm$}&  700          &$  +$0.21&0.32 \\\hline
  BHM800a$\mp$          &  800          &$\mp$0.20&0.25 \\\hline
  BHM800b\phantom{$\pm$}&  800          &$  +$0.20&0.27 \\\hline
\end{tabular}
  \caption{The benchmark points~(BP) of
    Refs.~\cite{LHCHiggsCrossSectionWorkingGroup:2016ypw,Robens:2016xkb}
    are shown in our conventions. The mass of the heavy Higgs boson is
    encoded in the BP's name. The BPs which have the same heavy
        Higgs-boson mass, but the name differs by the
    suffix a or b have the same (positive) value of $\sin\alpha$ and are
    distinguished by a tiny difference in the value of $\tan\beta$.
    \label{tab:BP}}
\end{table}\\
Furthermore we show various plots which describe the impact of the electroweak corrections 
in terms of the mass of the heavy Higgs boson~$M_{H_{h}}$ for different values of 
$\sin\alpha$ and $\tan\beta$ in the OS scheme. 
First results for the renormalization of the mixing
    angle~$\alpha$ in the $ZZ$-scheme were presented in Ref.~\cite{7254998}.
Finally plots with the comparison of the three renormalization schemes
for the mixing angle~$\alpha$ are shown in dependence of the heavy Higgs-boson
mass~$M_{H_{h}}$, for specific values of $\sin\alpha$ and $\tan\beta$.

For the numerical evaluation we use the following input parameters from
the PDG~\cite{Zyla:2020zbs} for the particle masses and their total decay widths:
\begin{align}
  \mt&=172.76\text{ GeV},& \Mz&=91.1876\text{ GeV},& \Mw&=80.379\text{ GeV},
  \nonumber\\
  \Gamma_t&=1.42\text{ GeV},& \Gamma_Z&=2.4952\text{ GeV},& \Gamma_W&=2.085\text{ GeV},
  \nonumber\\
  \Mhl&=125.25\text{ GeV},& \Gamma_{H_l}&={0.0032\text{ GeV}},& G_F&=1.1663787\cdot10^{-5}\text{ GeV}^{-2}, \nonumber \\
  \alpha_s(\Mz) &= 0.1179,& \alpha_{\text{em}}&= 1/137.035999084.
  \label{eq:inputparameters}              
\end{align}
We use the complex mass 
scheme~\cite{Denner:1999gp,*Denner:2005fg,*Denner:2006ic} for the 
renormalization of the $W$- and $Z$-boson masses as well as for the top 
quark 
mass. 
For the two processes which have
    the heavy Higgs boson as an external particle, i.e.\ for heavy
    Higgs-boson production in gluon fusion and for the heavy Higgs-boson
    decay into two photons, the light Higgs boson appears only as an
    internal particle and is thus also renormalized in the complex mass
    scheme.

\subsection{Higgs-boson production in gluon fusion in the \HSESM\label{sec:glugluHHSESM}}
\begin{figure}[!hb]
\begin{center}
\includegraphics[width=8.3cm]{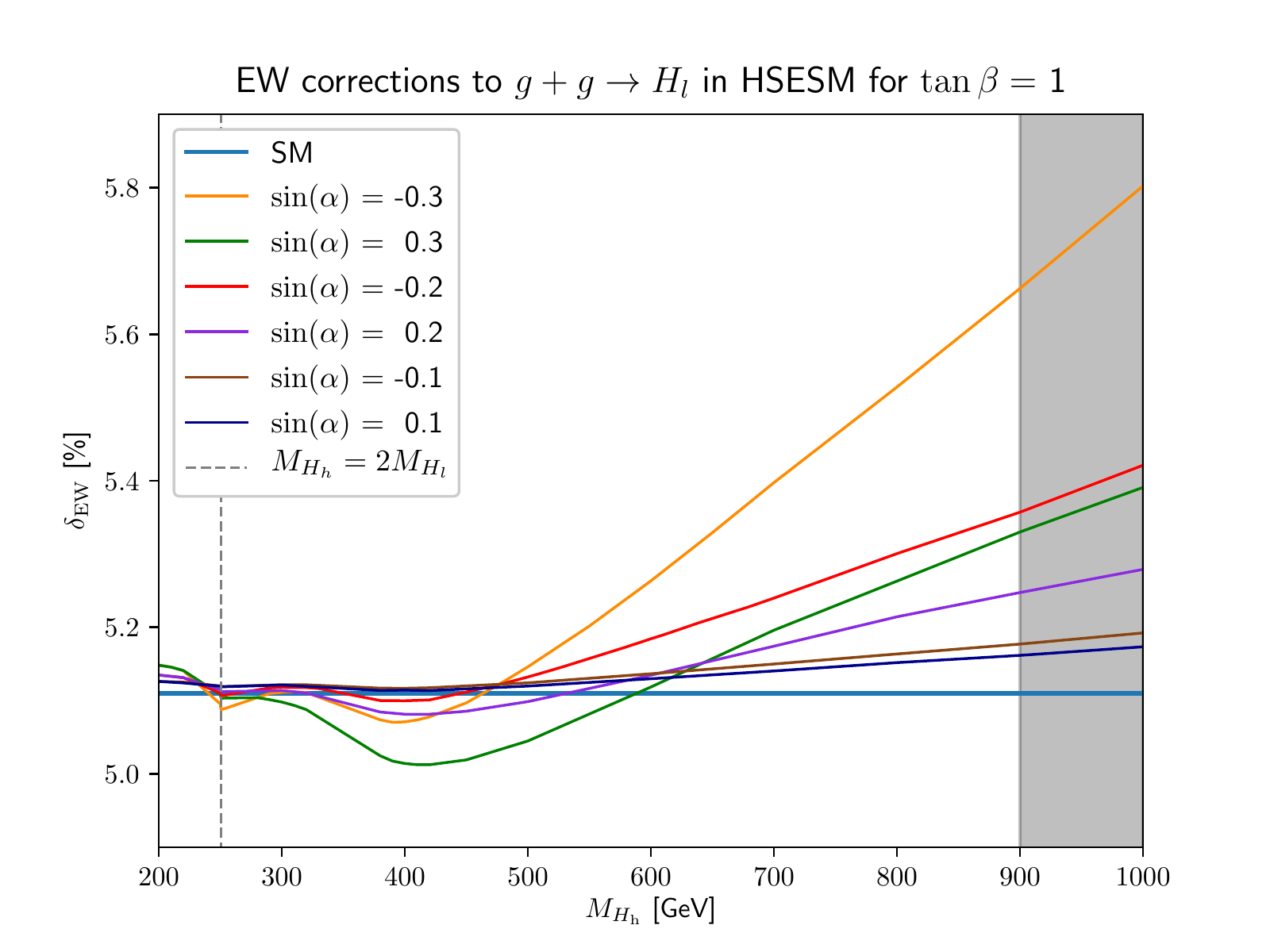}\\
\includegraphics[width=8.3cm]{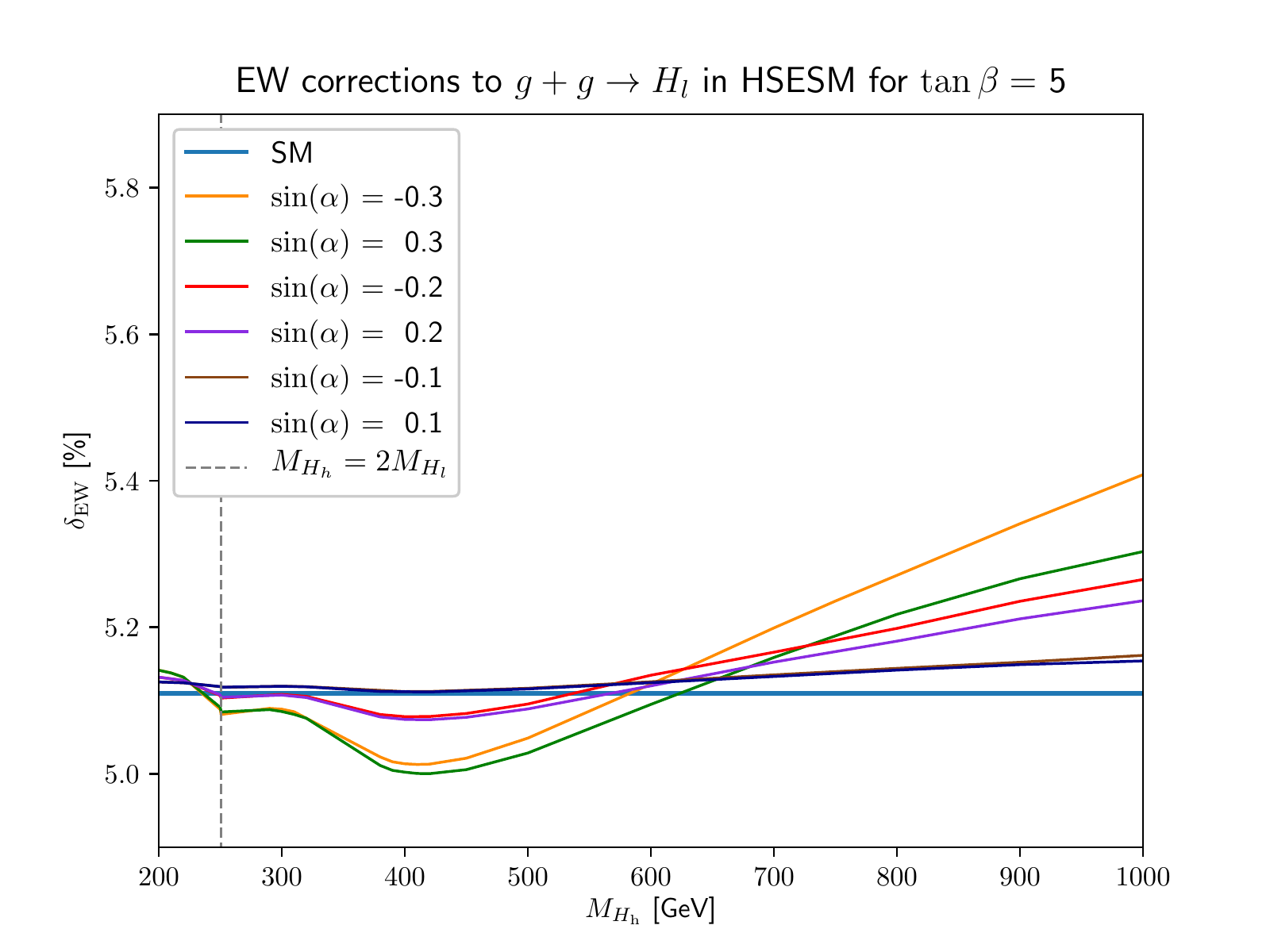}
\hspace*{-0.85cm}
\includegraphics[width=8.3cm]{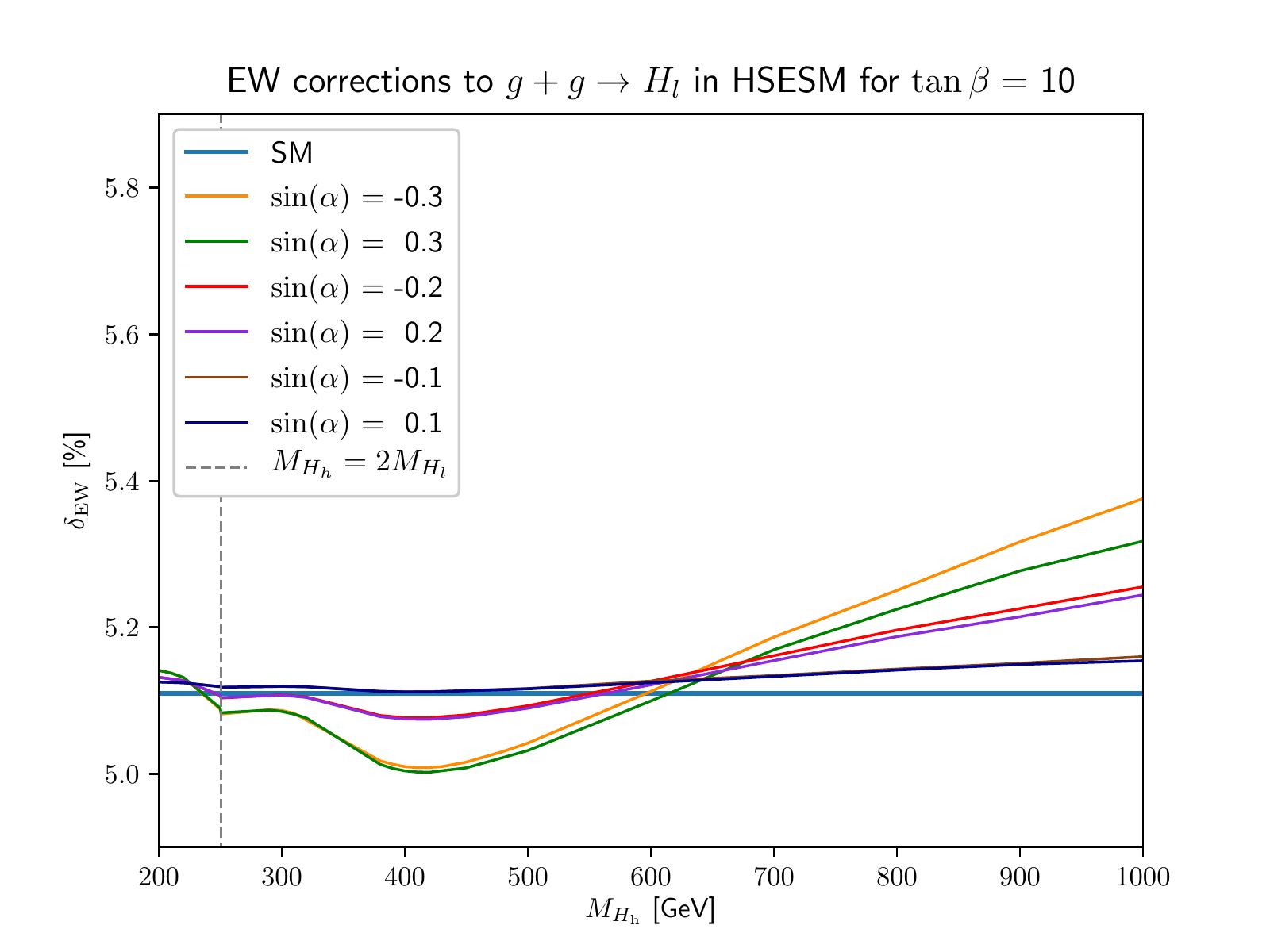}
\end{center}
\caption{Relative NLO electroweak percentage
  corrections~$\delta_{\text{EW}}$ in the OS scheme to the LO cross section
  of the process $g+g \to H_l$ in the
  \HSESM\ for $\tan \beta = 1,5,10$ with varying values of $\sin
  \alpha$. The SM limit is shown in blue.%
  \label{fig:Hgg_fixed_tb}}
\end{figure}

Choosing as reference renormalization scheme for the mixing
angle~$\alpha$ the OS scheme as defined in 
Section~\ref{sec:renormalization}, we show in
Fig.~\ref{fig:Hgg_fixed_tb} %
the dependence of the electroweak percentage
corrections~$\delta_{\text{EW}}$ defined in Eq.~(\ref{eq:deltaEW}) on
$M_{H_{h}}$ for the production of the light Higgs boson at 6 values of
$\sin\alpha$ for each plot.
The values of $\sin\alpha$ are chosen to cover the allowed region of
Ref.~\cite{Aad:2019uzh}. 
The three
plots of Fig.~\ref{fig:Hgg_fixed_tb} are for the three values of
$\tan\beta=1,5$ and~$10$.

\noindent
The grey shaded area in the plot for $\tan\beta=1$
indicates the onset of the non-perturbative regime, where at least one
of the parameters~$\frac{\lambda_i}{4\pi}$ ($i=1,2$ or $3$) of
Eq.~(\ref{eq:thepotential}) becomes larger than one. This only happens for small values of
$\tan\beta$, since the latter appears in the denominator of
Eqs.~(\ref{eq:lambda2}) and~(\ref{eq:lambda3}). The vertical dashed line indicates the location
where the heavy Higgs-boson mass is twice the light Higgs-boson mass.
All three plots of Fig.~\ref{fig:Hgg_fixed_tb}
show the same numerical range for the percentage correction and the heavy
Higgs-boson mass in order to allow for a better comparison of the results for the
three different values of $\tan\beta$. 
The electroweak corrections are very close to the ones in the SM (blue 
line), in the range of $5.0\%-5.8\%$, showing the largest deviations at high 
values of $M_{H_{h}}$, small values of $\tan\beta$ and large absolute 
values of $\sin\alpha$. 
The behaviour of $\delta_{\text{EW}}$ shows a small minimum around 
$M_{H_{h}}\approx400$~GeV and grows almost linearly afterwards. 
For $\tan\beta=1$ the corrections are a little more enhanced than at 
$\tan\beta=5$ and remain basically unchanged at larger values of 
$\tan\beta$. 
An interesting feature which we derive from the plots is the negligible dependence 
of the electroweak corrections on the sign of $\sin\alpha$ for large values of 
$\tan\beta$, revealing that the odd contributions in $\sin\alpha$ are 
$\tan\beta$-suppressed. This can be seen best in the last plot of
Fig.~\ref{fig:Hgg_fixed_tb} (for $\tan\beta=10$) where the lines for
$\pm\sin\alpha$ approach each other and almost coincide. This feature can also be
verified by inspecting the analytical expressions for the corrections.

For the BPs of Tab.~\ref{tab:BP} we show the electroweak percentage
corrections in Tab.~\ref{tab:BPsggHl} for the process~$g+g\to\hl$. The
    mixing angle~$\alpha$ has been renormalized in the OS scheme.
    All BPs are very close to the SM result. For high values of the
  heavy Higgs boson mass (larger than 600~GeV) one can see a small
  increase of the percentage correction compared to the SM result.
   Considering BPs with a heavy Higgs-boson mass close to the minimum of
   Fig.~\ref{fig:Hgg_fixed_tb} one can also see here a small decrease of
   the percentage correction. 
\begin{table}[ht]
  \centering
  \begin{tabular}{lcr}
  \begin{tabular}{|c||c|}
    \hline
BP                    &$\delta_{\text{EW}}$\\\hline
BHM200$-$\phantom{a}&  $5.2$            \\\hline
BHM200$+$\phantom{a}&  $5.1$                  \\\hline
BHM300$-$\phantom{a}&  $5.1$                  \\\hline
BHM300$+$\phantom{a}&  $5.1$                  \\\hline
BHM400a$-$          &  $5.1$                  \\\hline
BHM400a$+$          &  $5.0$                  \\\hline
    \end{tabular}
    &
  \begin{tabular}{|c||c|}
    \hline
BP                    &$\delta_{\text{EW}}$\\\hline
BHM400b\phantom{$\pm$}&  $5.0$          \\\hline
BHM500a$-$            &  $5.1$                \\\hline
BHM500a$+$            &  $5.1$                \\\hline
BHM500b\phantom{$\pm$}&  $5.1$                \\\hline
BHM600$-$\phantom{a}  &  $5.1$                \\\hline
BHM600$+$\phantom{a}  &  $5.1$                \\\hline
    \end{tabular}
    &
  \begin{tabular}{|c||c|}
    \hline
BP                    &$\delta_{\text{EW}}$\\\hline
BHM700a$-$            &    $5.2$        \\\hline
BHM700a$+$            &    $5.2$              \\\hline
BHM700b\phantom{$\pm$}&    $5.2$              \\\hline
BHM800a$-$            &    $5.2$              \\\hline
BHM800a$+$            &    $5.2$              \\\hline
BHM800b\phantom{$\pm$}&    $5.2$              \\\hline
    \end{tabular}
  \end{tabular}   
  \caption{The electroweak percentage
    corrections~$\delta_{\text{EW}}$~[\%] in the OS scheme are shown for the benchmark
    points of Tab.~\ref{tab:BP}.\label{tab:BPsggHl}}
\end{table}
The electroweak percentage corrections which are provided in
Fig.~\ref{fig:Hgg_fixed_tb} and Tab.~\ref{tab:BPsggHl} also apply to the
partial decay width of the process
$\hl\to g+g$ due to Eq.~(\ref{eq:widthHgg}).\\

\begin{figure}[!h]
\begin{center}
\includegraphics[width=8.30cm]{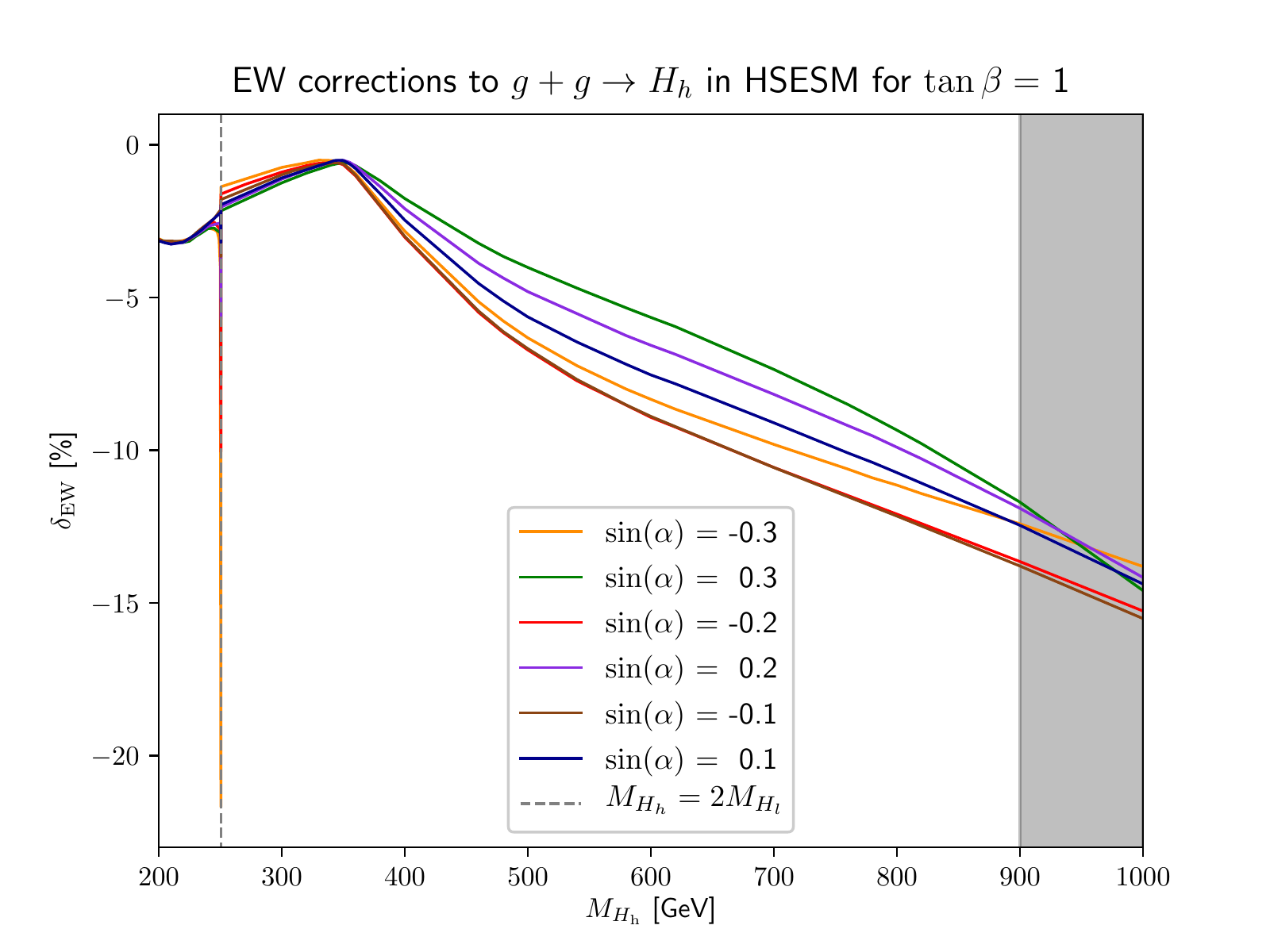}
\\
\includegraphics[width=8.30cm]{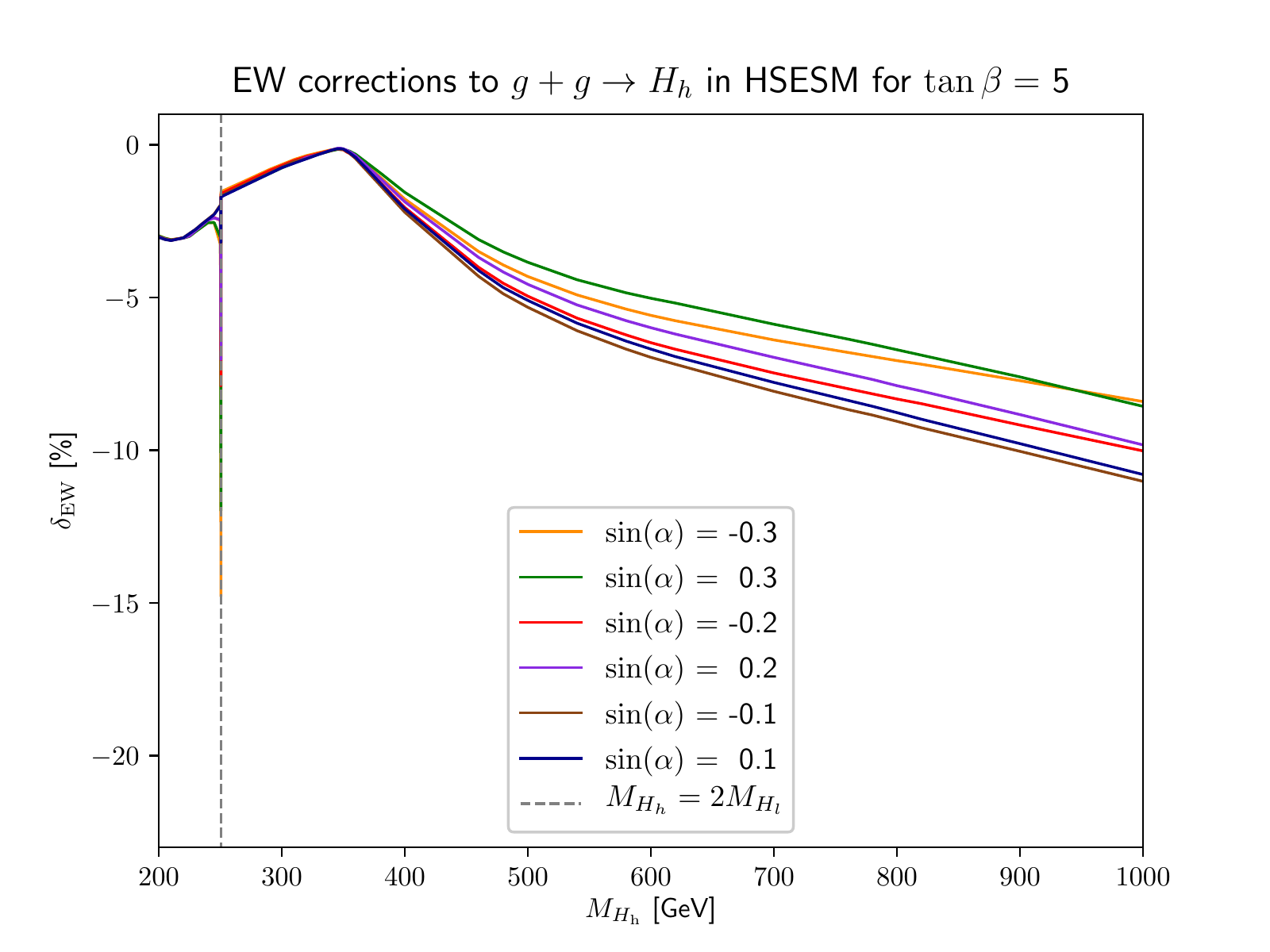}
\hspace*{-0.85cm}
\includegraphics[width=8.30cm]{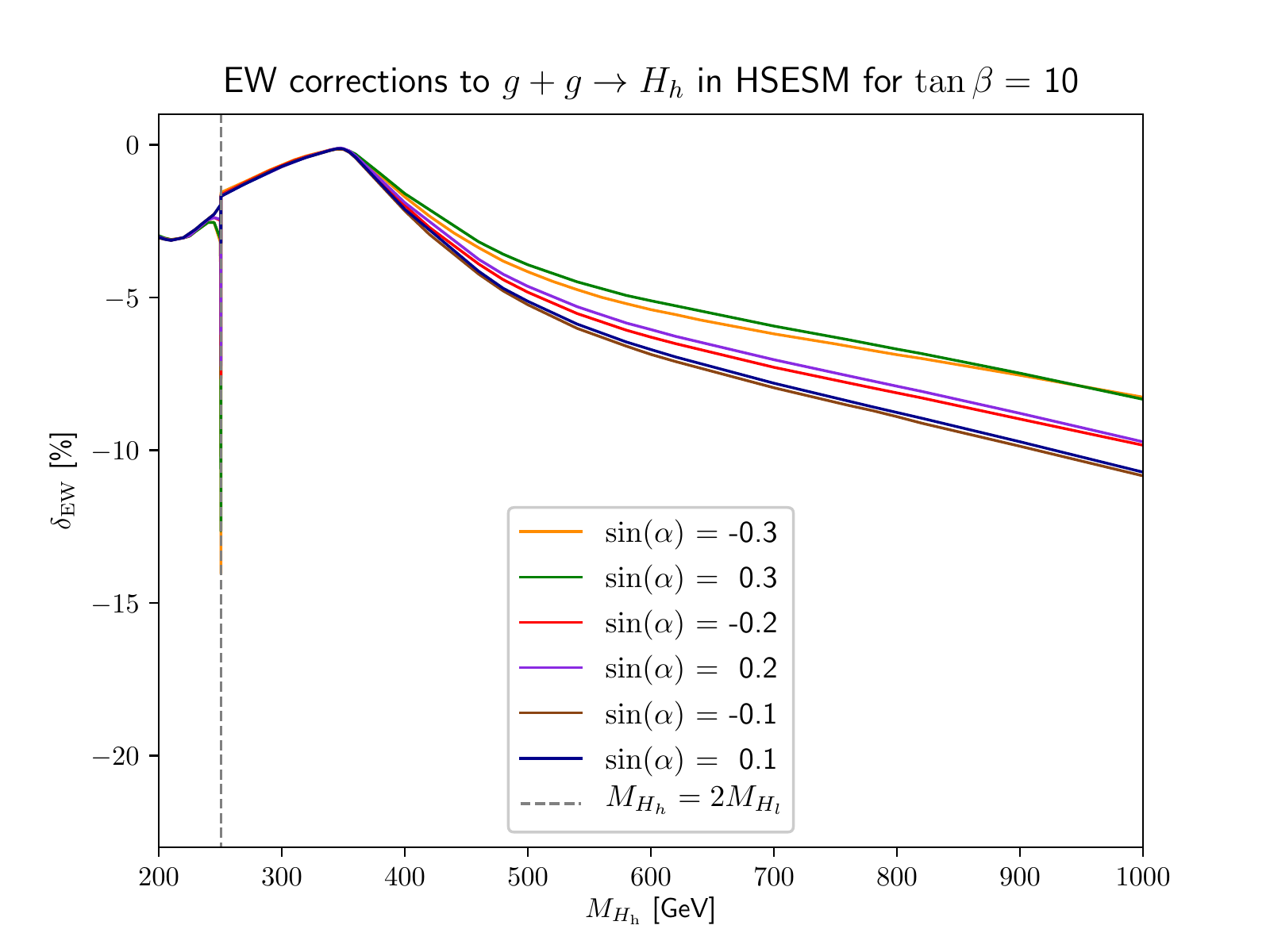}
\end{center}
\caption{Relative NLO electroweak percentage
  corrections~$\delta_{\text{EW}}$ in the OS scheme to the LO partonic
  cross section of the process $g+g \to H_h$ in the \HSESM\ for
  $\tan \beta = 1,5,10$ with varying values of
  $\sin \alpha$.\label{fig:Hhgg_fixed_tb}}
\end{figure}

The same plots for the production of the heavy Higgs boson are shown in
Fig.~\ref{fig:Hhgg_fixed_tb}, which features larger (negative)
electroweak percentage corrections compared to the production of the
light Higgs boson. The axis of ordinates shows again the electroweak
percentage correction~$\delta_{\text{EW}}$ as a function of the heavy
Higgs-boson mass~$M_{\hh}$ on the abscissa. Both axes show the
same range of values for all three plots.
Starting from around $-3\%$ for
$M_{\hh}\approx200$~GeV, the electroweak percentages 
corrections~$\delta_{\text{EW}}$ increase to a maximum close to
$\delta_{\text{EW}}\approx0\%$ around $M_{\hh}\approx350$~GeV and decrease
then almost linearly to reach $\delta_{\text{EW}}\approx-15\%$ for
$M_{\hh}\approx1000$~GeV (and $\tan\beta=1$).  We see again a comparable
behaviour for $\tan\beta=5$ and $\tan\beta=10$, while the corrections
are more sizable for $\tan\beta=1$.  Also for the production of a heavy
Higgs boson we see that the electroweak corrections are basically even
in $\sin\alpha$ for large $\tan\beta$.

For the production of a heavy Higgs boson~$\hh$ we observe in
Fig.~\ref{fig:Hhgg_fixed_tb} cusps for a heavy Higgs-boson mass
of $M_{H_{h}}=2M_{H_{l}}$. These cusps are threshold singularities
which arise from the external heavy Higgs-boson wave function
renormalization factor.
Such threshold singularities of the processes considered here 
have been analyzed in detail in the SM in Ref.~\cite{Actis:2008uh}.
Here, these singularities are regularized in the
complex mass scheme by the complex light Higgs-boson mass. Since the
total decay width of the light Higgs-boson is, however, very small, see
Eq.~(\ref{eq:inputparameters}), the cusps do not completely disappear
and remain visible in the electroweak percentage corrections.

For the BPs of Tab.~\ref{tab:BP} we show the electroweak percentage
corrections in Tab.~\ref{tab:BPsggHh} for the process~$g+g\to\hh$. 
The mixing angle~$\alpha$ has been renormalized in the OS scheme.
The BPs follow the shape of the plots of
Fig.~\ref{fig:Hhgg_fixed_tb}. For Higgs-boson masses going from 200
GeV to 300 GeV the percentage correction increases, reaching a
maximum close to zero for the benchmark point BHM300.
For BPs with a larger value of the heavy Higgs-boson mass the electroweak
percentage correction decreases.
Comparing for a fixed value of the heavy Higgs-boson mass the BPs
  with suffix a (BHM400a$\pm$, BHM500a$\pm$, BHM700a$\pm$  and BHM800a$\pm$) to those BPs with suffix b (BHM400b, BHM500b, BHM700b
  and BHM800b), respectively, we see that the electroweak percentage corrections for $g+g\to H_h$
  are more sensitive to the sign of $\sin \alpha$ than to small changes in $\tan \beta$.

\begin{table}[!ht]
  \centering
  \begin{tabular}{lcr}
  \begin{tabular}{|c||c|}
    \hline
BP                  & $\delta_{\text{EW}}$\\\hline
BHM200$-$\phantom{a}& $-3.1$         \\\hline
BHM200$+$\phantom{a}& $-3.1$         \\\hline
BHM300$-$\phantom{a}& $-0.7$                  \\\hline
BHM300$+$\phantom{a}& $-1.1$                  \\\hline
BHM400a$-$          & $-2.3$         \\\hline
BHM400a$+$          & $-1.7$                  \\\hline
    \end{tabular}
    &
  \begin{tabular}{|c||c|}
    \hline
BP                    & $\delta_{\text{EW}}$\\\hline
BHM400b\phantom{$\pm$}& $-1.7$                  \\\hline
BHM500a$-$            &  $-5.2$        \\\hline
BHM500a$+$            &  $-4.2$                 \\\hline
BHM500b\phantom{$\pm$}&  $-4.2$                 \\\hline
BHM600$-$\phantom{a}  & $-6.7$         \\\hline
BHM600$+$\phantom{a}  & $-5.8$                  \\\hline
    \end{tabular}
    &
  \begin{tabular}{|c||c|}
    \hline
BP                    & $\delta_{\text{EW}}$\\\hline
BHM700a$-$            & $-7.6$         \\\hline
BHM700a$+$            & $-6.9$                  \\\hline
BHM700b\phantom{$\pm$}& $-6.9$                  \\\hline
BHM800a$-$            & $-8.2$         \\\hline
BHM800a$+$            & $-7.9$                  \\\hline
BHM800b\phantom{$\pm$}& $-7.9$                  \\\hline
    \end{tabular}
  \end{tabular}
  \caption{The electroweak percentage
    corrections~$\delta_{\text{EW}}$~[\%] in the OS scheme are shown for
    the benchmark points of Tab.~\ref{tab:BP}.\label{tab:BPsggHh}}

\end{table}

The electroweak percentage corrections which we provide in
Fig.~\ref{fig:Hhgg_fixed_tb} and Tab.~\ref{tab:BPsggHh} also apply to the
partial decay width of the process
$\hh\to g+g$ due to Eq.~(\ref{eq:widthHgg}).\\

\begin{figure}[!h]
\begin{center}
\includegraphics[width=8.3cm]{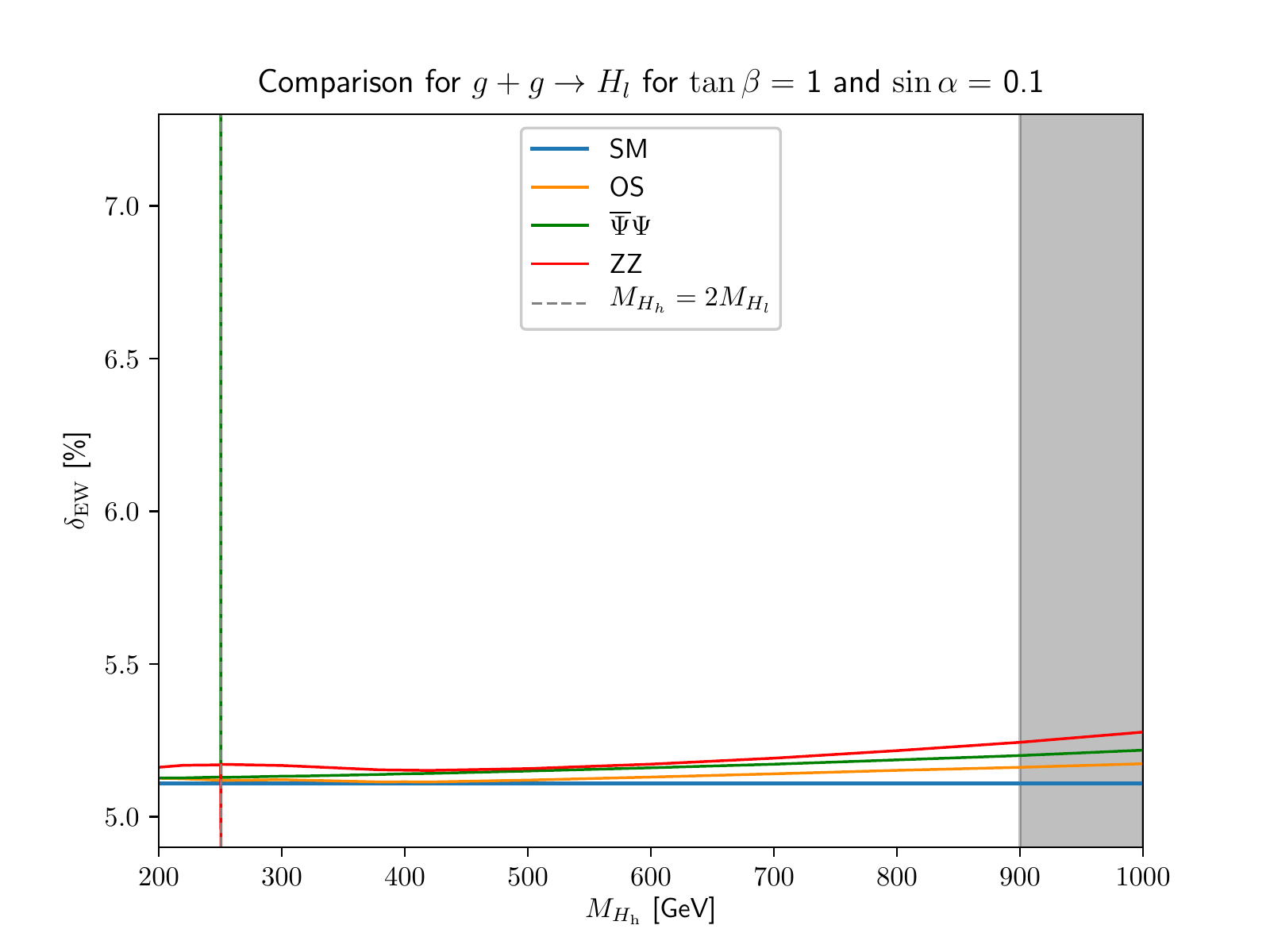}
\hspace*{-0.9cm}
\includegraphics[width=8.3cm]{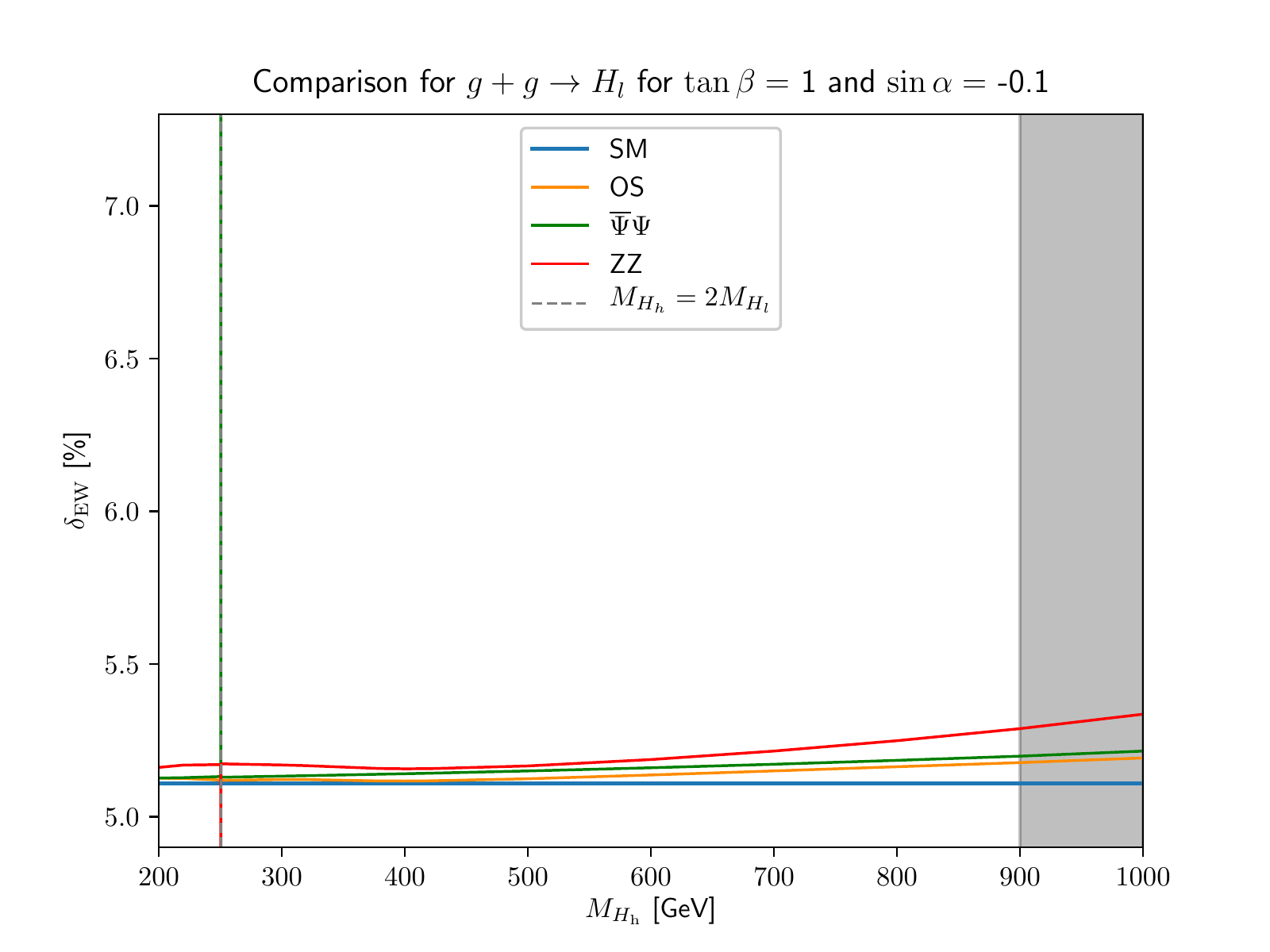}\\
\includegraphics[width=8.3cm]{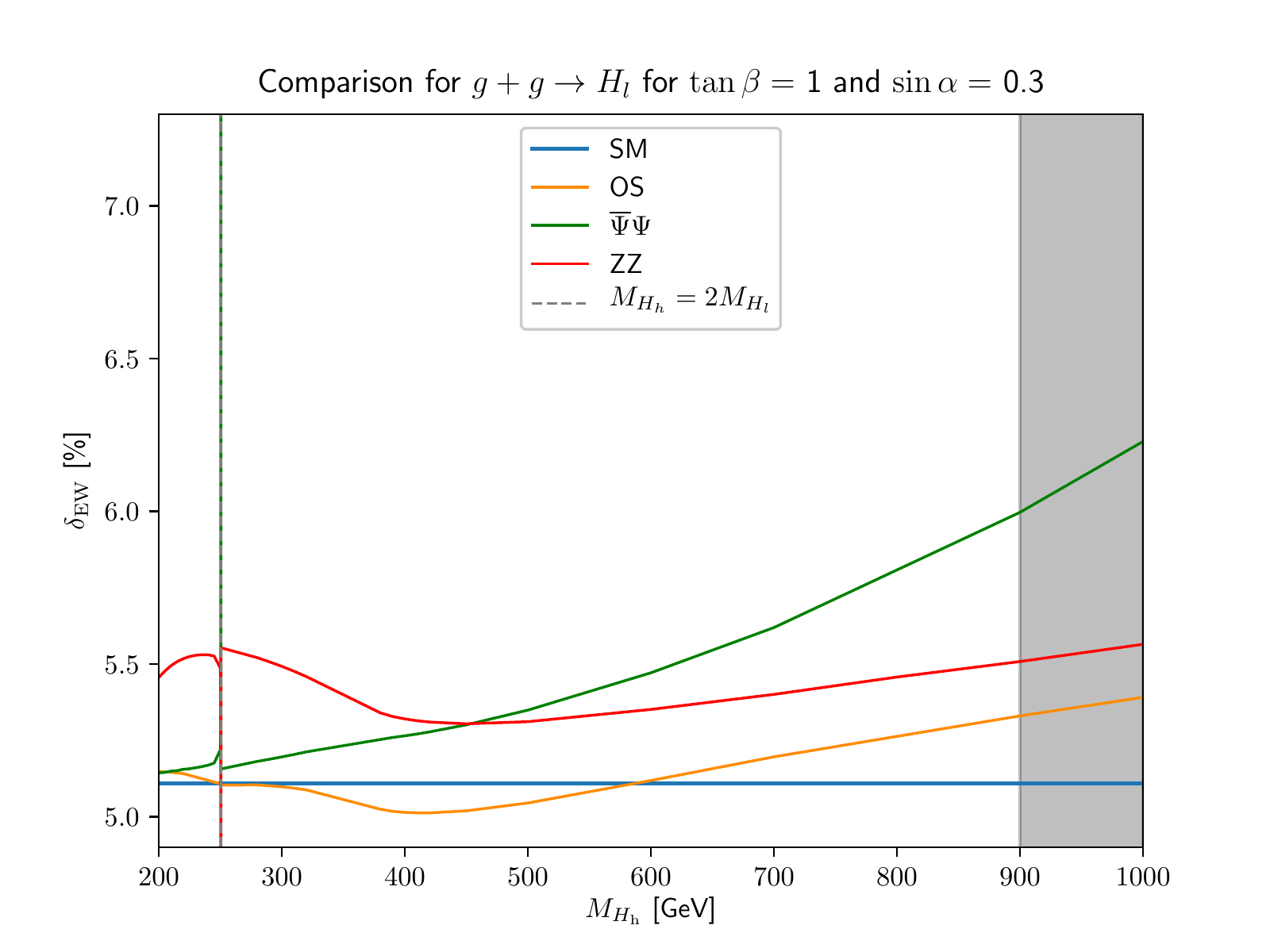}
\hspace*{-0.9cm}
\includegraphics[width=8.3cm]{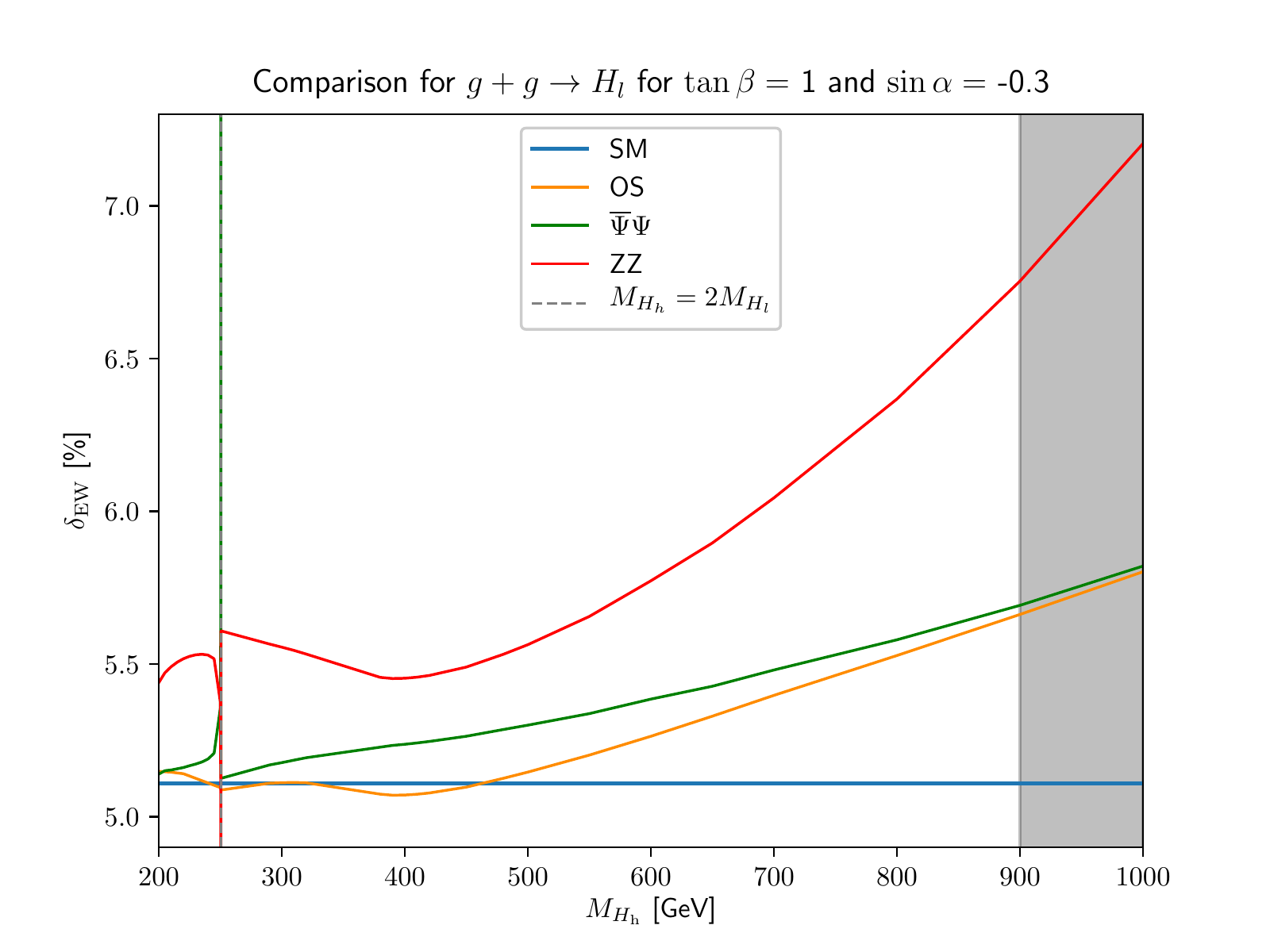}
\\
\includegraphics[width=8.3cm]{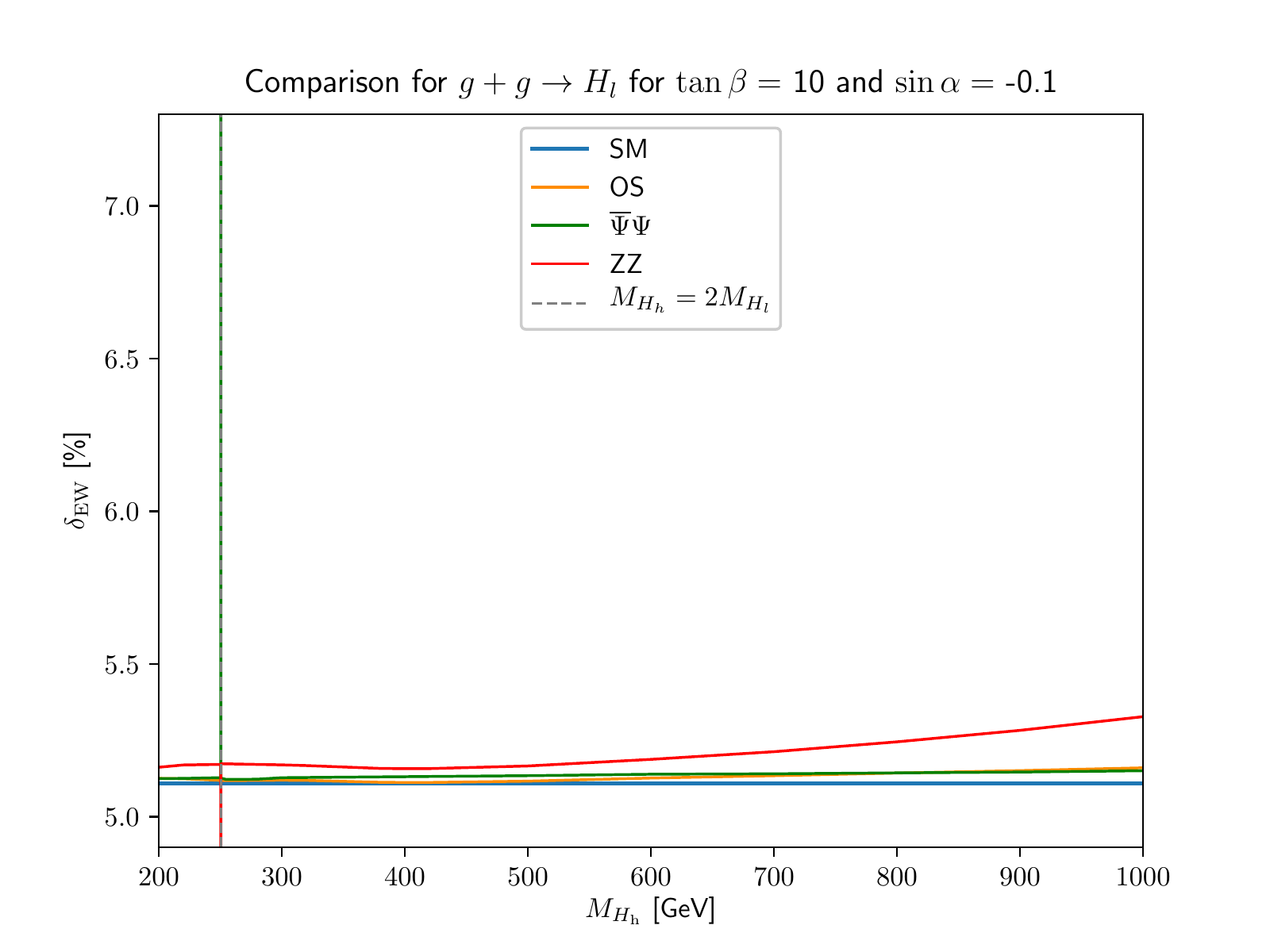}
\hspace*{-0.9cm}
\includegraphics[width=8.3cm]{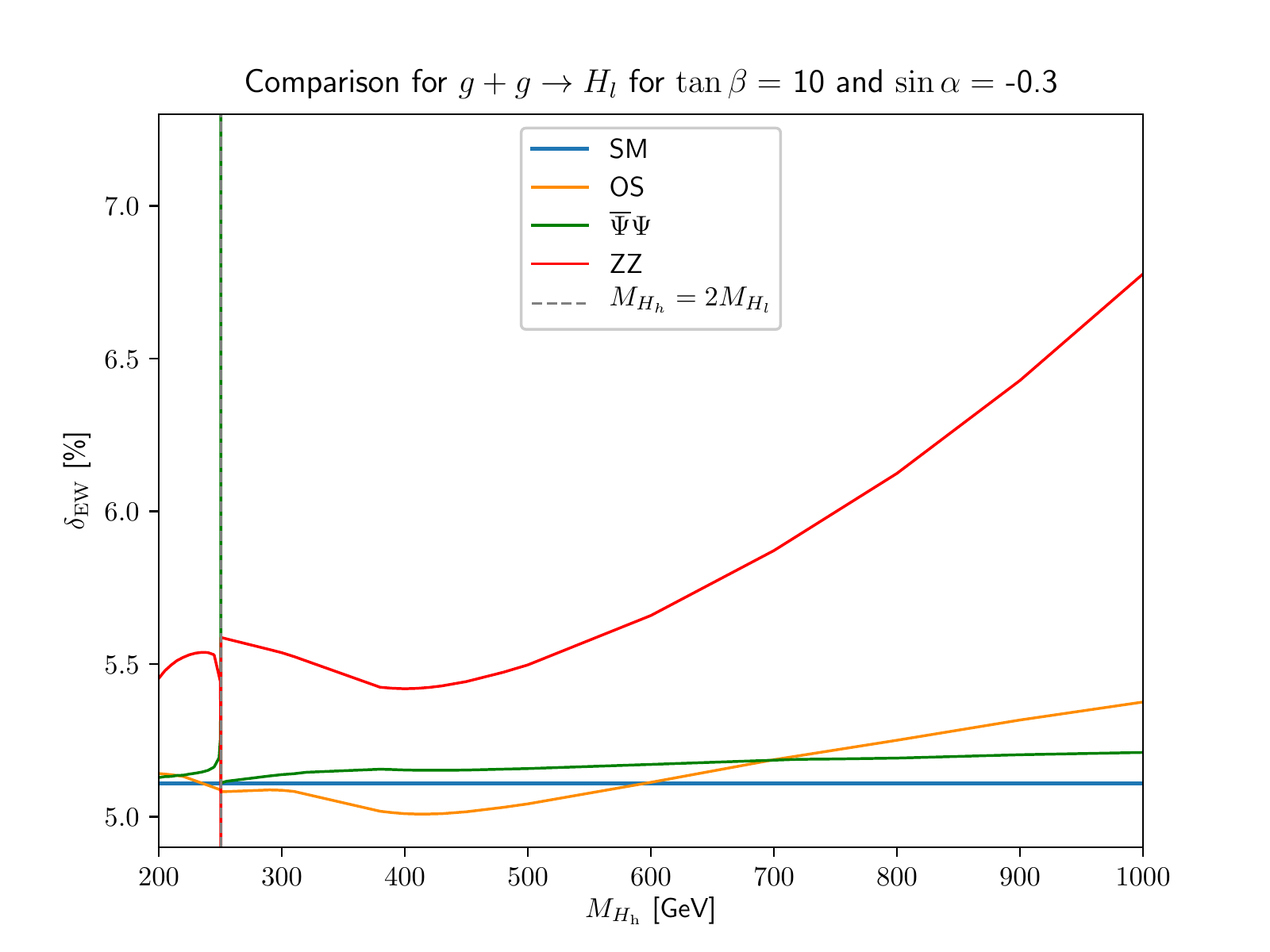}
\end{center}
\caption{Comparison of the renormalization schemes for $g+g \to H_l$ in
  the \HSESM\ for $\tan \beta = 1,10$ with varying values of $\sin
  \alpha$. The SM limit is shown in blue.\vspace{1.0cm}\mbox{}\label{fig:Hgg_schemes}}
\end{figure}
\begin{figure}[!ht]
\begin{center}
\includegraphics[width=8.3cm]{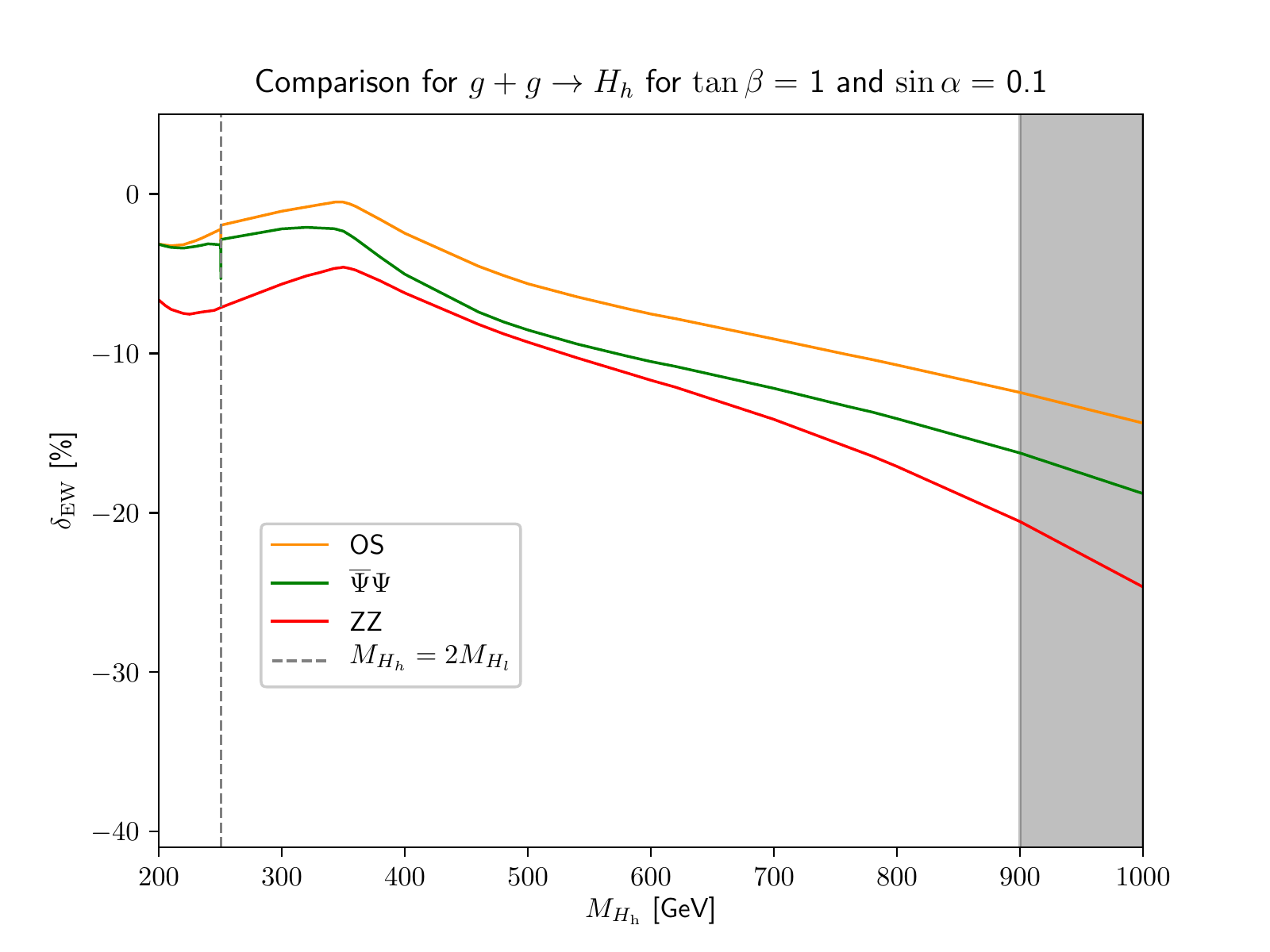}
\hspace*{-0.9cm}
\includegraphics[width=8.3cm]{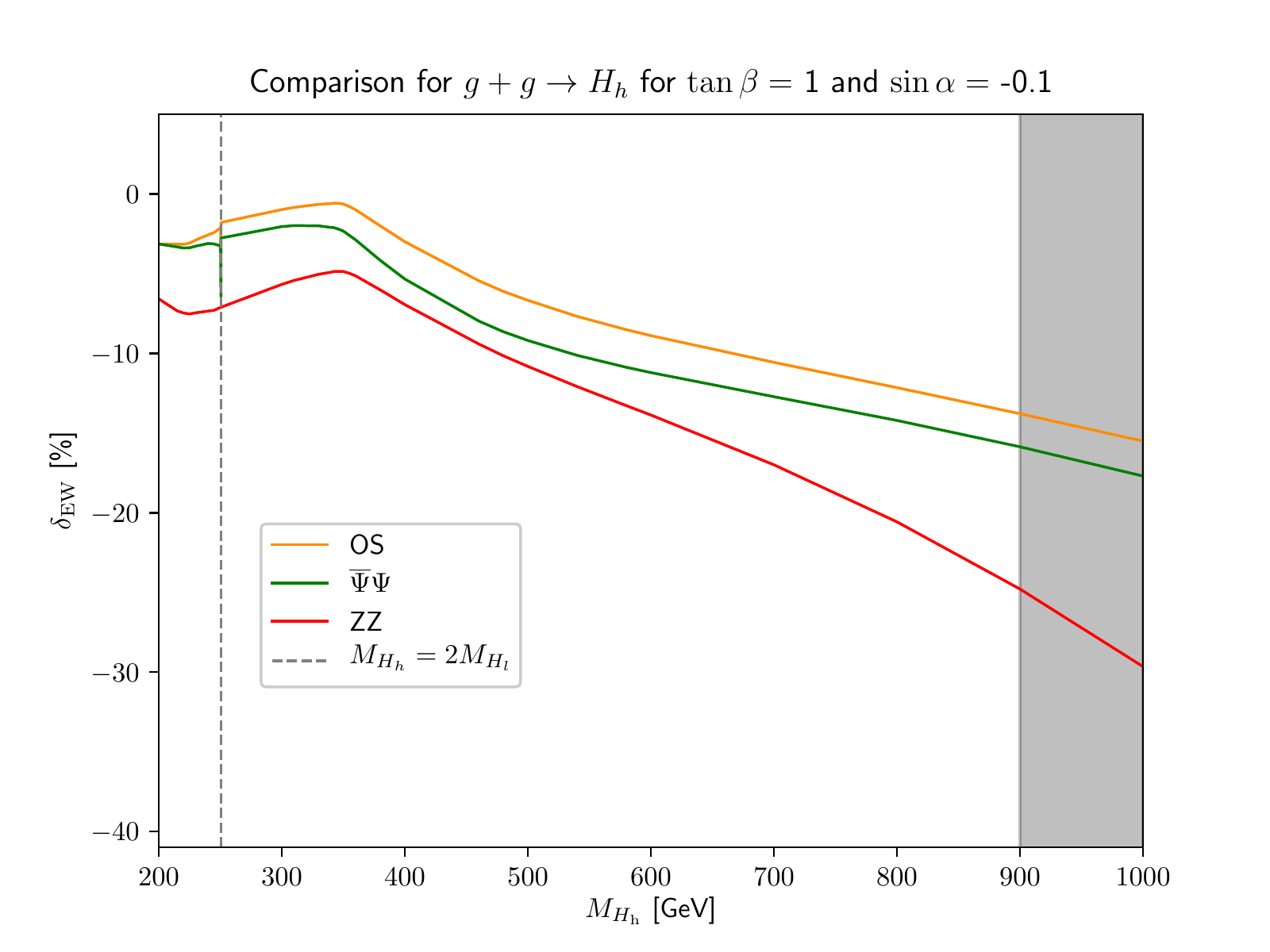}\\
\includegraphics[width=8.3cm]{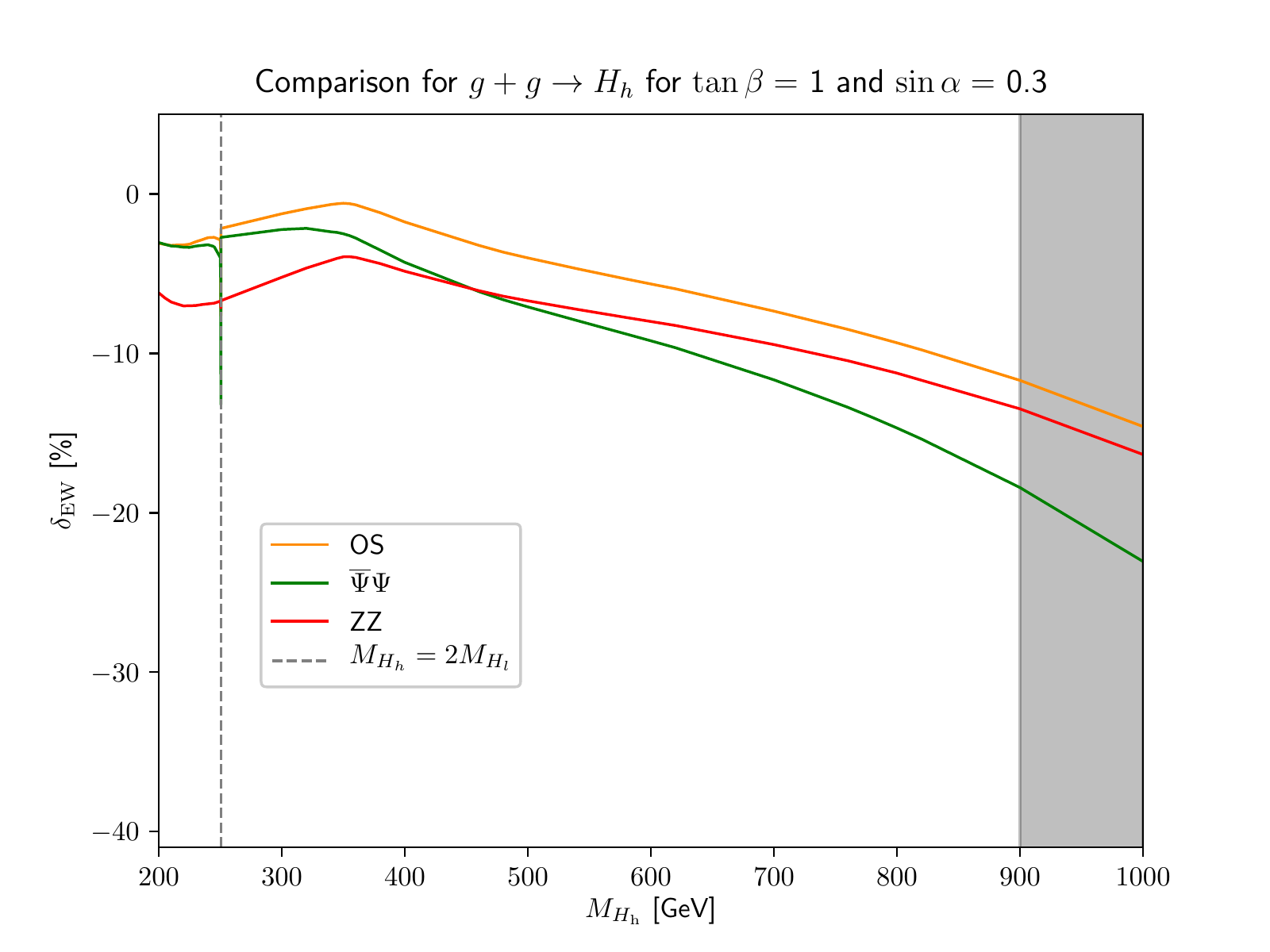}
\hspace*{-0.9cm}
\includegraphics[width=8.3cm]{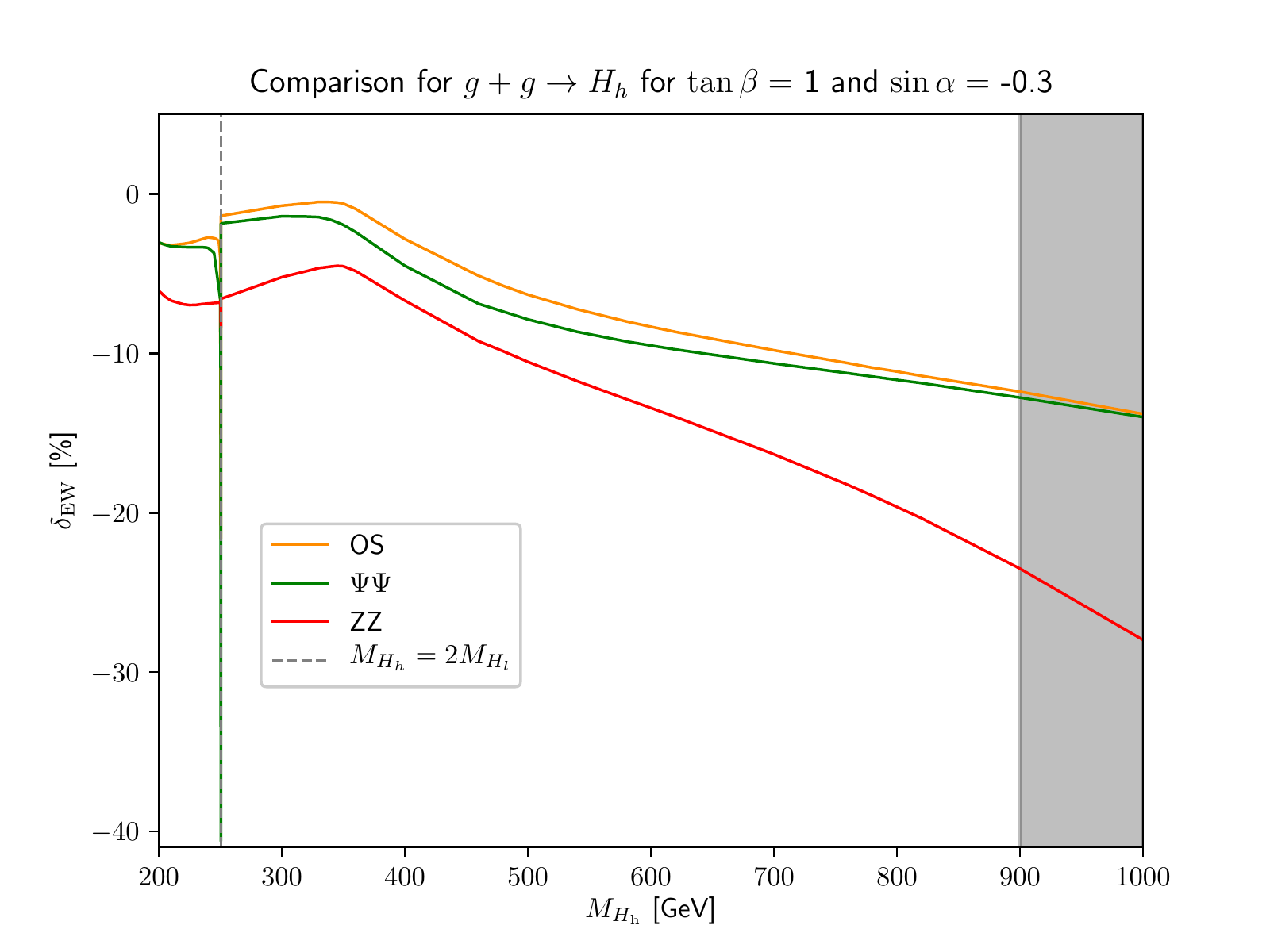}\\
\includegraphics[width=8.3cm]{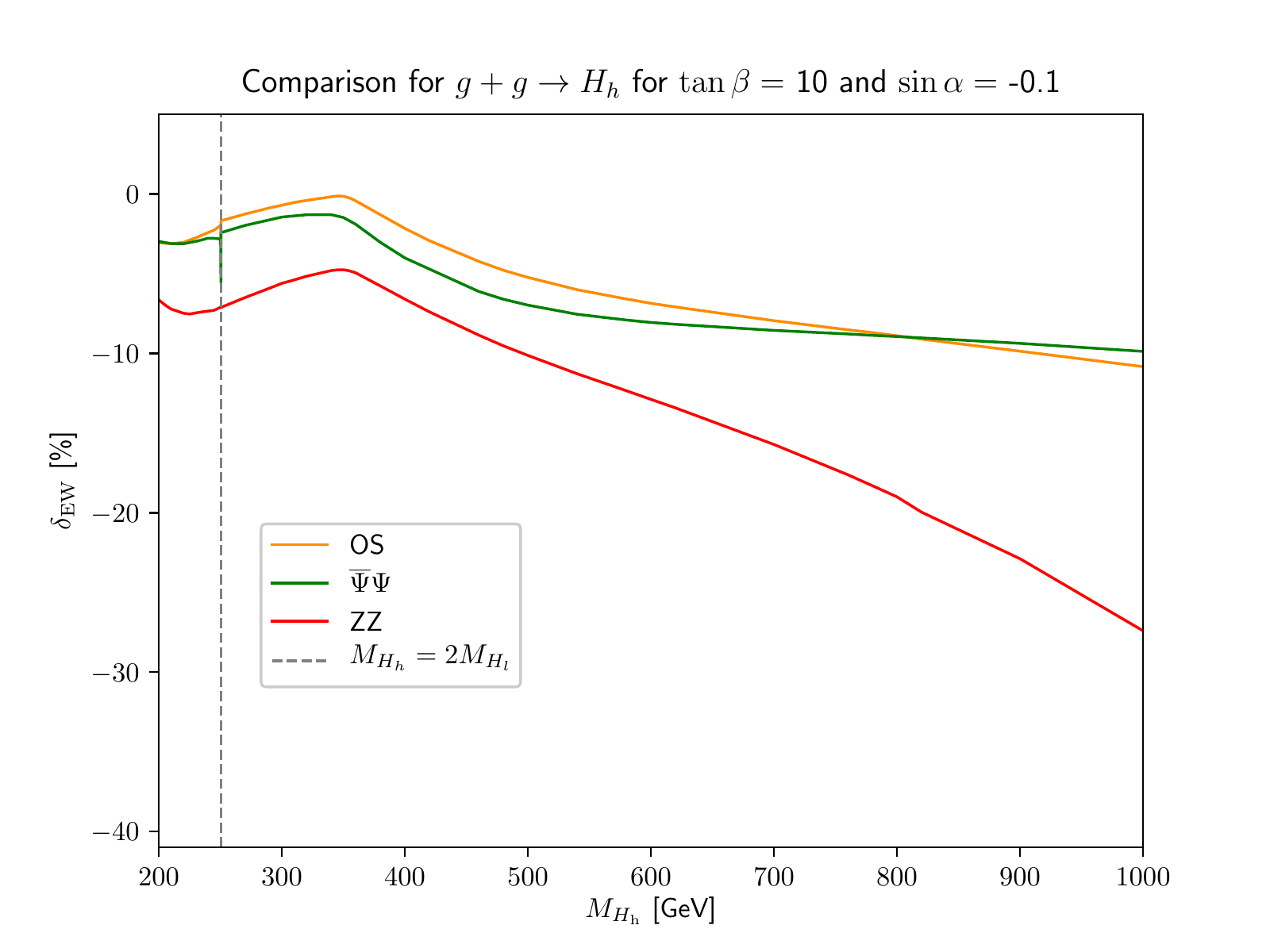}
\hspace*{-0.9cm}
\includegraphics[width=8.3cm]{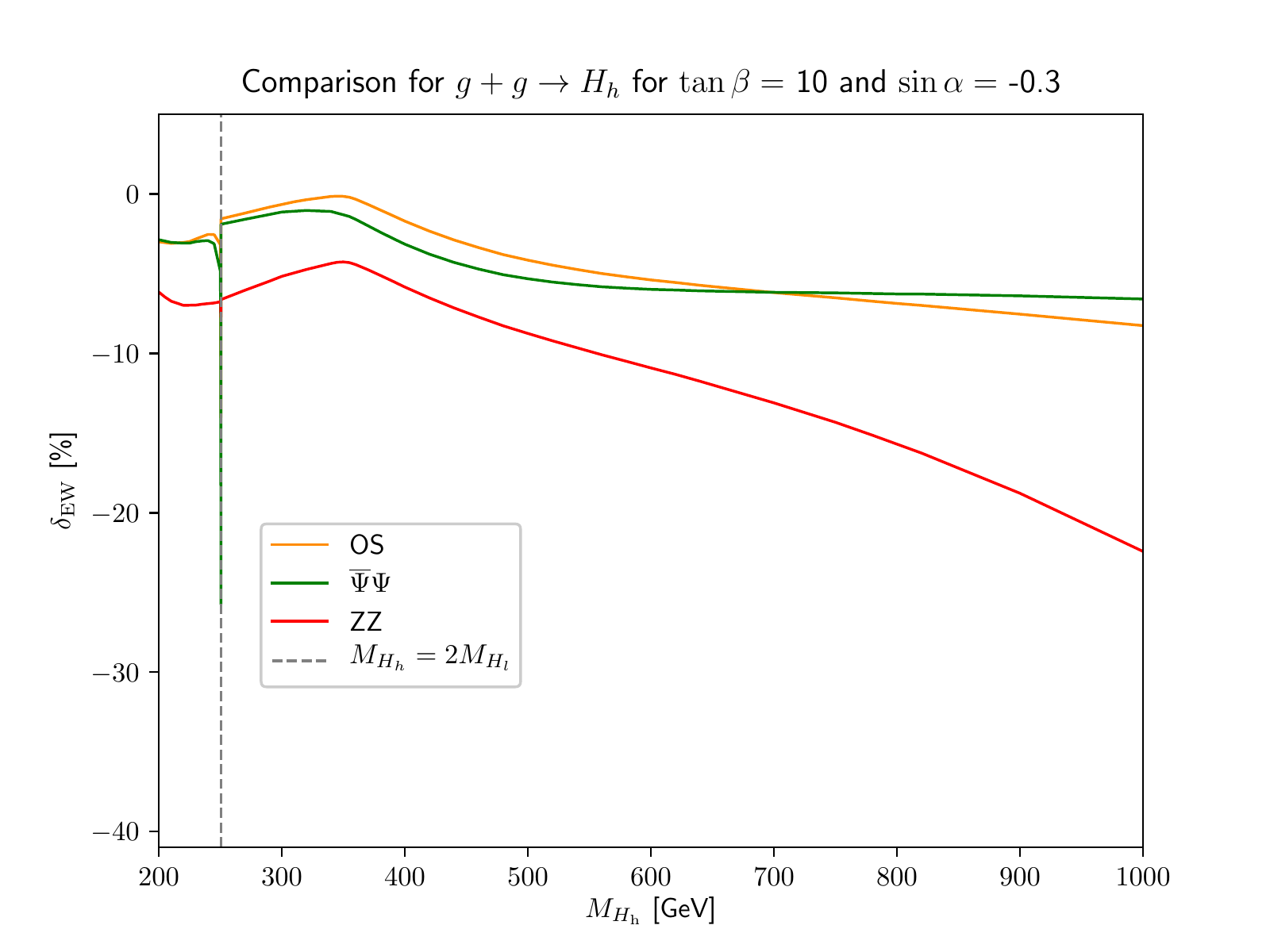}
\end{center}
\caption{Comparison of the renormalization schemes for $g+g \to H_h$ in
  the \HSESM\ for $\tan \beta = 1,10$ with varying values of $\sin
  \alpha$.\vspace{1.0cm}\mbox{}\label{fig:Hhgg_schemes}}
\end{figure}

Now we want to compare the three renormalization schemes under consideration,
i.e.\ the OS scheme, the $\overline{\Psi}\Psi$ scheme and the $ZZ$ scheme
as introduced in Section~\ref{sec:renormalization}.  
In Fig.~\ref{fig:Hgg_schemes} we report the corrections for the light
Higgs-boson production for $\sin\alpha=\pm 0.1$ and $\pm 0.3$.
We consider only $\tan\beta=1$ and $10$, since the
behaviour for $\tan\beta=5$ is close to the one for $\tan\beta=10$. For large values of $\tan\beta$ we
show just the behaviour for negative $\sin\alpha$, since the 
corrections are essentially even in $\sin\alpha$ in this case.

The electroweak corrections remain stable (and close to the SM) 
for small values of $\sin\alpha$, however, it can be seen that the $ZZ$ scheme
is slightly shifted upwards from the SM line for the whole range of heavy Higgs-boson
masses under consideration. This feature is even 
enhanced for $\sin\alpha=\pm 0.3$. In particular for $\sin \alpha = -0.3$ large differences
between the $ZZ$ scheme and the other two schemes exist and the $ZZ$ scheme shows a large dependence on the
heavy Higgs-boson mass~$M_{H_{h}}$, which enhances the correction at
large $M_{H_{h}}$ values. For $\tan\beta=1$, going from negative $\sin\alpha=-0.3$ to positive $\sin\alpha=+0.3$, the
percentage correction of the $\overline{\Psi}\Psi$ scheme is shifted upwards by
approximately the same amount as the OS scheme is shifted downwards. 
Fixing instead now $\sin\alpha=-0.3$ and going from $\tan\beta=1$ to
$\tan\beta=10$ the $\overline{\Psi}\Psi$ scheme is shifted downwards
stronger than the OS scheme. Since for large $\tan\beta$ the corrections
are insensitive to the sign of $\sin\alpha$, the corrections for
$\tan\beta=10$ and $\sin\alpha=-0.3$ are comparable to the corrections
for $\tan\beta=10$ and $\sin\alpha=+0.3$. Thus the shift in the
$\overline{\Psi}\Psi$ scheme is even larger when going from $\tan\beta=1$
to $\tan\beta=10$ for the positive value of $\sin\alpha=+0.3$. 
Hence the $\overline{\Psi}\Psi$ scheme displays
a stronger dependence on $\tan \beta$ than the OS scheme.
The difference in the three schemes is the renormalization of the mixing
angle~$\alpha$ as shown by the explicit expressions for the 
counterterm~$\delta\alpha$ in Section~\ref{sec:renormalization}.  A peculiarity of the $ZZ$ scheme is
that its counterterm~$\delta\alpha$ of Eq.~(\ref{eq:deltaalphaZZ}) has
an additional term, $c_\alpha s_\alpha\left(\delta_{H_lZZ}-\delta_{H_hZZ}\right)$, 
which 
incorporates the information about
the amplitudes of the
processes $H_l\to Z+Z$ and $H_h\to Z+Z$. 
Such a term is absent in the $\overline{\Psi}\Psi$ and
OS schemes of Eqs.~(\ref{eq:deltaalphaPsiPsi})
and~(\ref{eq:deltaalphaOS}),
causing the overall shift of the percentage correction in the $ZZ$
scheme compared to the other schemes as well as its stronger $M_{H_h}$
dependence (for negative values of $\sin\alpha$). Furthermore, both the
counterterm corresponding to the $\overline{\Psi}\Psi$ scheme and the
$ZZ$ scheme depend on all Higgs field strength renormalization factors
of Eq.~(\ref{eq:Higgsfieldrenorm}), whereas the counterterm
corresponding to the OS scheme only depends on the off-diagonal
ones. Since the $\tan\beta$ dependence originates from the trilinear
Higgs couplings, the presence of the additional diagonal field strength
renormalization factors in the $\overline{\Psi}\Psi$ scheme and the $ZZ$
scheme introduces also an additional $\tan \beta$ dependence.

For the process of light Higgs-boson production there is no 
physical threshold singularity at $M_{H_{h}}=2M_{H_{l}}$. %
As a result of this there are no cusps in Fig.~\ref{fig:Hgg_fixed_tb} in
the OS scheme, but only a small kink is visible for
$M_{H_{h}}=2M_{H_{l}}$. Analyzing Fig.~\ref{fig:Hgg_schemes} we observe
for the $ZZ$ and $\overline{\Psi}\Psi$~schemes that also for the process
$g+g\to\hl$ threshold singularities appear for
$M_{H_{h}}=2M_{H_{l}}$. These threshold singularities are introduced
artificially through the two process dependent renormalization schemes as already
mentioned in Section~\ref{sec:renormalization}. Both schemes rely on
processes, where the decay of a heavy Higgs-boson enters, i.e.
$\hh\to Z+Z$ for the $ZZ$ scheme and $\hh\to\Psi+\overline{\Psi}$ for
the $\overline{\Psi}\Psi$ scheme.  In these two processes the heavy
Higgs-boson wave function factor, which contains the threshold
singularity, enters again.
Thus these two schemes introduce the threshold singularities
artificially into the process $g+g\to\hl$. Also for that reason we consider
the result in the OS-scheme as our final result, which is free of this
drawback.

The dependence on the renormalization scheme for the heavy Higgs-boson
production in gluon fusion is shown in Fig.~\ref{fig:Hhgg_schemes}.  The
results in the $ZZ$ scheme appear shifted compared to the OS scheme,
being always at least $4\%$ different. The reason for this shift is
again the first term of Eq.~(\ref{eq:deltaalphaZZ}) as already described
in the discussion of the light Higgs-boson production in gluon fusion
earlier in this section. For negative values of $\sin\alpha$ the $ZZ$
scheme shows in addition a strong dependence on the heavy Higgs-boson
mass, arriving at differences of the order of $15\%$ for
$M_{H_{h}}=1000$~GeV between the $ZZ$ scheme and the other
schemes. Going for the fixed value of $\tan\beta=1$ from negative
$\sin\alpha=-0.3$ to positive $\sin\alpha=+0.3$, the percentage
correction of the $\overline{\Psi}\Psi$ scheme is shifted downwards by
approximately the same amount as the $ZZ$ scheme is shifted upwards. For
large values of $|\sin\alpha|=0.3$ the $\overline{\Psi}\Psi$ scheme
shows again a stronger dependence on $\tan \beta$ than the other
schemes. In all three schemes the EW percentage corrections have a
maximum between $\Mhh = 300$~GeV and $\Mhh = 400$~GeV, but the precise
location and its shape are scheme dependent. In
Fig.~\ref{fig:Hhgg_schemes} one can see the threshold singularities for
$M_{H_{h}}=2M_{H_{l}}$, which are present in the plots for all schemes
as expected. They are again regularized in the complex mass scheme by
the complex light Higgs-boson mass, but remain as cusps since the width
of the light Higgs boson is very small. The depth of the cusp caused by
this threshold is strongly scheme dependent, with a deep cusp in the
$\overline{\Psi}\Psi$ scheme and a hardly visible one in the $ZZ$
scheme.
\subsection{Higgs-boson decay into two photons in the \HSESM\label{sec:gamgamHHSESM}}
In this section we present results for the light and heavy Higgs-boson
decay into two photons.
Like in Section~\ref{sec:glugluHHSESM} for the Higgs-boson production in
gluon fusion, we choose also here for the Higgs-boson decay into two
photons the OS scheme as defined in Section~\ref{sec:renormalization} as
reference renormalization scheme for the mixing angle~$\alpha$.

\begin{figure}[!ht]
\begin{center}
\includegraphics[width=8.3cm]{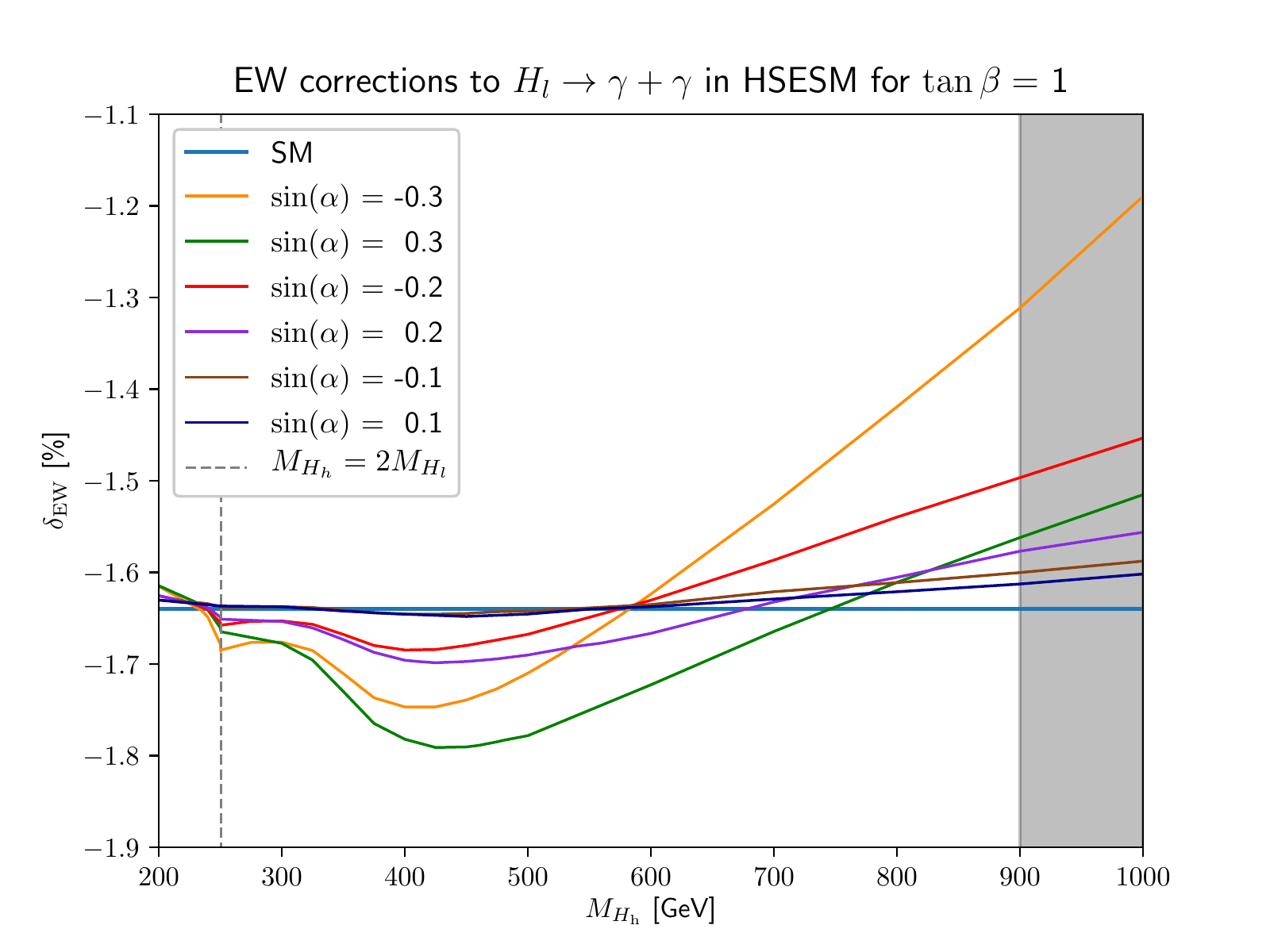}\\
\includegraphics[width=8.3cm]{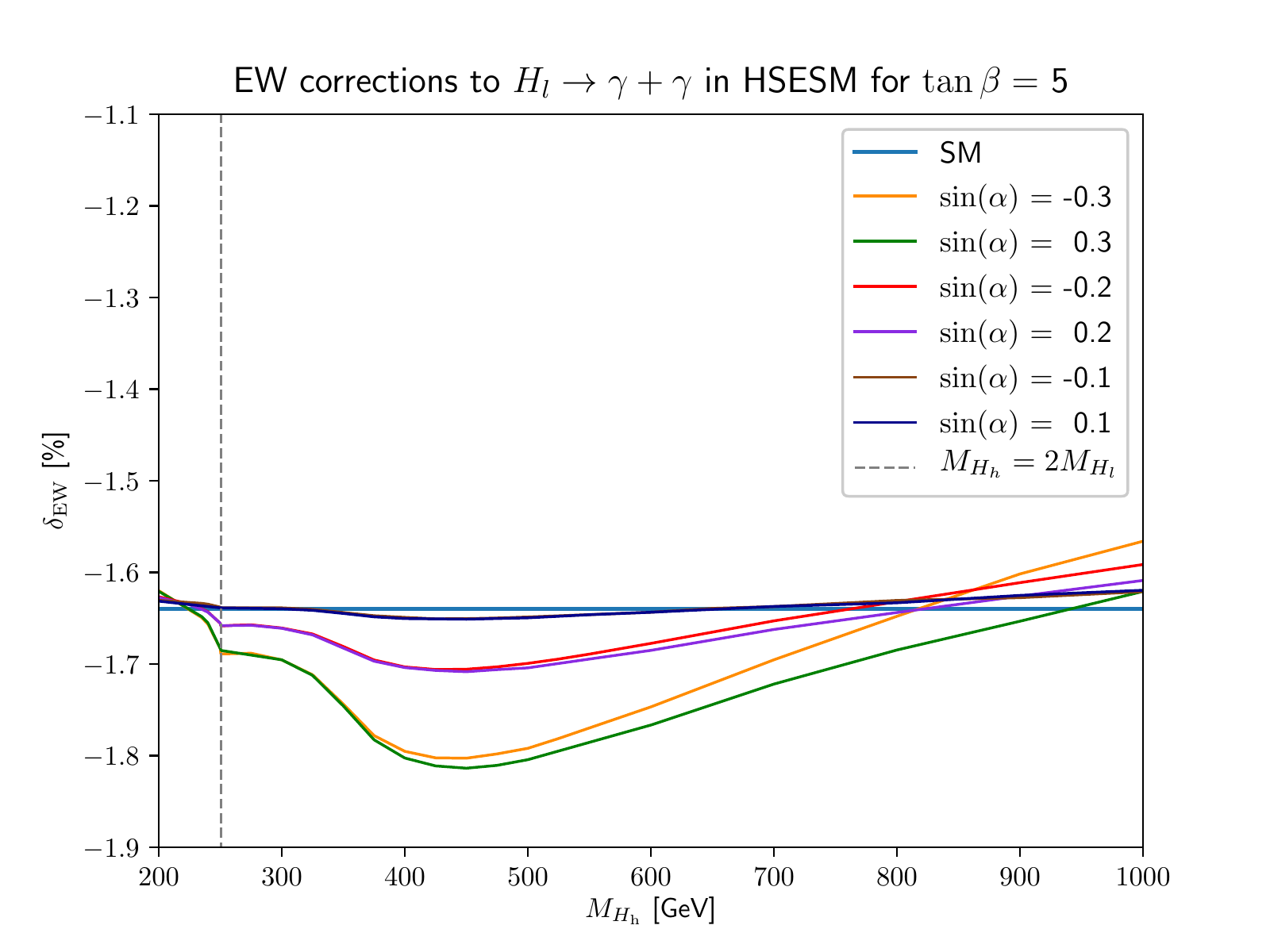}
\hspace*{-0.89cm}
\includegraphics[width=8.3cm]{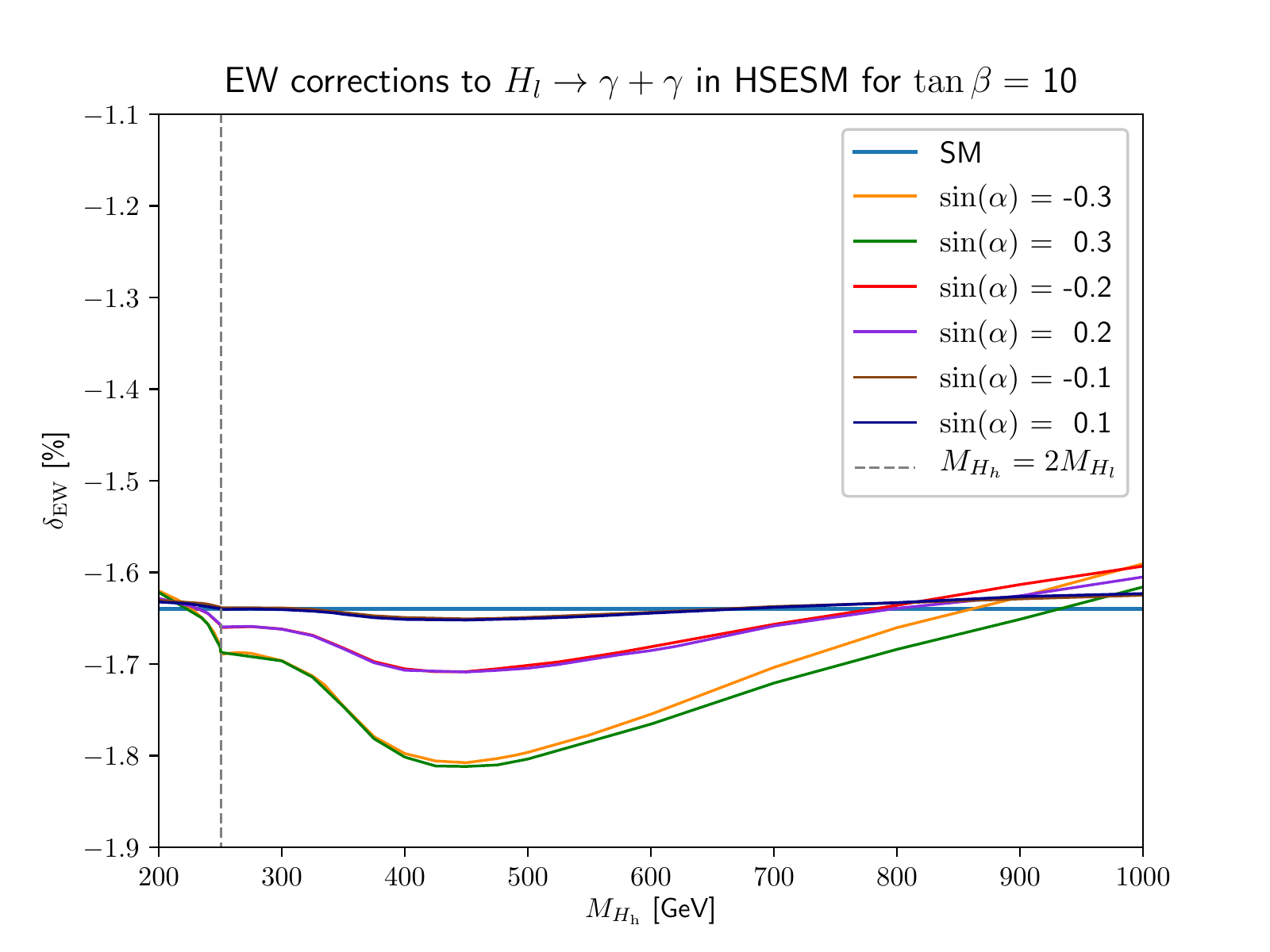}
\end{center}
\caption{NLO electroweak percentage
  corrections~$\delta_{\text{EW}}$ in the OS scheme relative to the LO partial decay
  width of the process $H_l \to \gamma+\gamma$ in the \HSESM\ for
  $\tan \beta = 1,5,10$ with varying values of $\sin \alpha$.  The
  SM limit is shown in blue.\label{fig:HAA_fixed_tb}}
\end{figure}
In Fig.~\ref{fig:HAA_fixed_tb} we show the electroweak percentage
corrections~$\delta_{\text{EW}}$ defined in Eq.~(\ref{eq:deltaEW}) as a
function of the heavy Higgs-boson mass~$M_{\hh}$ for the decay of a
light Higgs boson for 6 values of $\sin\alpha$.
The three plots of 
Fig.~\ref{fig:HAA_fixed_tb}, for the three values of
$\tan\beta = 1, 5$ and 10, show again the same numerical range for the
percentage correction and the heavy Higgs-boson mass in order to allow
to better compare the results for the three different values of
$\tan\beta$.
The SM result is again shown as the solid
horizontal blue line in each plot.  The electroweak corrections are
negative and very close to the SM case.  They range between about
$-1.8$\% to about $-1.2$\% and  are thus hardly to distinguish from the
SM case. The largest deviations from the SM result
can be observed again for high values of $M_{\hh}$, small values of
$\tan\beta$ and large absolute values of $\sin\alpha$.
Similarly to light Higgs-boson production in gluon fusion of
Section~\ref{sec:glugluHHSESM} the electroweak percentage 
corrections~$\delta_{\text{EW}}$ for the light Higgs-boson decay into two
photons shows also here a minimum between around
$M_{\hh}\approx400-500$~GeV. Afterwards the percentage corrections grow 
again almost linearly.
For the small value of $\tan\beta = 1$ in the first plot of
Fig.~\ref{fig:HAA_fixed_tb} the different lines for the various values
of $\sin\alpha$ differ more from each other than for larger
values of  $\tan\beta = 5,10$, where the dependence of the
electroweak corrections on the sign of $\sin\alpha$ becomes again
negligible due to the $\tan\beta$-suppression already discussed in the
previous Section~\ref{sec:glugluHHSESM}. This can be seen best again in the last plot of
Fig.~\ref{fig:HAA_fixed_tb} (for $\tan\beta = 10$), where the lines for
$\pm\sin\alpha$  approach each other and almost coincide.

\begin{figure}[!ht]
\begin{center}
\includegraphics[width=8.3cm]{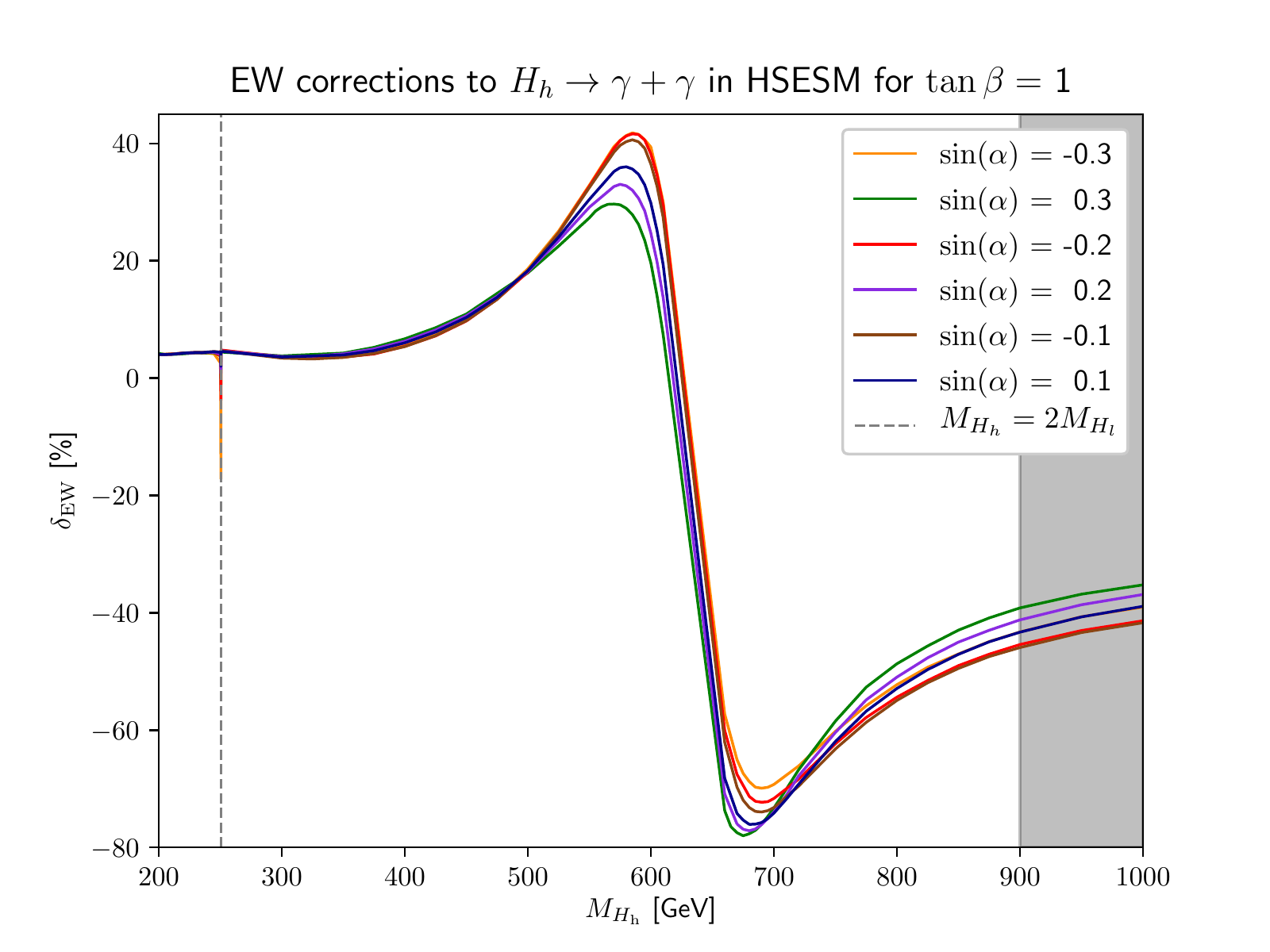}\\
\includegraphics[width=8.3cm]{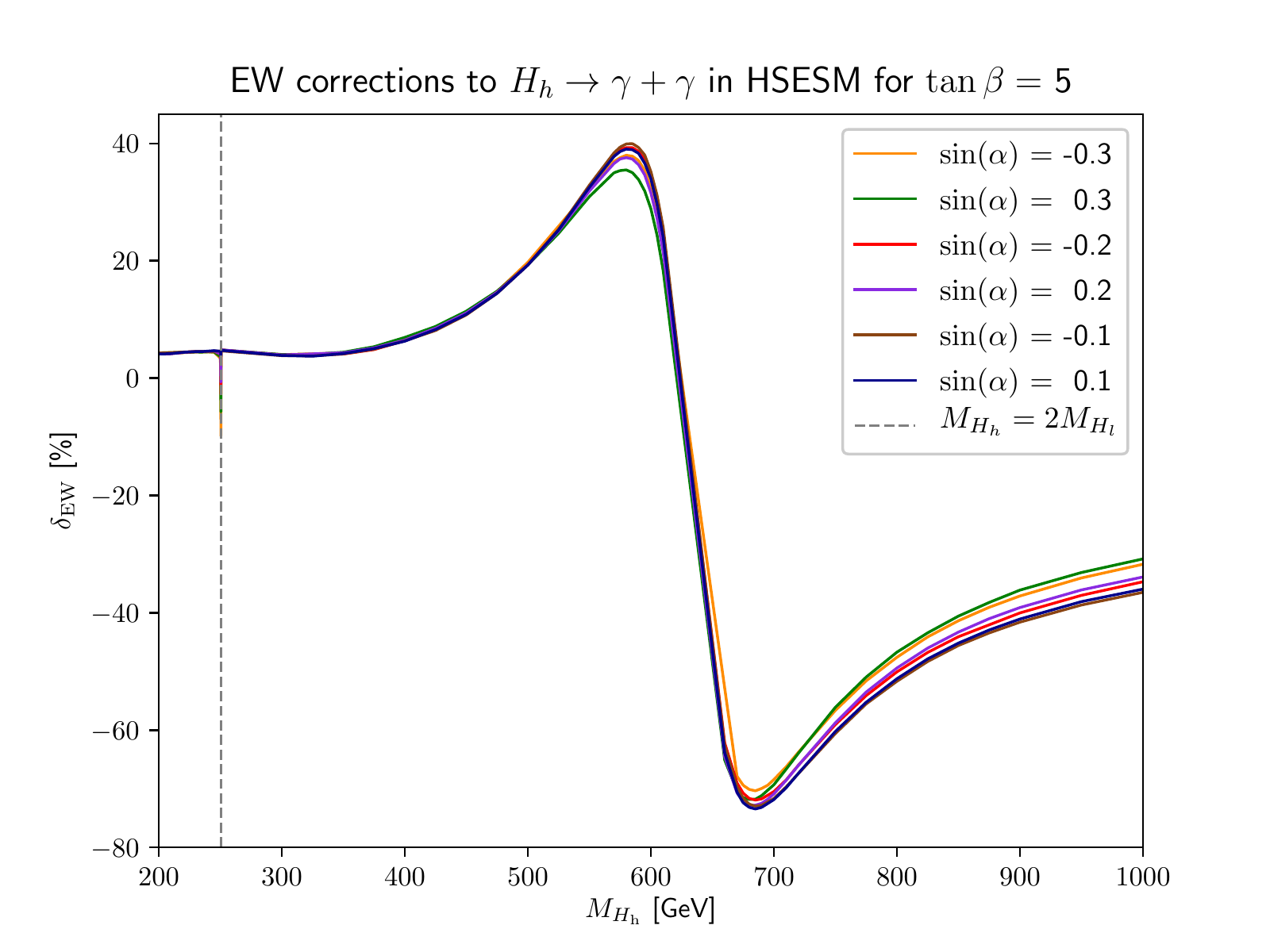}
\hspace*{-0.9cm}
\includegraphics[width=8.3cm]{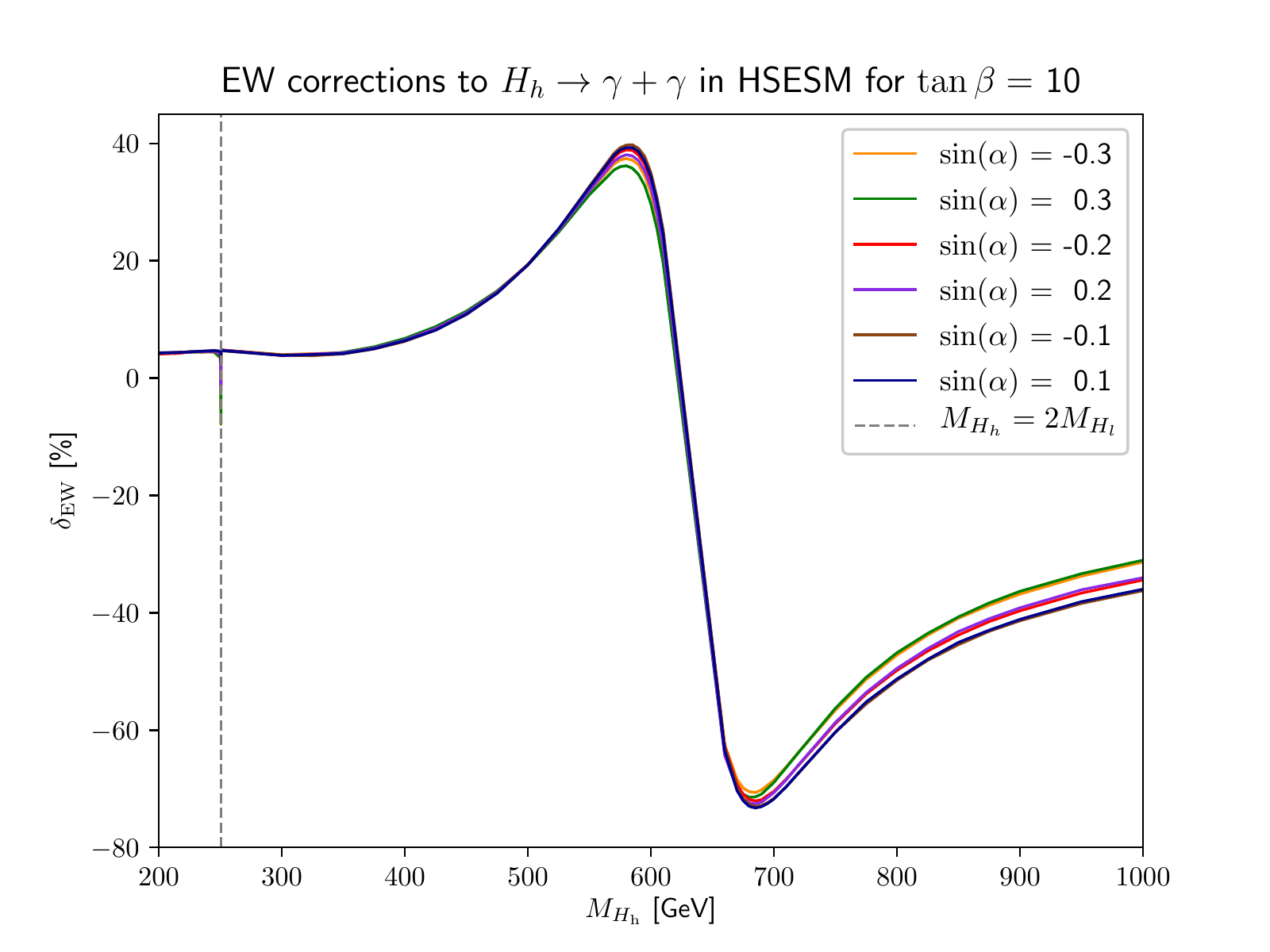}
\end{center}
\caption{NLO electroweak percentage corrections~$\delta_{\text{EW}}$
  relative to the LO partial decay width of the process
  $H_h \to \gamma+\gamma$ in the \HSESM\ for $\tan \beta = 1,5,10$ with varying values of
  $\sin\alpha$. The mixing angle~$\alpha$ is renormalized in the OS
  scheme.\label{fig:HhAA_fixed_tb}} 
\end{figure}

For the BPs of Tab.~\ref{tab:BP} we show the electroweak percentage
corrections for the process~$\hl\to\gamma+\gamma$ in
Tab.~\ref{tab:BPsHlAA}, where the mixing angle~$\alpha$ has been
renormalized in the OS scheme. The corrections are very close to the
result of the SM and follow the shape of the plots in
Fig.~\ref{fig:HAA_fixed_tb} with a slight minimum around
$M_{\hh}\approx400-500$~GeV.
\begin{table}[ht]
  \centering
  \begin{tabular}{lcr}
  \begin{tabular}{|c||c|}
    \hline
BP                    &$\delta_{\text{EW}}$\\\hline
BHM200$-$\phantom{a}&  $-1.6$            \\\hline
BHM200$+$\phantom{a}&  $-1.6$                  \\\hline
BHM300$-$\phantom{a}&  $-1.7$                  \\\hline
BHM300$+$\phantom{a}&  $-1.7$                  \\\hline
BHM400a$-$          &  $-1.7$                  \\\hline
BHM400a$+$          &  $-1.8$                  \\\hline
    \end{tabular}
    &
  \begin{tabular}{|c||c|}
    \hline
BP                    &$\delta_{\text{EW}}$\\\hline
BHM400b\phantom{$\pm$}&  $-1.8$          \\\hline
BHM500a$-$            &  $-1.7$                \\\hline
BHM500a$+$            &  $-1.8$                \\\hline
BHM500b\phantom{$\pm$}&  $-1.8$                \\\hline
BHM600$-$\phantom{a}  &  $-1.7$                \\\hline
BHM600$+$\phantom{a}  &  $-1.7$                \\\hline
    \end{tabular}
    &
  \begin{tabular}{|c||c|}
    \hline
BP                    &$\delta_{\text{EW}}$\\\hline
BHM700a$-$            &   $-1.7$        \\\hline
BHM700a$+$            &   $-1.7$               \\\hline
BHM700b\phantom{$\pm$}&   $-1.7$               \\\hline
BHM800a$-$            &   $-1.6$               \\\hline
BHM800a$+$            &   $-1.6$               \\\hline
BHM800b\phantom{$\pm$}&   $-1.6$               \\\hline
   \end{tabular}
  \end{tabular}
  \caption{The electroweak percentage
    corrections~$\delta_{\text{EW}}$~[\%] for the decay process
    $\hl\to\gamma+\gamma$ are shown for the benchmark points of
    Tab.~\ref{tab:BP} for the mixing angle~$\alpha$ in the OS scheme.\label{tab:BPsHlAA}}
\end{table}
The electroweak percentage corrections for the heavy Higgs-boson decay
into two photons in the \HSESM\ are shown in
Fig.~\ref{fig:HhAA_fixed_tb} in three plots as a function of the heavy
Higgs-boson mass~$M_{\hh}$. 
The three plots are again for the three values of $\tan\beta=1$, $5$, 
$10$ and show the six values of $\sin\alpha=\pm0.1$, $\pm0.2$, 
$\pm0.3$.  All three plots cover again the same range for the heavy
Higgs-boson mass and the electroweak percentage correction, which allows
to compare the three scenarios.  While the electroweak percentage corrections for small heavy
Higgs-boson masses around 200-300~GeV are only of the order of a few
percent, they range from about $-80\%$ to about $+40\%$ for heavy
Higgs-boson masses around 600~GeV and can thus become very large. 
Due to the wide range from large positive to large negative electroweak corrections, the
different curves for the different values of $\sin\alpha$ are hard to distinguish.

The electroweak corrections
have a maximum of $30\%$--$42$\% for a heavy Higgs-boson mass between
$570$~GeV and $585$~GeV, a turning point for a heavy Higgs-boson mass of
$M_{\hh}\approx630$~GeV and reach a minimum of $-70$\% to $-78$\% for a heavy
Higgs-boson mass in the range of $675$~GeV to $690$~GeV. The exact size and location of the
extrema depend on $\tan \beta$ and $\sin \alpha$.
Again, for a small value of $\tan\beta=1$ the curves for different
values of $\sin\alpha$ are further apart than for large values of
$\tan\beta=5,10$, where the curves with identical~$|\sin\alpha|$ approach each other and essentially
coincide for $\tan\beta=10$ due to the $\tan\beta$-suppression. 
For $\tan\beta=1$ the maximum for $\sin \alpha = -0.3$ is located at
$M_{\hh} = 585$~GeV with a value of around $42\%$, whereas the maximum
for $\sin \alpha =+0.3$ is located at $M_{\hh} = 570$~GeV with a value
of around $30\%$. The location and size of the minimum behave in a
similar way with the minimum of $-70$\% being located at
$M_{\hh} = 690$~GeV for $\tan \beta = 1$ and $\sin \alpha = -0.3$ and
the minimum of $-78$\% being located at $M_{\hh} = 675$~GeV for
$\tan \beta = 1$ and $\sin \alpha =+0.3$. The reason for the fast change
from large positive to large negative corrections within a few GeV of
the heavy Higgs-boson mass can be understood when taking into account
Fig.\ref{fig:ampsquared}(b), where the squared modulus of the LO
amplitude~$|A_{\gamma\gamma}^{\text{LO}}|^2$ is shown. There the squared
modulus of the amplitude strongly decreases for Higgs-boson masses
larger than about 300~GeV and reaches a minimum between 630-640~GeV,
where the squared modulus becomes almost zero. After the minimum it
increases only slightly (see inset of Fig.\ref{fig:ampsquared}(b)), but
stays close to zero. The electroweak percentage
corrections~$\delta_{\text{EW}}$ of Fig.~\ref{fig:HhAA_fixed_tb} being
according to Eq.~(\ref{eq:deltaEW}) an interference between the LO and
NLO amplitude normalized to the squared modulus of the LO amplitude inherit
this feature from the LO amplitude in the form of a large enhancement
factor. In fact, the real and imaginary parts of the pure two-loop
amplitude only change moderately for Higgs-boson masses between
600--700~GeV. However, the LO-NLO interference is such that it changes
sign in this region. This change of sign paired with the large
enhancement leads to a very sharp drop.
We observe a cusp for a heavy Higgs-boson mass of
$M_{\hh}=2M_{\hl}$ in Fig.~\ref{fig:HhAA_fixed_tb}; this time also 
for a renormalization of the mixing angle~$\alpha$ in the OS scheme.
The cusp at the threshold of $M_{\hh}=2M_{\hl}$
originates from the heavy Higgs-boson wave function factor. This
threshold is here in this process $\hh\to\gamma+\gamma$ a real physical
threshold and not introduced artificially due to the choice of the
renormalization scheme, like for the processes $g+g\to\hl$ and
$\hl\to\gamma+\gamma$ in the $ZZ$ or $\overline{\Psi}\Psi$ schemes. 
This threshold singularity is  here again regularized by the complex
mass of the light Higgs boson.

For the BPs of Tab.~\ref{tab:BP} we show the electroweak percentage
corrections in Tab.~\ref{tab:BPsHhAA} for the
process~$\hh\to\gamma+\gamma$. The mixing angle $\alpha$ is renormalized
in the OS scheme. For smaller values of the heavy Higgs-boson mass
between 200 to 400~GeV the electroweak corrections of the BPs are of
moderate size and are of the order of a few percent. They coincide with what one
expects from Fig.~\ref{fig:HhAA_fixed_tb} in this mass
range. Approaching heavy Higgs-boson masses of 500 to 600~GeV the
electroweak corrections of the BPs are large and positive in accordance with
Fig.~\ref{fig:HhAA_fixed_tb}. 
All BPs which only differ in the sign of $\sin\alpha$ are very close
to each other, except for the benchmark point BHM600$\pm$. The latter BP is close to
the location where in Fig.~\ref{fig:HhAA_fixed_tb} the percentage
correction changes from large positive to large negative corrections and
the curve has a steep slope. Thus the percentage correction vastly
changes in a small window of a few GeV explaining the different percentage correction for 
BHM600$-$ and BHM600$+$. The large negative size of the electroweak
corrections for the benchmark points BHM700 and BHM800 has the same
reason as the large negative corrections in
Fig.~\ref{fig:HhAA_fixed_tb}, i.e. the squared modulus of the LO amplitude
becomes small. Studying the benchmark points BHM400a$\pm$, BHM400b, BHM500a$\pm$, BHM500b, BHM700a$\pm$, BHM700b, BHM800a$\pm$
  and BHM800b we see that the electroweak percentage corrections for $H_h \to \gamma+\gamma$
  are more sensitive to the sign of $\sin \alpha$ than to small changes in $\tan \beta$ as was already
  the case for $g+g\to H_h$.
\begin{table}[ht]
  \centering
  \begin{tabular}{lcr}
  \begin{tabular}{|c||c|}
    \hline
BP                    &$\delta_{\text{EW}}$\\\hline
BHM200$-$\phantom{a}&  $4.0$          \\\hline
BHM200$+$\phantom{a}&  $4.0$          \\\hline
BHM300$-$\phantom{a}&  $3.8$          \\\hline
BHM300$+$\phantom{a}&  $3.9$                  \\\hline
BHM400a$-$          &  $6.0$          \\\hline
BHM400a$+$          &  $6.7$                  \\\hline
    \end{tabular}
    &
  \begin{tabular}{|c||c|}
    \hline
BP                    &$\delta_{\text{EW}}$\\\hline
BHM400b\phantom{$\pm$}& $6.7$                 \\\hline
BHM500a$-$            & $19.2$        \\\hline
BHM500a$+$            & $19.1$                 \\\hline
BHM500b\phantom{$\pm$}&  $19.1$                \\\hline
BHM600$-$\phantom{a}  &  $35.1$       \\\hline
BHM600$+$\phantom{a}  &  $29.6$          \\\hline
    \end{tabular}
    &
  \begin{tabular}{|c||c|}
    \hline
BP                    &$\delta_{\text{EW}}$\\\hline
BHM700a$-$            & $-70.3$       \\\hline
BHM700a$+$            & $-71.0$                 \\\hline
BHM700b\phantom{$\pm$}& $-71.0$                 \\\hline
BHM800a$-$            & $-50.0$       \\\hline
BHM800a$+$            & $-49.4$                 \\\hline
BHM800b\phantom{$\pm$}& $-49.4$                 \\\hline
    \end{tabular}
  \end{tabular}

  \caption{The electroweak percentage
    corrections~$\delta_{\text{EW}}$~[\%] for the decay process
    $\hh\to\gamma+\gamma$ are shown for a renormalization of the mixing
    angle~$\alpha$ in the OS scheme.\label{tab:BPsHhAA}} 
\end{table}

\begin{figure}[!h]
\begin{center}
\includegraphics[width=7.9cm]{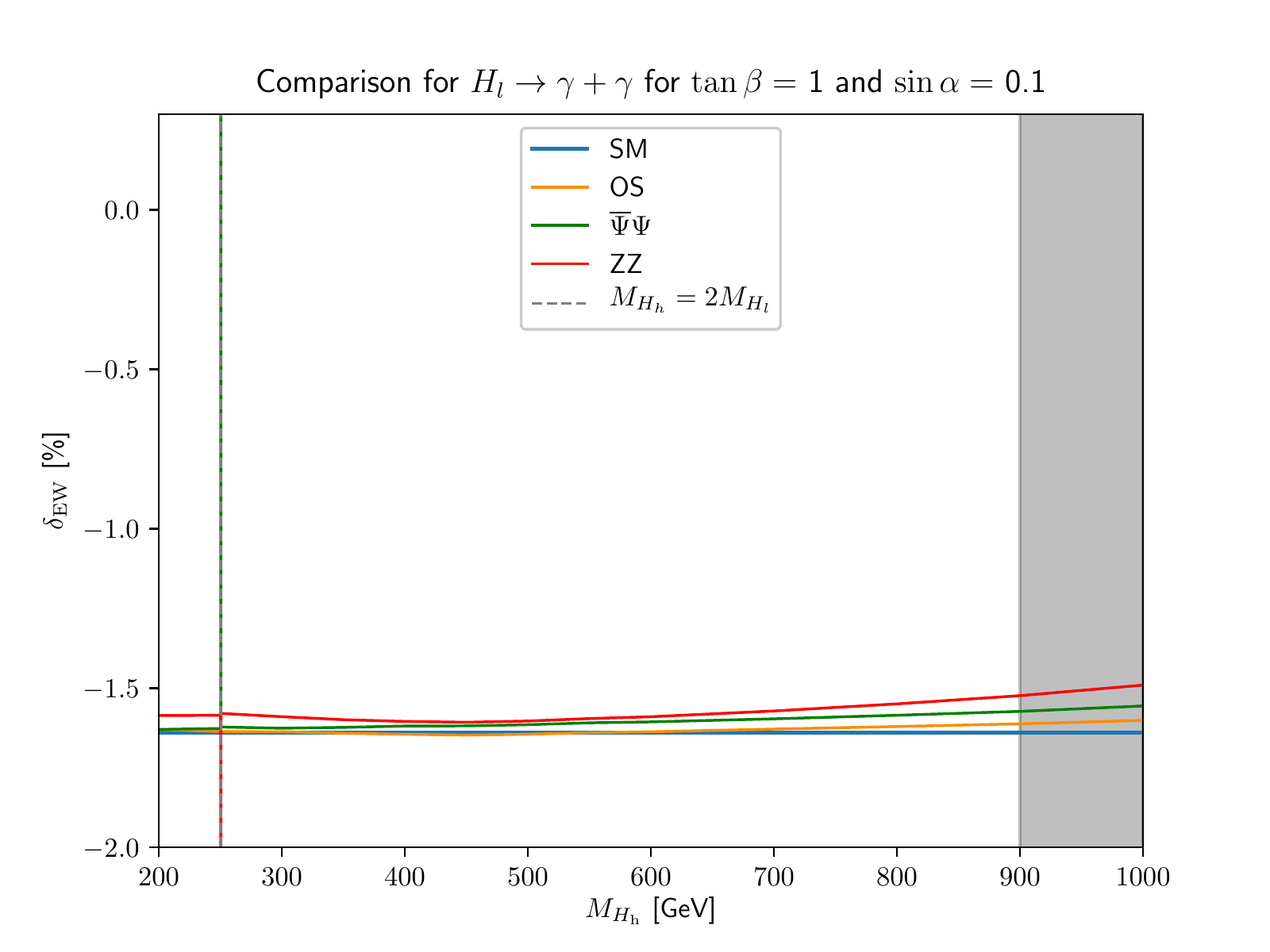}
\includegraphics[width=7.9cm]{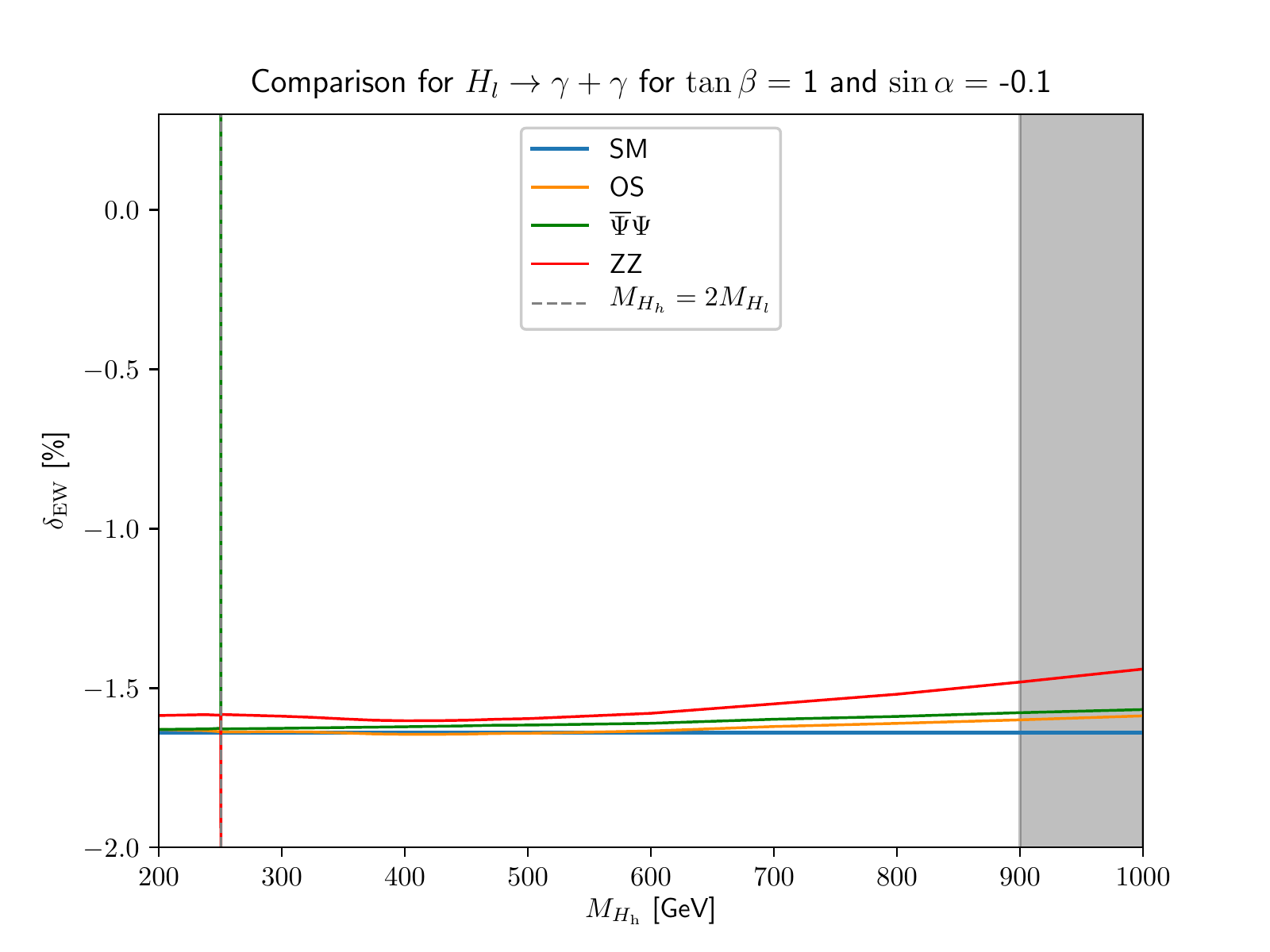}\\
\includegraphics[width=7.9cm]{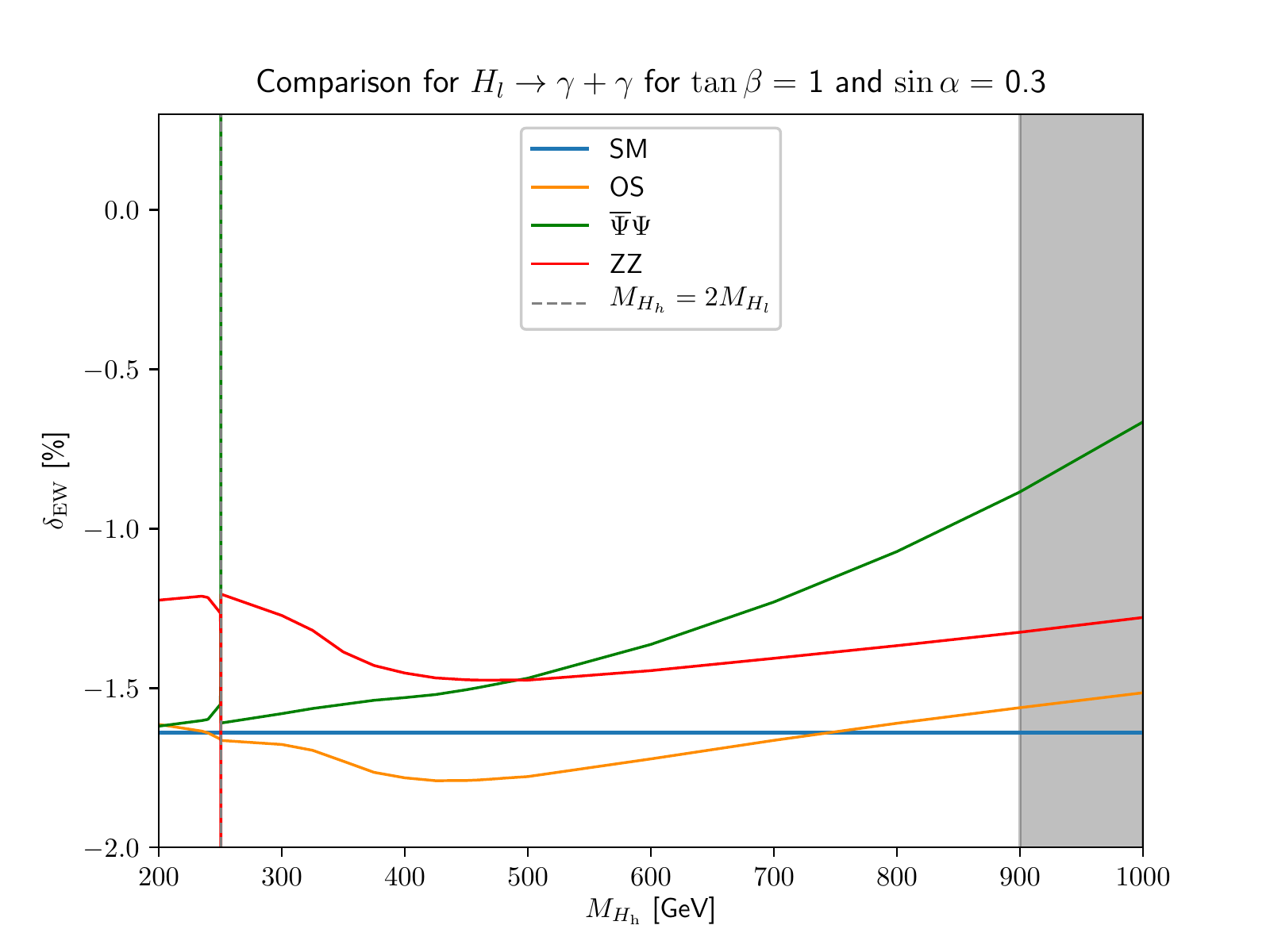}
\includegraphics[width=7.9cm]{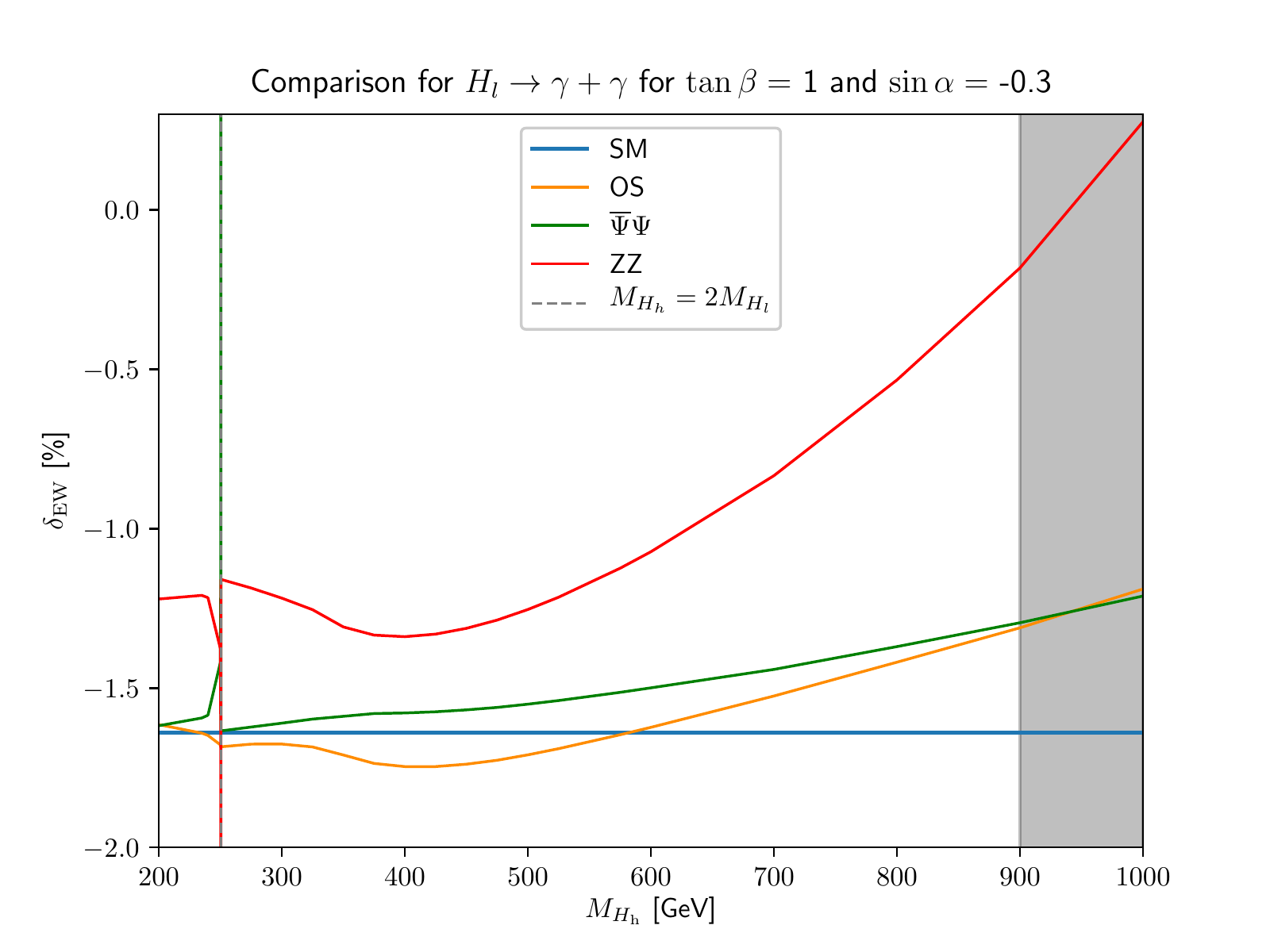}\\
\includegraphics[width=7.9cm]{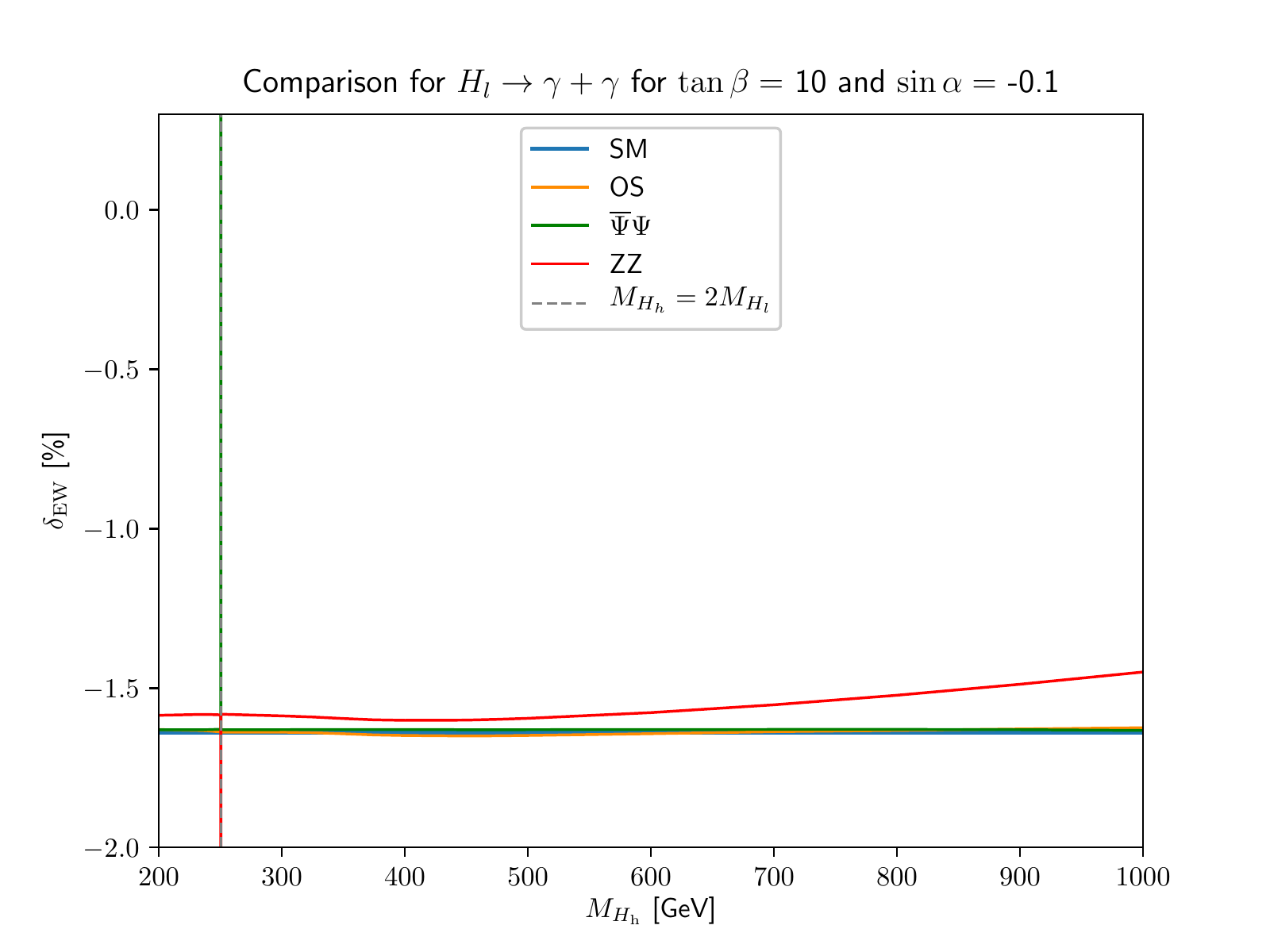}
\includegraphics[width=7.9cm]{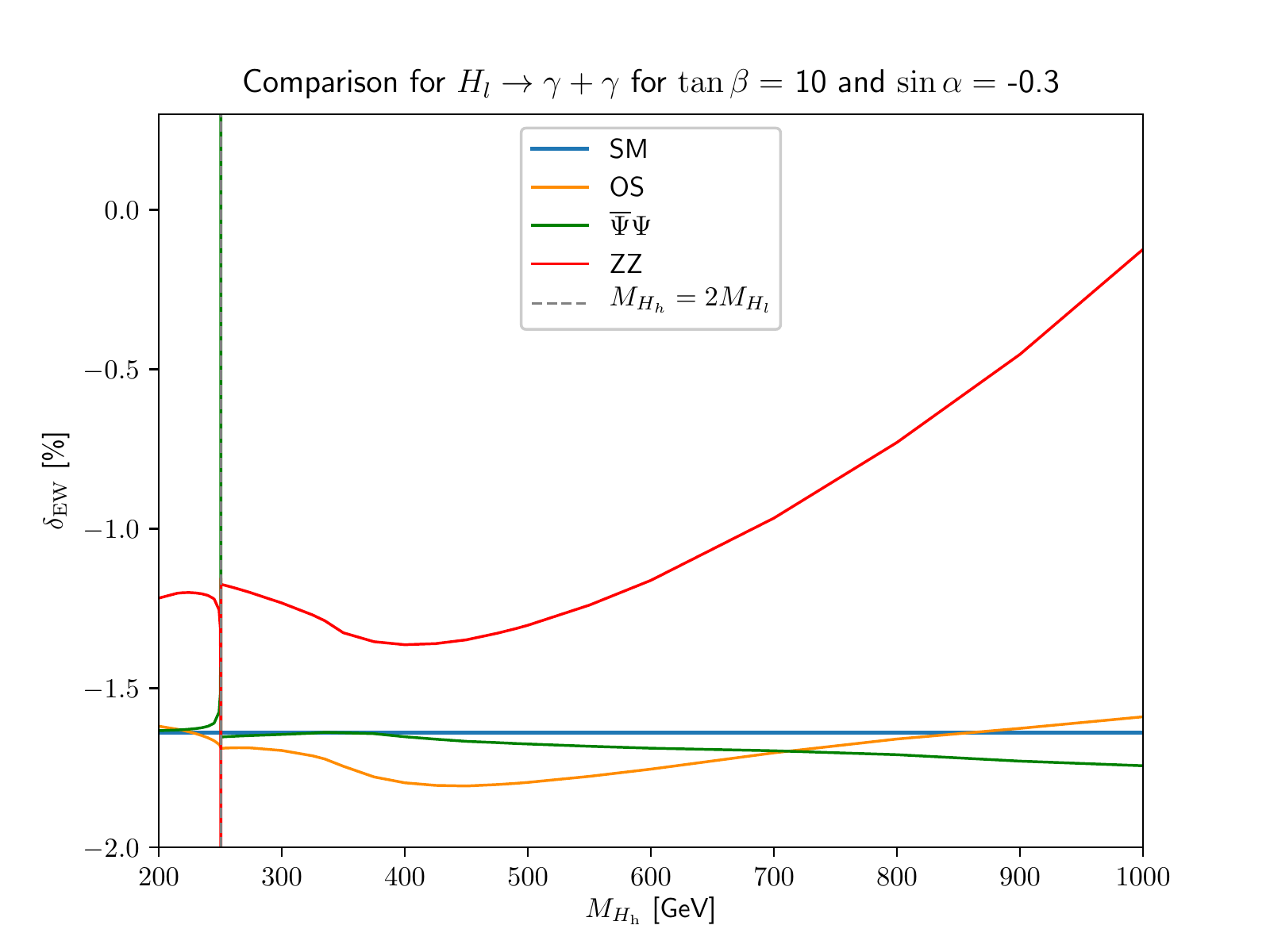}
\end{center}
\caption{Comparison of the different renormalization schemes for the
  mixing angle~$\alpha$ for the decay process $H_l \to \gamma+\gamma$ in the
  \HSESM\ for $\tan \beta = 1,10$ with varying values of $\sin \alpha$.
  The SM limit is shown by the blue line.\vspace{1.0cm}\mbox{}\label{fig:HAA_schemes}}
\end{figure}

\begin{figure}[!ht]
\begin{center}
\includegraphics[width=7.9cm]{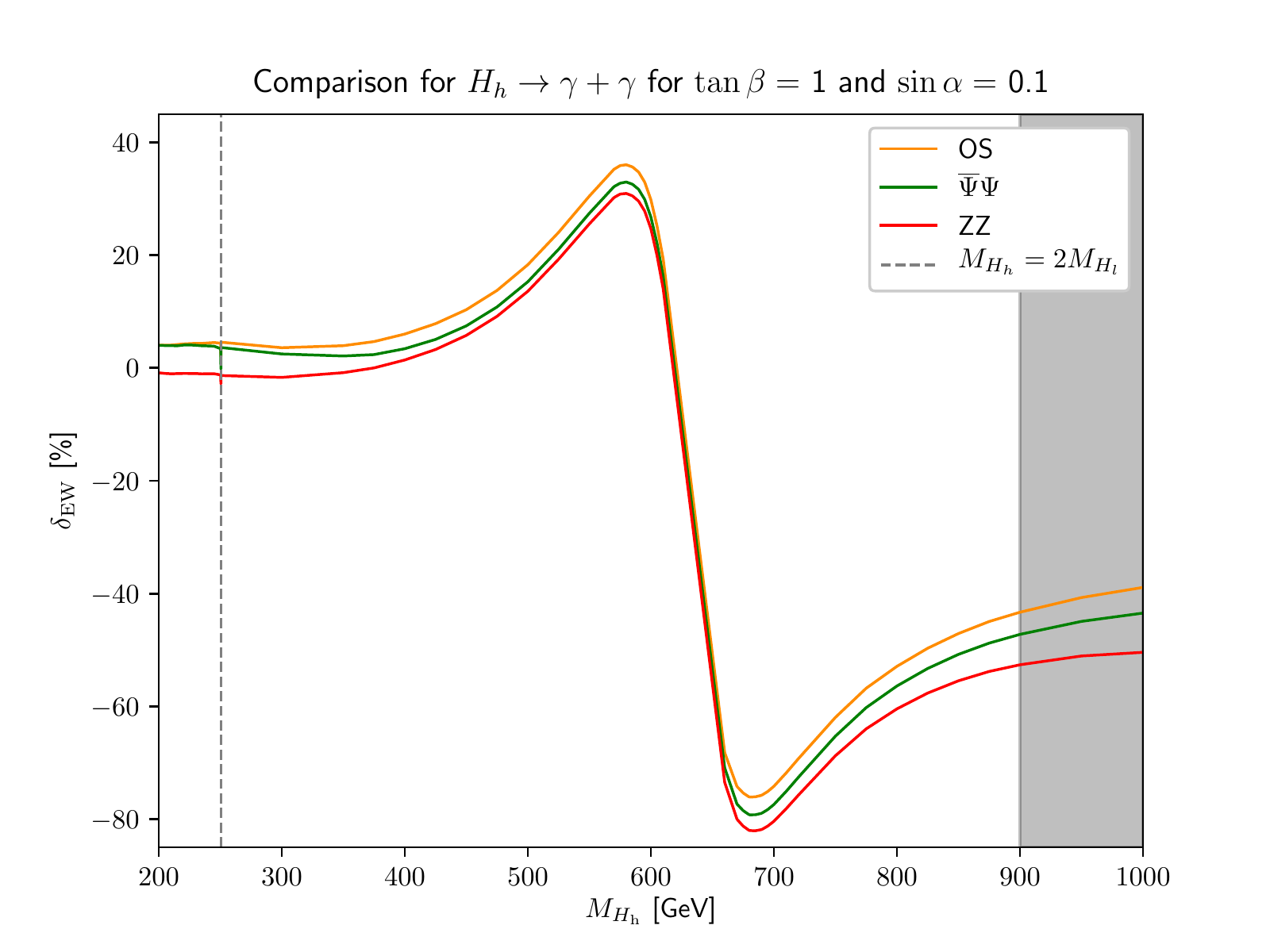}
\includegraphics[width=7.9cm]{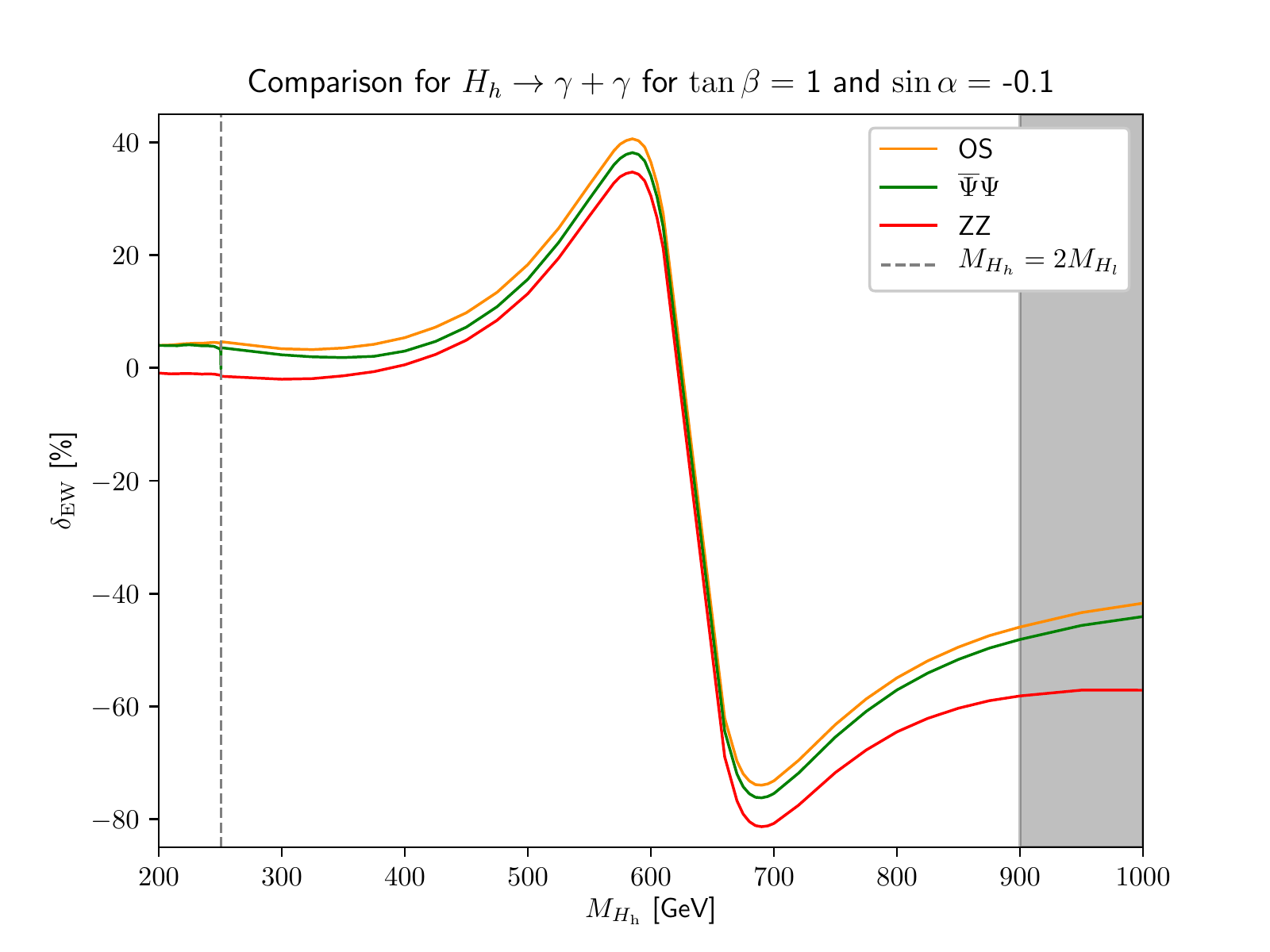}\\
\includegraphics[width=7.9cm]{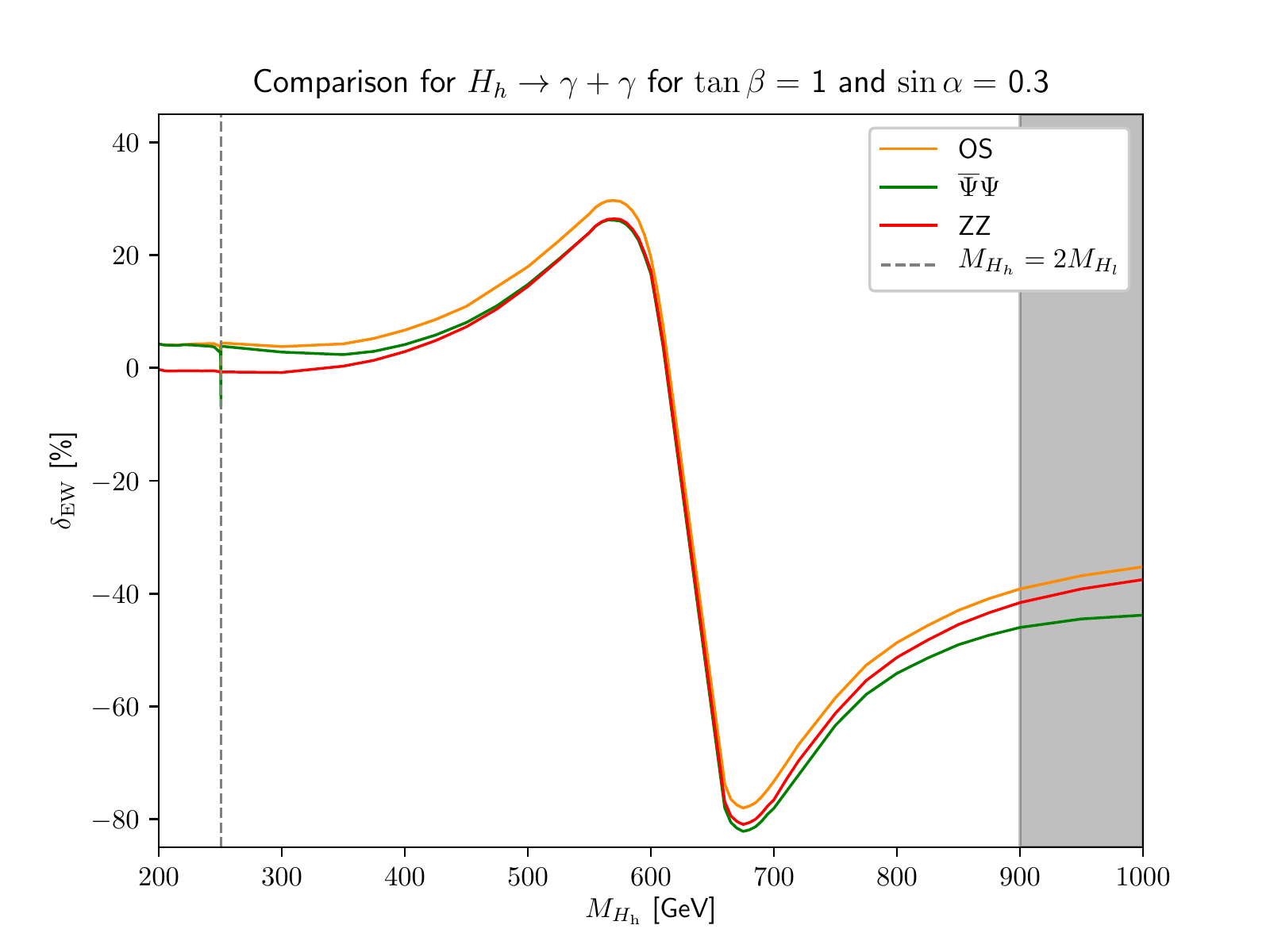}
\includegraphics[width=7.9cm]{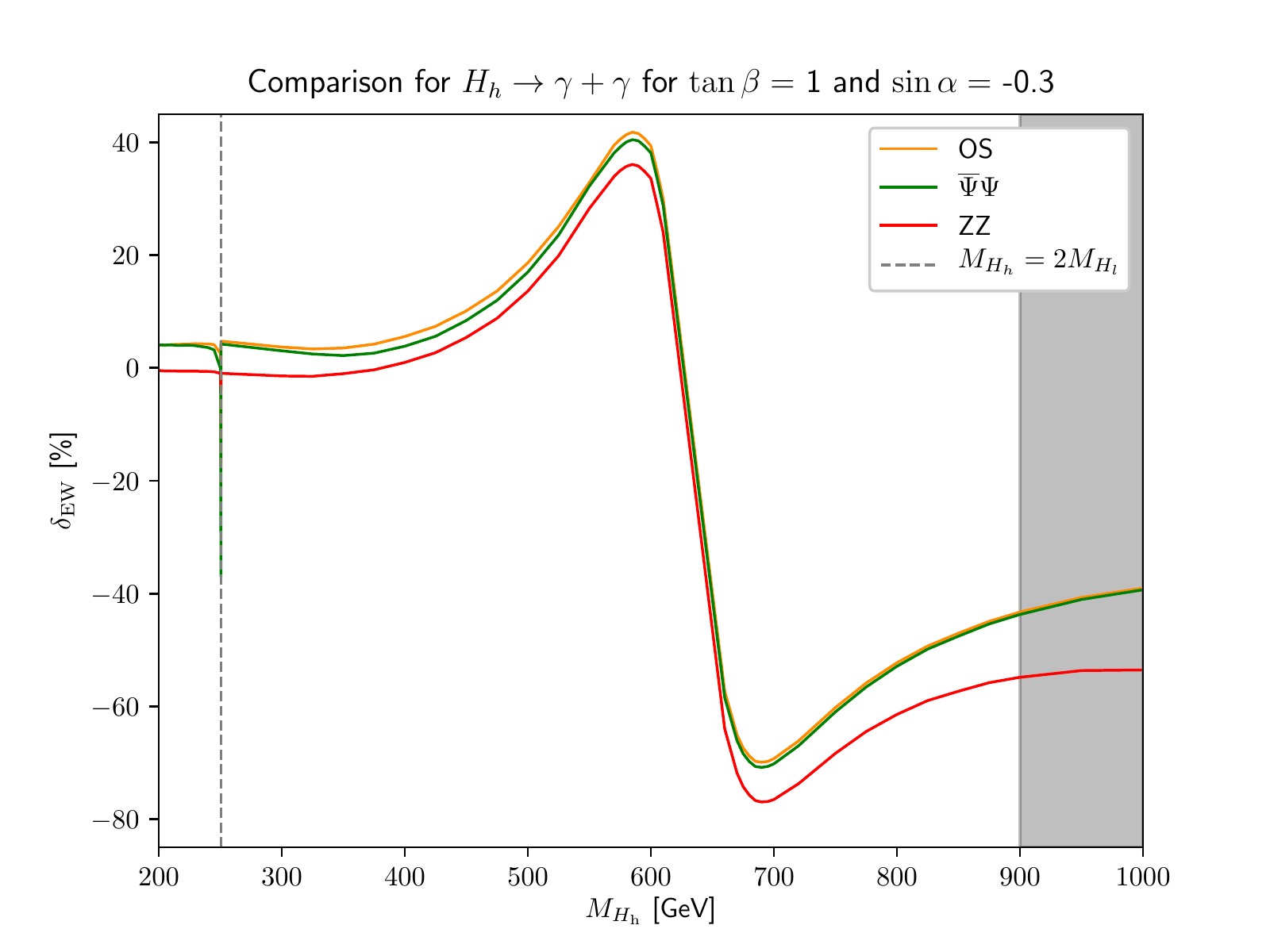}\\
\includegraphics[width=7.9cm]{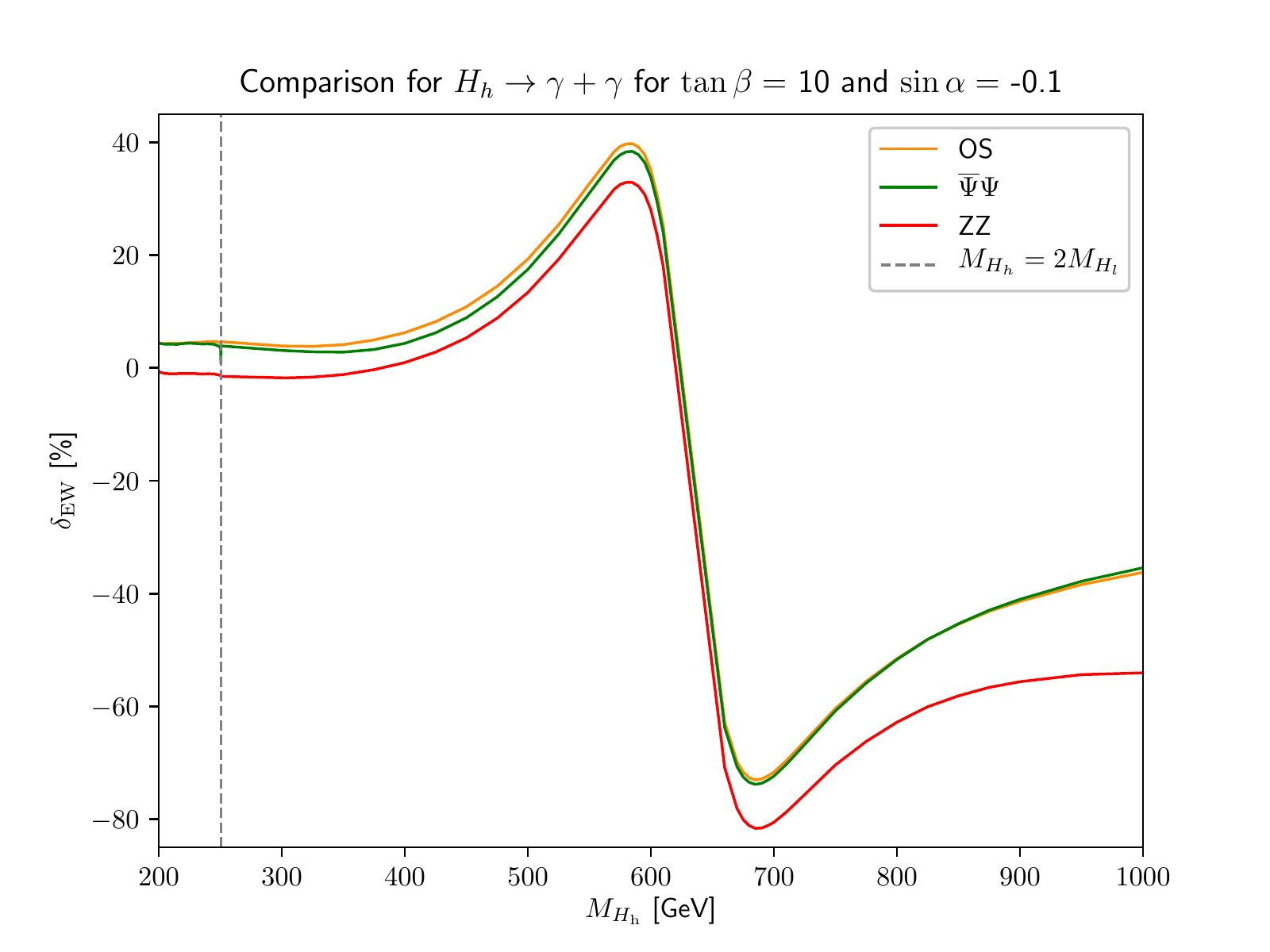}
\includegraphics[width=7.9cm]{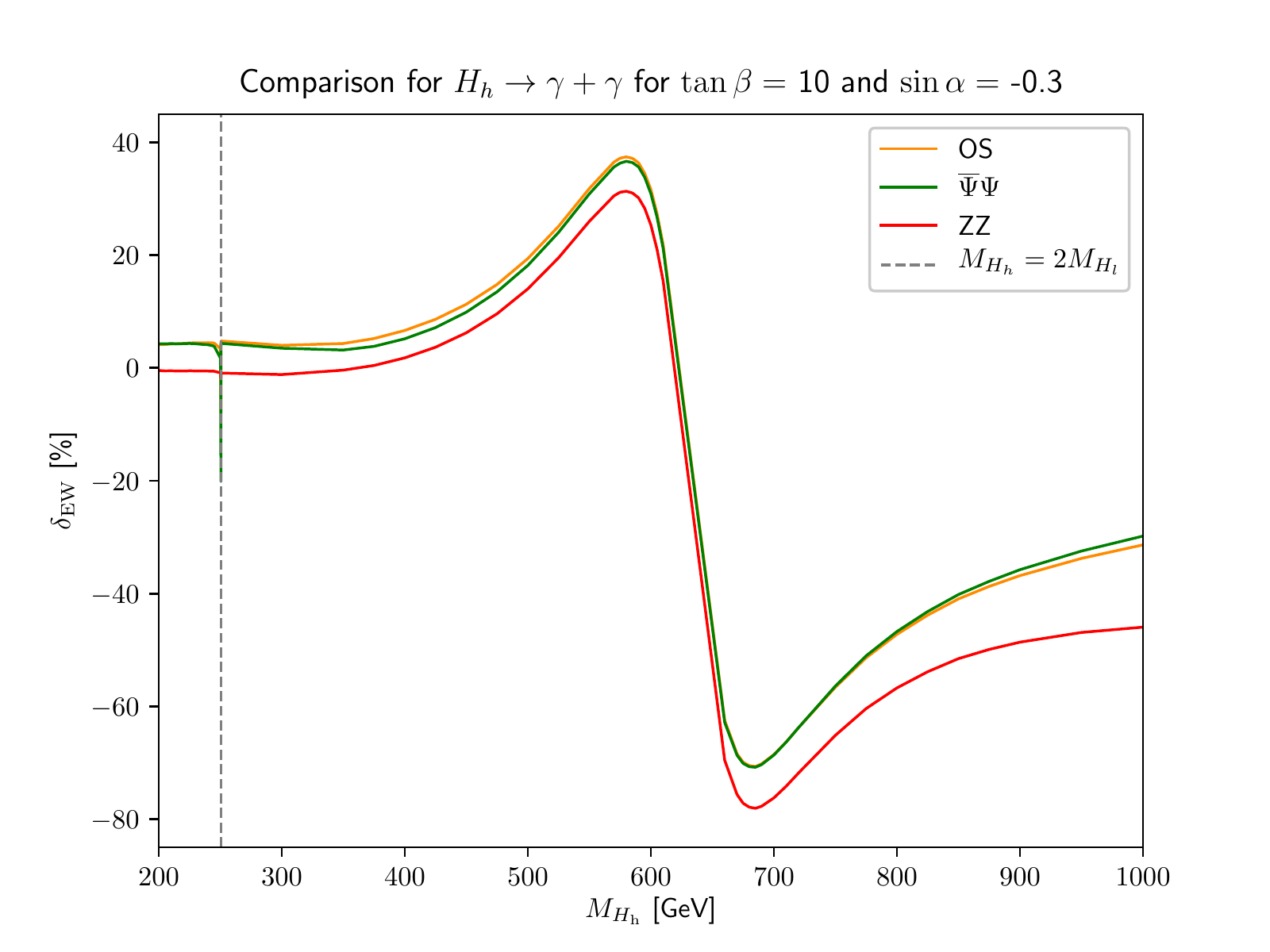}
\end{center}
\caption{Comparison of the different renormalization schemes for the
  mixing angle~$\alpha$ for the decay $H_h \to \gamma+\gamma$ in the \HSESM\ for
  $\tan \beta = 1,10$ with varying values of
  $\sin \alpha$.\vspace{1.5cm}\mbox{}\label{fig:HhAA_schemes}}
\end{figure}

Now we compare again the results for the electroweak percentage
corrections in the three different renormalization schemes for the mixing angle~$\alpha$ under consideration in this
paper for the two decay processes $\hl\to\gamma+\gamma$ and
$\hh\to\gamma+\gamma$.
The reasoning for the choices of $\sin\alpha$ and $\tan\beta$ in
Fig.~\ref{fig:HAA_schemes} and also in Fig.~\ref{fig:HhAA_schemes} is
the same as the one described already in Section~\ref{sec:glugluHHSESM}
for Figs.~\ref{fig:HAA_fixed_tb} and~\ref{fig:HhAA_fixed_tb}.


We start in Fig.~\ref{fig:HAA_schemes} with the NLO electroweak percentage
corrections as a function of the heavy Higgs-boson mass~$M_{\hh}$ for
the process $\hl\to\gamma+\gamma$.
All six plots show again the same
range for the electroweak percentage corrections and the heavy Higgs-boson mass in
order to allow for an easy comparison of the different scenarios.
The comparison plots show features similar to the case of $g+g \to
\hl$. As before, the two process dependent 
$ZZ$- and $\overline{\Psi}\Psi$-schemes contain the artificially
introduced threshold singularities for $M_{\hh}=2M_{\hl}$. For the values $\sin\alpha=\pm0.1$ the three schemes
are hardly distinguishable since we are close to the SM case. However, also for this process the $ZZ$ scheme is
shifted upwards from the SM for all values of $\tan \beta$ and $\sin \alpha$. Its sensitivity to the sign of $\sin \alpha$ is
demonstrated by comparing the plots with $\tan \beta = 1$ and $\sin \alpha =\pm0.3$. In particular,
for $\sin \alpha = -0.3$ we observe again a strong dependence of the $ZZ$ scheme on $\Mhh$. In addition,
the $\overline{\Psi}\Psi$ scheme shows the strongest dependence on $\tan
\beta$. 
Despite the fact that the differences for $\sin\alpha=-0.3$ look
large we want to note that the percentage corrections only cover a small
range of about 1.5\%, so that also the differences between the schemes
are at most of the same size.

In Fig.~\ref{fig:HhAA_schemes} we compare likewise the three different
renormalization schemes for the NLO electroweak percentage corrections 
of the decay $\hh\to\gamma+\gamma$. All three schemes show the
threshold singularity at $M_{\hh}=2M_{\hl}$, with a strong dependence of the
size of the cusp on the scheme. As it was the case for heavy Higgs-boson production in gluon fusion
the lowest value of the electroweak percentage corrections at this threshold is
obtained in the $\overline{\Psi}\Psi$ scheme, whereas barely any cusp is visible 
in the $ZZ$ scheme. Another similarity to the process
$g+g\to \hh$ is the fact that the $ZZ$ 
scheme is shifted downwards compared to the OS scheme. In contrast to all other processes, the $ZZ$
scheme here does not show the strongest $\Mhh$ dependence for negative
values of $\sin \alpha$ and large heavy Higgs-boson mass~$\Mhh$. Indeed
in this mass region the dependence on $\Mhh$ is milder in this case than for positive $\sin
\alpha$. For $\sin \alpha = -0.3$ large differences
between the $ZZ$ scheme and the other two schemes exist and the $ZZ$
scheme shows a weak dependence on the
heavy Higgs-boson mass~$M_{H_{h}}$.  The $\overline{\Psi}\Psi$
scheme again shows the strongest dependence on $\tan \beta$ for
$\sin\alpha=+0.3$ when going from $\tan \beta=1$ to $\tan \beta=10$.
Since for large $\tan\beta$ the corrections are insensitive to the sign of $\sin\alpha$, the corrections
for $\tan\beta=10$ and $\sin\alpha =+0.3$ are comparable to the corrections for $\tan\beta = 10$ and
$\sin\alpha =-0.3$.\\

Among the considered schemes the OS-scheme is the one which does not
introduce artificial threshold singularities for $M_{\hh}=2M_{\hl}$ for
the light Higgs-boson production and decay processes studied in this
paper. The $ZZ$ scheme can in addition lead to large corrections
due to the existence of finite process dependent contributions (first
term of Eq.~(\ref{eq:deltaalphaZZ})). 
As a result of this we consider the results in the OS scheme as
our final results.

\newpage
\section{Summary and conclusion\label{sec:summary}}
We have computed the two-loop electroweak corrections for four
loop-induced processes in the real Higgs-Singlet Extensions of
the Standard Model~(HSESM). For the Higgs-boson production we considered
the light and heavy Higgs-boson production in gluon fusion. 
For the Higgs-boson decay we considered the light and heavy Higgs-boson
decays into two photons. The HSESM has three additional new input parameters
compared to the Standard Model~(SM) which are the new heavy
Higgs-boson mass, the mixing angle~$\alpha$ and the ratio of the two
vacuum expectation values~$\tan\beta$. The latter needs not to be 
renormalized for the above processes, while for the mixing
angle~$\alpha$ we consider three renormalization schemes from literature
among which we choose the on-shell scheme for our final results.

The few new parameters of the HSESM allow us to provide the electroweak
percentage corrections to these processes for scans over a wide range of
the new input parameters. In addition we also provide the results for
benchmark scenarios which have been collected by the Higgs cross section working group.

The electroweak corrections for light Higgs-boson production through
gluon fusion and for the decay into two photons are in both cases very
close to the SM.  The electroweak corrections for heavy Higgs-boson
production are negative and can reach up to about $-10\%$ in the
on-shell scheme for large heavy Higgs-boson masses. The heavy
Higgs-boson decay into two photons shows an interesting feature. The
electroweak percentage corrections change here from large positive
corrections to large negative corrections within about $\pm50$~GeV for
heavy Higgs-boson masses of around 630~GeV.  This feature is inherited
from the leading-order amplitude which becomes very small in this region
explaining thus the large size of the electroweak corrections.

\vspace{2ex}
\noindent
{\bf{Acknowledgments}}\\
We would like to thank Ansgar Denner, Stefan Dittmaier, Jean-Nicolas
Lang and Giampiero Passarino for valuable discussions. 
The work of B.S. and C.S. was supported by the Deutsche
Forschungsgemeinschaft~(DFG) under contract STU 615/2-1.

\cleardoublepage
\begin{appendix}

\clearpage
%
%
%
\providecommand{\href}[2]{#2}\begingroup\raggedright\endgroup
%
%
%

\end{appendix}
\end{document}